\documentclass[aps,prb,reprint,floatfix,longbibliography]{revtex4-2}
\usepackage{amsmath,mathtools}
\usepackage{amssymb}
\usepackage{array}
\usepackage{graphicx}
\usepackage{float}
\usepackage[caption=false]{subfig}
\usepackage{adjustbox}
\usepackage{xcolor}
\usepackage{hyperref}
\usepackage{listings}
\usepackage[T1]{fontenc} 
\usepackage[utf8]{inputenc}
\usepackage[normalem]{ulem}
\def\1{\mathbbm{1}} % central operator
\def\I{\mathrm{i}} % the imaginary i
\def\rvec{{\bf r}}

\def\evec{{\bf e}}

\def\pvec{{\bf p}}

\def\Jvec{{\bf J}}
\def\kvec{{\bf k}}

\def\Avec{{\bf A}}

\def\Bvec{{\bf B}}
\def\ket#1{\bigl|{ #1} \bigr\rangle}

\def\ovlp#1#2{\bigl\langle{ #1}\big|{#2} \bigr\rangle}

\def\kdotp{\kvec\cdot\pvec}

\newcolumntype{P}[1]{>{\centering\arraybackslash}p{#1}}

\begin{document}

\title{Hole spin splitting in a Ge quantum dot with finite barriers}
\author{Jiawei Wang, Xuedong Hu, Herbert F Fotso}
\affiliation{Department of Physics, University at Buffalo, The State University of New York, Buffalo, NY 14260-1500}
\date{\today}
\begin{abstract}
    We study the low-energy spectrum of a single hole confined in a planar Ge quantum dot (QD) within the effective-mass formalism. The QD is sandwiched between two GeSi barriers of finite potential height grown along the [001] direction. To treat this finite barrier problem, we adopt an independent-band approach in dealing with the boundary conditions.  The effects of different system parameters are investigated, including the width of the out-of-plane confining well, the size of the dot, and silicon concentration in the confining layers. 
    The more accurate finite-barrier model results in a non-trivial dependence of the anisotropic $g$-factor on the silicon concentration in the barrier, and the choice of boundary conditions can have a non-negligible impact on its predicted value.
    On the other hand, the corresponding $g$-factor predicted by the hard-wall model only depends on the silicon concentration monotonically. The comparison shows that the hard-wall model falls short in capturing the interplay between strain and band offset.
    Furthermore, while the ideal model of a planar dot with a square-well heterostructure already has an intrinsic spin-orbit coupling, realistic effects arising from the experimental setup may give rise to additional contributions. 
    We investigate the impact from the top-gate electric field and the residual tensile strain on the qubit states. 
    The results indicate that these effects are important contributions to the total spin-orbit coupling, which enables fast electric control.
\end{abstract}

\maketitle

\section{Introduction}
\label{sec:Intro}

Spin qubits in semiconductors, especially those based on electron spin, have attracted sustained attention due to their promise for scalability, backed by the sophisticated semiconductor fabrication technologies ~\cite{1998LossDiVincenzo,1998KaneNat393,2005PettaSci309,2007HansonRMP79,2008HansonNat453,2013ZwanenburgRMP85,2021ChatterjeeNatRevPhys3,2021ScappucciNatRevMat6,2023FangMatQuantTech3}. Group IV semiconductors, such as Si and Ge, are preferred host materials due to the abundance of zero-spin isotopes and possible isotopic enrichment, which suppresses nuclear spin-induced decoherence.  
The electron spins in these systems are also controllable both magnetically, via traditional electron spin resonance, and electrically, via electric dipole spin resonance (EDSR) ~\cite{1965Rashba}.
While conduction electron spins in Si quantum dots (QD) have been shown to be promising ~\cite{2018YonedaNatNanoTech13, 2018ZajacSci359, 2022NoiriNat601, 2022MillsSciAdv8, 2022XueNat601, 2023BurkardRMP95}, one potentially significant challenge in the long term is the need for micromagnets in order to enable fast EDSR because of the relatively weak spin-orbit coupling. This adds to the complexities of device fabrication and multi-qubit control. As an alternative, hole spins have recently emerged as a viable qubit candidate due to the potential for intrinsically fast electrical control and the added benefit that the underlying atomic $p$ orbital helps eliminate the contact hyperfine interaction with nuclear spins ~\cite{2004WoodsPRB69, 2005BulaevPRL95, 2006DanneauPRL97, 2006GvozdicPhysScr2006, 2007BulaevPRL98, 2007HeissPRB76, 2007KlauserSemiQuantBits, 2008FischerPRB78, 2008RoddaroPRL101, 2009HsiehPRB80, 2009ZwanenburgNanoLett9, 2013KloeffelPRB88, 2016WatzingerNanoLett16, 2018KloeffelPRB97, 2018Abadillo-UrielAPL113, 2019MachnikowskiPRB100, 2022WangNatComm13, 2024RussSpinless}. 
There has also been experimental evidence that hole spin qubits could have a relatively long coherence time, especially compared to their fast operations ~\cite{2010FischerPRL105, 2012SzumniakPRL109, 2021KobayashiNatMat20}. Recent experiments have already demonstrated multi-qubit processors and strong coupling to a superconducting resonator based on hole spins ~\cite{2021HendrickxNat591, 2024BorsoiNatNanoTech19, 2023YuNatNanoTech18, 2024ZhangNatNanoTechnol}. 

The effective spin-orbit coupling of holes that enables fast electrical control mostly originates from the strong mixing of valence bands at finite momentum. 
This mixing has strong $k$ dependence and thus is sensitive to system dimensions including the thickness of the heterostructure and the strength of the electrical confinement of the quantum dot. It can also enhance the mixing of quantum dot orbitals and lead to spin splitting beyond the linear regime. Therefore, a comprehensive study of hole spin properties must account for its dependence on system geometry and dimension, as well as the applied electric and magnetic fields.

Theoretical studies of hole spin qubits are usually based on the $\kdotp$ method, as the states of the confined holes are typically dominated by contributions from near the first Brillouin zone center \cite{2010FischerPRL105, 2011KloeffelPRB84, 2021TerrazosPRB103, 2022BoscoPRAppl18, 2021WangNpjQuantInf7, 2023SarkarPRB108, 2024WangPRB109, 2024WangNpjQuantumInf10}.  In planar germanium structures, the four-band Luttinger-Kohn model is considered adequate because the most important multi-band effect is the coupling between heavy holes (HH) and light holes (LH)\cite{1955LuttingerKohn}. The more recently developed Burt-Foreman model \cite{1987BurtSemiSciTech2, 1988BurtSemiSciTech3a, 1988BurtSemiSciTech3b, 1992BurtJoPCondMat4, 1993ForemanPRB48} can further improve the accuracy of the theoretical prediction, as it avoids the use of artificial symmetrization in heterostructures \cite{2005GaleriuThesis}.  Spectral properties of a hole in a QD can then be calculated by expanding the Hamiltonian over single-band single-hole orbitals (the so-called Linear Combination of Atomic Orbitals, or LCAO, approach).  

A convenient assumption in choosing the orthonormal basis orbitals for the LCAO calculation is that the Ge quantum dot is sandwiched between infinite barriers.  
This approximation simplifies the calculations by considering material properties of the confined layer only, neglects the fact that the number of quantized orbitals inside the well is finite, and most conveniently, has the wave function vanish at the two boundaries of the quantum well.
In reality, however, in a Ge$_{1-\alpha}$Si$_{\alpha}$-Ge-Ge$_{1-\alpha}$Si$_{\alpha}$ heterostructure, with a typical 20\% Si in the barrier, the barrier height is only about 132 meV.  Considering that the typical heavy-hole-light-hole (HH-LH) splitting ($\Delta_{\rm HL}$) in a Ge quantum well is several tens of meV, and the confinement energy for light holes along the growth direction is also in that order, the barriers are far from infinite.  If a strong electric field is applied along the growth direction, a confined hole tunneling out of the QD becomes a realistic possibility if the potential barrier is finite, while it is not possible in the case of an infinite barrier no matter how large the electric field is.  With finite barriers, the abrupt change of Luttinger/Foreman parameters at the well-barrier interfaces can couple boundary conditions of different bands \cite{1990BastardHabilitation, 1983AltarelliPRB28, 1987EppengaPRB36}, and spurious solutions are known to appear when solving for the eigenenergies \cite{1987AndreaniPRB36, 2003CartoixaJoAP93}.  In short, to lay a solid foundation for our understanding of hole spin qubits,
a systematic treatment of a Ge QD that adopts realistic parameters and conditions, and accounts for the boundary conditions properly, is still needed.

Here we present a theoretical analysis of the spectra of a single hole in a planar quantum dot based on a Ge$_{1-\alpha}$Si$_{\alpha}$-Ge-Ge$_{1-\alpha}$Si$_{\alpha}$ heterostructure.  We modeled the heterostructure using a finite square well combined with an in-plane harmonic potential. Calculations are done using the LCAO technique with single-band QD orbitals.  Our focus is on the spin splitting of the lowest heavy-hole states, which form the hole spin qubit.  A uniform magnetic field is applied in the in-plane or out-of-plane directions. We study the dependence of the qubit spin splitting on silicon concentration in the barrier material at different dot sizes, as well as quantum well widths and magnetic field directions. We also evaluate the impacts of different treatments of the boundary conditions.  Furthermore, we quantitatively study how the top-gate electric field and the residual strain in the barriers affect the spin splitting.  The analysis presented in this paper represents a reliable starting point for further theoretical studies of hole spin properties.

\section{Model and Method}
\label{sec:theory}

\subsection{System Hamiltonian}
\begin{figure}[ht]%[htbp]
    \centering
    \subfloat{\includegraphics[width=0.4\textwidth]{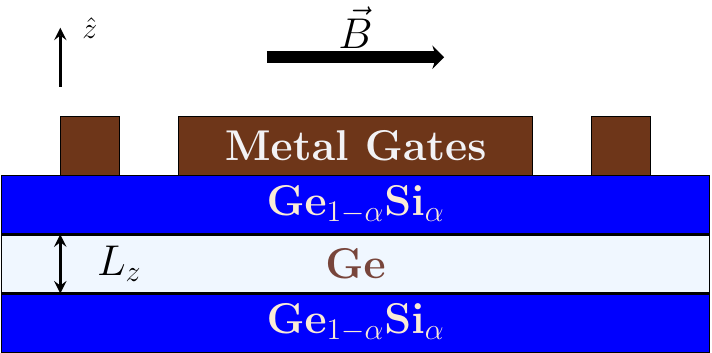}}
    \hfill
    \subfloat{\includegraphics[width=0.2\textwidth]{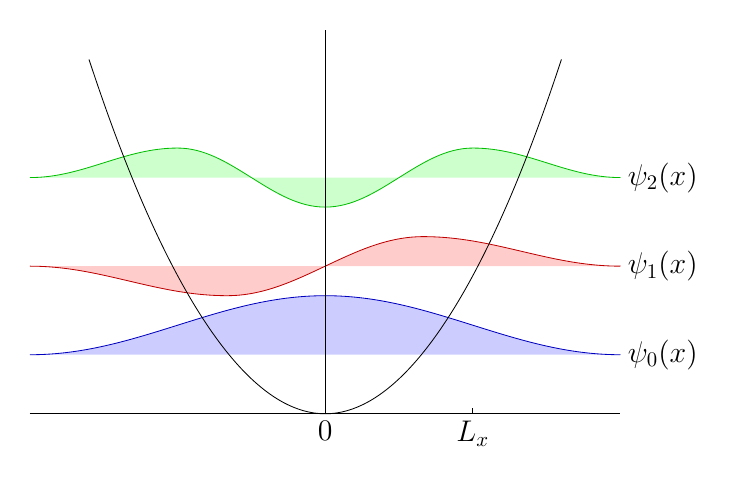}}
    \subfloat{\includegraphics[width=0.2\textwidth]{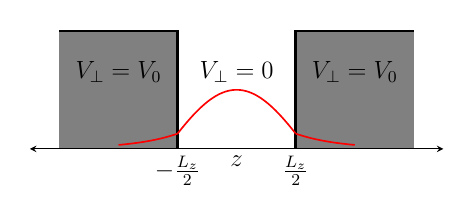}}
    \caption{Schematic figure of a planar Ge quantum dot in an in-plane magnetic field: (upper) Illustration of the heterostructure; (lower left) The in-plane confinement with harmonic potential; (lower right) The out-of-plane confinement is modeled with a finite square well.}
    \label{fig:schematic}
\end{figure}

The Ge planar quantum dot is shown schematically in Fig. \ref{fig:schematic}. Along the growth direction, the strained germanium layer is sandwiched between two silicon-germanium alloys, thus forming a quantum well for holes in the valence band. The top metal gates provide the in-plane confinement for the holes. The total confinement potential is thus separated into the in-plane part and the out-of-plane part.
The total Hamiltonian of our system in a magnetic field is given by
\begin{equation}\label{eq:Htotal}
    H = H_{\rm BF}\left(\kvec + \frac{e \Avec}{\hbar}\right) + E_{\rm V_b}(z) + V_\parallel(x,y) + H_{\rm strain} + H_{\rm Zeeman}\,.
\end{equation}
The gauge used in this study is the symmetric gauge
\begin{equation}\label{eq:symgauge}
    \Avec = \frac{1}{2}(B_y z - B_z y) \evec_x + \frac{1}{2}(B_z x - B_x z) \evec_y + \frac{1}{2}(B_x y - B_y x) \evec_z\,.
\end{equation}

% Burt-Foreman
Confinement in the growth direction determines the quantization axis for the holes. Therefore, the four-band Burt-Foreman Hamiltonian in the $\ket{j,j_z}$ basis with $j=3/2$ takes the form 
\begin{equation}\label{eq:bf4b} %heavy-hole light-hole matrix in bulk case
H_{\rm  BF}(\kvec) = 
\begin{pmatrix} P + Q  & S_- & R & 0\\
S_-^\dagger & P - Q & C & R \\
R^\dagger & C^\dagger  & P - Q & -S_+\\
0 & R^\dagger & -S_+^\dagger & P + Q \end{pmatrix}\,.
\end{equation}
Conventionally, heavy holes (HH) refer to the two $j_z=|3/2|$ states, while light holes (LH) refer to the two $j_z=|1/2|$ states. The matrix elements are defined by:
\begin{eqnarray}
    P &=& \frac{\hbar^2\gamma_1}{2m_0}(k_x^2+k_y^2+k_z^2)\,,\nonumber\\
    Q &=& \frac{\hbar^2\gamma_2}{2m_0} (k_x^2+k_y^2-2k_z^2)\,,\nonumber\\
    R &=& -\frac{\sqrt3\hbar^2}{2m_0}(\bar \gamma k_-^2 - \mu k_+^2)\,,\label{eq:bfelements}\\
    S_\pm &=& -\frac{\sqrt3\hbar^2}{m_0}[k_\pm(\sigma-\delta)k_z + k_z\pi k_\pm]\,,\nonumber\\
    C &=& -\frac{\hbar^2}{m_0}[k_-(\sigma-\delta-\pi)k_z - k_z(\sigma-\delta-\pi)k_-]\,,\nonumber
\end{eqnarray}
where $k_\pm=k_x\pm \I k_y$, and $\bar \gamma$, $\mu$, $\sigma$, $\pi$, $\delta$ are Foreman parameters obtained from linear combinations of Luttinger parameters $\gamma_1$, $\gamma_2$, $\gamma_3$, as shown in Appendix \ref{app:parameters}.  Matrix element $C$, which does not appear in the Luttinger-Kohn Hamiltonian, is the LH-LH coupling originating from the exact envelope-function treatment. It does not affect the bulk dispersion of a material, but may have a non-negligible contribution if boundary conditions introduce coupling between different bands \cite{1993ForemanPRB48}. Nevertheless, this LH-LH coupling may be negligible in a realistic strained heterostructure. 
There is a sign difference between \eqref{eq:bf4b} and the original Foreman representation. It arises from the choice of different zone center states as shown in Appendix \ref{app:signBF}.

% Strain
In our system, the germanium layer is sandwiched between two layers of relaxed silicon-germanium alloy in the out-of-plane direction. The germanium lattice is thus uniaxially strained, compressed in the in-plane directions by the relaxed Ge$_{1-\alpha}$Si$_{\alpha}$ due to the lattice-matched growth process, and stretched in the growth direction.  Consequently, there are finite strain contributions to the diagonal matrix elements of $H_{\rm BF}$ via the Bir-Pikus Hamiltonian \cite{1974BirPikus,2009kpMethod}
\begin{equation}\label{eq:BP4b} %heavy-hole light-hole matrix in bulk case
H_{\rm strain}(\kvec) = 
\begin{pmatrix} P_\varepsilon + Q_\varepsilon  & S_{\varepsilon} & R_\varepsilon & 0\\
S_{\varepsilon}^\dagger & P_\varepsilon - Q_\varepsilon & 0 & R_\varepsilon \\
R_\varepsilon^\dagger & 0  & P_\varepsilon - Q_\varepsilon & -S_{\varepsilon}\\
0 & R_\varepsilon^\dagger & -S_{\varepsilon}^\dagger & P_\varepsilon + Q_\varepsilon \end{pmatrix}\,,
\end{equation}
where in our case,
\begin{eqnarray}
    P_\varepsilon = -a_v(\varepsilon_{xx}+\varepsilon_{yy}+\varepsilon_{zz})\,,\nonumber\\
    Q_\varepsilon = -\frac{b_v}{2}(\varepsilon_{xx}+\varepsilon_{yy}-2\varepsilon_{zz})\,,
\end{eqnarray}
with other matrix elements vanishing. Here, $a_v$ is the hydrostatic deformation potential, $b_v$ is the uniaxial deformation potential, and $\varepsilon_{ij}$ are components of the strain tensor. 
The strain profile in such a Ge quantum well increases the splitting between the HH and LH bands at the $\Gamma$ point. 

% Out-of-plane
In the envelope-function approximation, the out-of-plane confinement is provided by the valence-band offset between the strained Ge layer and the relaxed GeSi alloy, together with the uniaxial strain contribution, and is modeled by a one-dimensional finite square well.  The band offset contribution takes the form
\begin{equation}\label{eq:qw}
    E_{\rm V_b}(z) = \begin{cases}
        0\,,\quad & |z|\le\frac{L_z}{2}\,\\
        V_0\,,\quad & |z|>\frac{L_z}{2}\,
    \end{cases}\,,
\end{equation}
where $L_z$ is the well width. The barrier height $V_0$ is given in Appendix \ref{app:parameters}.
\begin{figure}[ht]%[htbp]
    \centering
    \includegraphics[width=0.4\textwidth]{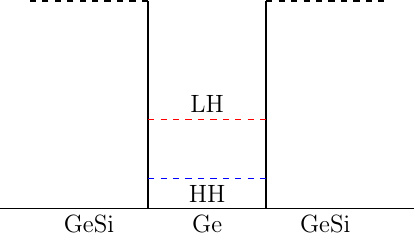}
    \caption{Band alignment of the Ge/GeSi heterostructure with a combination of the valence-band offset and the unixial strain by the lattice matching mechanism.}
    \label{fig:paradigm1}
\end{figure}
Such a band offset, combined with the uniaxial strain, will give an effective barrier height
\begin{equation}
    V_{\rm eff}(z) = E_{\rm V_b}(z) + H_{\rm strain}\,.
\end{equation}
Notice that the barrier height is different for HH and LH bands, as shown in Fig.~\ref{fig:paradigm1}.

% In-plane
The in-plane confinement is modeled as a two-dimensional harmonic potential,
\begin{equation}\label{eq:ip}
    V_\parallel(x,y) = \frac{1}{2} m_\parallel(\omega_x^2 x^2 + \omega_y^2 y^2)\,.
\end{equation}
The harmonic potential is a good approximation for the region around the minimum form by the confinement gates, which contains enough states for our LCAO calculations \cite{1996TaruchaPRL77}.
Here $m_\parallel$ refers to the in-plane effective mass of $j_z=|3/2|$ states $m_\parallel=m_0/(\gamma_1+\gamma_2)$. The in-plane confinement is often described in terms of effective dot sizes $L_x$ and $L_y$, where $L_{x,y}=\sqrt{\hbar/(m_\parallel\omega_{x,y})}$. In this work, we focus on the case of circular confinement $L_x=L_y$.

Since the in-plane confinement is defined by the electrostatic potential produced by the top gates, it is the same for HH and LH states. Consequently, the effective dot sizes are different for HH and LH states because of their different in-plane masses.  The different length scales mean that HH and LH orbitals are generally not orthogonal to each other, complicating the calculation of HH-LH coupling matrix elements, as shown in Appendix \ref{app:cpinplane}.

Lastly, the Zeeman Hamiltonian for $j=3/2$ states is
\begin{equation}
    H_{\rm Zeeman} = 2\kappa\mu_B\Bvec\cdot\Jvec + 2q\mu_B\Bvec\cdot\mathcal{J}\,,
\end{equation}
where $\Jvec = (J_x,J_y,J_z)$ is the total angular momentum with matrix representations \eqref{eq:Jmatrices} and $\mathcal{J} = (J_x^3, J_y^3, J_z^3)$,
$\mu_B$ is the Bohr magneton ($\approx57.88$ ${\rm \mu eV\cdot T^{-1}}$), while the isotropic and anisotropic $g$-factors $\kappa$ and $q$ are material-dependent.

\subsection{Boundary conditions}
\label{ssec:boundary}

One of the most challenging aspects of studying hole properties in a finite-barrier quantum well is to determine the appropriate boundary conditions.  In the envelope-function approximation, the total wave function takes the form
\begin{equation}
    \Psi(x,y,z) = \sum_l \psi_l(x,y)\phi_l(z)u_{l0}(\rvec)\,,
\end{equation}
where $\psi_l(x,y)$ and $\phi_l(z)$ are the in-plane and out-of-plane components of the envelope function in band $l$, and $u_{l0}$ is the corresponding Bloch function at the zone center.
In general, the total wave function and its first derivative in the growth direction should be continuous at the boundaries of a finite well,
\begin{eqnarray}
    \Psi(x,y,z_{0+}) &=& \Psi(x,y,z_{0-})\,,\\
    \partial_z \Psi(x,y,z)|_{z_{0+}} &=& \partial_z \Psi(x,y,z)|_{z_{0-}}\,,
\end{eqnarray}
where $z_0$ refers to either of the edges of the quantum well $z_0 = \pm L_z/2$.
The second boundary condition is usually in the form of continuity of probability current in the $z$ direction
\begin{equation}
    j(z) = \int \int \Psi^* \partial_z \Psi dx dy\,.
\end{equation}

Heterostructures, such as the Ge/SiGe system we study here, are built from layers of materials with similar properties.  Therefore, the $\kdotp$ formalism assumes that the $\Gamma$-point Bloch functions remain the same in different layers. Such an assumption leads to the continuity of the envelope functions in each band
\begin{equation}\label{eq:1stbc}
    \phi_w(z_0) = \phi_b(z_0)\,,
\end{equation}
where the subscripts $w$ and $b$ refer to ``well'' and ``barrier''.
On the other hand, in the $\kdotp$ formalism, the second boundary condition is subtle due to considerations regarding the effective mass in different materials. Below, we consider two scenarios.

\subsubsection{One-band model}
\label{sssec:1bbc}

Conduction electrons in semiconductors with weak spin-orbit coupling are usually described with a one-band model. In the case of holes in a quantum dot, the one-band model is also the usual choice for basis states in the LCAO calculations. The second boundary condition within the one-band model is the BenDaniel-Duke type \cite{1966BenDanielDukePR152}:
\begin{equation}\label{eq:1bbc}
    \left.\frac{1}{m_w}\frac{\partial \phi_w}{\partial z}\right|_{z_{0+}} = \left.\frac{1}{m_b}\frac{\partial \phi_b}{\partial z}\right|_{z_{0-}}\,.
\end{equation}
For a finite square well, the $z$ component of the envelope function is a sinusoidal wave (with wave vector $k_z$) in the well and an evanescent wave (with inverse decay length scale $\chi_z$) in the barrier. The Schr\"{o}dinger equation in different layers then leads to the relationship
\begin{equation}\label{eq:Energyeq}
    \frac{\hbar^2 k_z^2}{2m_w} = -\frac{\hbar^2 \chi_z^2}{2m_b} + V_{\rm eff}(z)\,.
\end{equation}
Combining all the equations, we can solve the problem similar to the textbook finite square well problem. 

In general, material properties are different across the layers in a heterostructure, which means all Luttinger/Foreman parameters as well as material $g$-factors in the Zeeman term should be layer-dependent. In order to be consistent with the abrupt-barrier model \eqref{eq:qw}, material parameters should be step functions at each boundary. This sudden change should be relatively small considering that layers in a heterostructure have similar compositions and are lattice-matched. Therefore, the lowest-order approximation here is to neglect the small changes, which is also consistent with the assumption that the zone center Bloch states are the same across the boundaries. 
Considering the zone-center Bloch states are the same across the boundaries within the effective-mass approximation, it is reasonable to neglect the small changes in material parameters and take them as constants across the boundaries \cite{2004WoodsPRB69,2024WangNpjQuantumInf10}.

Alternatively, if one ignores the same-Bloch-state assumption and emphasizes the fact that materials in different layers are different (here it is Ge vs Ge$_{1-\alpha}$Si$_\alpha$), it also seems reasonable to assume a change of effective mass along the growth direction.
This assumption leads to modifications in the quantized solutions such as the effective wave vector and energy since the second boundary condition Eq.~\eqref{eq:1bbc} is now different.

\subsubsection{Multi-band model: the coupled boundary conditions}
\label{sssec:mbbc}

One-band model treats the boundary conditions at the lowest-order approximation, though such an independent-band assumption is questionable.  Instead, the four-band Burt-Foreman Hamiltonian is the minimal model for holes in a quantum dot due to the strong HH-LH coupling away from the Brillouin zone center.  The boundary conditions for a confined hole are thus in general in the matrix form.  Specifically, assuming that Luttinger/Foreman parameters undergo sudden changes at a boundary, the strong HH-LH coupling in $H_{\rm BF}$ would then lead to coupled boundary conditions in the four-band description of the holes. In other words, the continuity of probability current becomes a matrix equation
\begin{equation}\label{eq:mbbc}
    \mathcal{D}_w \ket{\phi_w} = \mathcal{D}_b \ket{\phi_b}\,,
\end{equation}
where the solution
\begin{equation}
    \ket{\phi} = \biggl( \phi^{\scriptscriptstyle{3/2}}(z) \quad  \phi^{\scriptscriptstyle{1/2}}(z) \quad \phi^{\scriptscriptstyle{-1/2}}(z)  \quad \phi^{\scriptscriptstyle{-3/2}}(z) \biggr)^T\,,
\end{equation}
contains envelope functions from all four bands.  The derivative operator in the Luttinger-Kohn model or in the Burt-Foreman model is \cite{1987AndreaniPRB36,1993ForemanPRB48}
\begin{widetext}
\begin{equation}\label{eq:D_4band}
\begin{split}
    \mathcal{D}_{\rm LK} &= \begin{pmatrix}
        (\gamma_1 - 2\gamma_2)\frac{\partial }{\partial z} &  -\sqrt3\I \gamma_3 k_- & 0 & 0 \\
        \sqrt3\I \gamma_3 k_+ & (\gamma_1 + 2\gamma_2)\frac{\partial }{\partial z} & 0 & 0 \\
        0 & 0 & (\gamma_1 + 2\gamma_2)\frac{\partial }{\partial z} & \sqrt3\I \gamma_3 k_- \\
        0 & 0 & -\sqrt3\I \gamma_3 k_+ & (\gamma_1 - 2\gamma_2)\frac{\partial }{\partial z}
    \end{pmatrix}\\
    \mathcal{D}_{\rm BF} &= \begin{pmatrix}
        (\gamma_1 - 2\gamma_2)\frac{\partial }{\partial z} &  -2\sqrt3\I\pi k_-  & 0 & 0 \\
        2\sqrt3\I (\sigma-\delta) k_+ & (\gamma_1 + 2\gamma_2)\frac{\partial }{\partial z} & 2\I(\sigma-\delta-\pi)k_- & 0 \\
        0 & -2\I(\sigma-\delta-\pi)k_+ & (\gamma_1 + 2\gamma_2)\frac{\partial }{\partial z} & 2\sqrt3\I \pi k_- \\
        0 & 0 & -2\sqrt3\I (\sigma-\delta) k_+ & (\gamma_1 - 2\gamma_2)\frac{\partial }{\partial z}
    \end{pmatrix}
\end{split}
\end{equation}
\end{widetext}
where the Luttinger parameters $\gamma_1$, $\gamma_2$, and $\gamma_3$ are different across the boundary
\begin{equation}
    \gamma_i(z) = \begin{cases}
        \gamma_{iw}\,, &|z|\le \frac{L_z}{2}\\
        \gamma_{ib}\,, &|z|> \frac{L_z}{2}
    \end{cases}\,,
\end{equation}
The Foreman parameters $\sigma$, $\delta$, $\pi$ are similarly defined.

Solving such an eigenvalue problem becomes highly non-trivial. Early studies include calculations of band structures in superlattices \cite{1981BastardPRB24, 1983AltarelliPRB28} and quantum wells \cite{1984EkenbergPRB30,1985SchuurmansPRB31,1985BroidoPRB31,1987EppengaPRB36,1987AndreaniPRB36, 1996WinklerPRB53}. However, the eigenvalue techniques adopted in these studies are not straightforwardly applicable in the current work: in quantum dots the in-plane wave vectors $k_\pm$ become operators. In this case, one has to solve a large set of coupled differential equations, and many spurious solutions may appear even if the problem is solvable \cite{1987AndreaniPRB36, 2003CartoixaJoAP93}.

In appendix \ref{app:cpsubbands}, we propose a possible treatment to modify the single-band basis states of the quantum well in an effort to construct a total envelope function that satisfies the coupled boundary conditions.  Interestingly, we find that the coupled boundary conditions do not modify the single-band basis states significantly from the uncoupled ones.  As such, our current LCAO calculations are still constructed upon the basis states from a one-band model.

\subsection{Numerical Method}
The four-band problem for a single hole confined in a quantum dot can only be solved numerically. We first solve the in-plane and the out-of-plane effective mass states individually in the corresponding confining potentials, obtaining the single-band basis states as products of the two. We then construct the matrix representation of the total Hamiltonian over these single-band orbitals, and diagonalize the Hamiltonian to obtain its energy spectrum.

The out-of-plane states are solutions of the finite square well. Calculations are done self-consistently by considering all the bound states in the quantum well. Therefore, the number of quantum-well sub-bands depends on the system setup. For example, for a typical silicon concentration of $\alpha=0.2$,the effective barrier heights are 142 meV for HH and 81 meV for LH. In this situation, a 14-nm-wide quantum well has 5 bound states for each HH band and 2 bound states for each LH band.
We have tested that the highest two HH states can only affect the results up to 0.2\%. We have also tested the convergence by manually lifting the barrier height by 100 meV. The results show that the change of $k_z$ by the lifted barriers contributes the most to the numerical difference, and the higher orbitals have a very small effect.

The in-plane states are solutions of a 2D harmonic oscillator if an in-plane $\Bvec$ field is applied, or Fock-Darwin states if the field is along the $z$ direction. The number of in-plane orbitals is 15 for each band in our calculations. This number of orbitals can already provide satisfactory convergence: adding more orbitals will not significantly change the result.

\section{Spectral studies of a hole spin}
\label{sec:result}

In this section, we discuss our numerical analysis of the spin spectra of the ground heavy-hole state in a single-hole Ge planar quantum dot.  Our main objective is to obtain hole spectral properties with a realistic treatment of the quantum dot confinement, particularly along the growth direction.  By exploring such an effective finite well, we can relate hole properties to material properties and experimental conditions such as the content of Si in the GeSi barrier (which directly relates to the height of the finite barrier), the degree of strain in the Ge well, the width of the quantum well, and the applied electric and magnetic fields.

As discussed in the previous section, there are certain ambiguities with respect to the finite well boundary condition, such as whether we should assume that the effective mass changes across the boundary or not.  Consequently, we perform our calculations under both conditions.  Results obtained with a sudden change of effective masses are labeled as ``shifted-$m^*$'', while results obtained neglecting this sudden change are labeled as ``fixed-$m^*$''. The ultimate assessment of these two approaches can only be given experimentally.

In all the calculations, material properties in the Ge$_{1-\alpha}$Si$_{\alpha}$ alloy are based on linear interpolation of parameters in Si and Ge. In general, the larger $\alpha$ is (the larger concentration of Si), the higher the barrier.  On the other hand, larger $\alpha$ also means larger lattice mismatch between the alloy (which is relaxed) and the Ge quantum well.  To have a sufficiently high barrier to confine the holes, while also allowing high-quality lattice-matched growth of strained Ge quantum well, we limit our calculations to a range of $\alpha$ from 0.15 to 0.25. This provides a reasonable barrier height, and the alloy is germanium-like. All the other relevant parameters used in this paper can be found in Appendix \ref{app:parameters}.

\subsection{Energy spectra in an applied $\Bvec_\parallel$ field}
\label{ssec:Bip}

\begin{figure}[ht] %[htbp]
\centering
\includegraphics*[width=8.0cm]{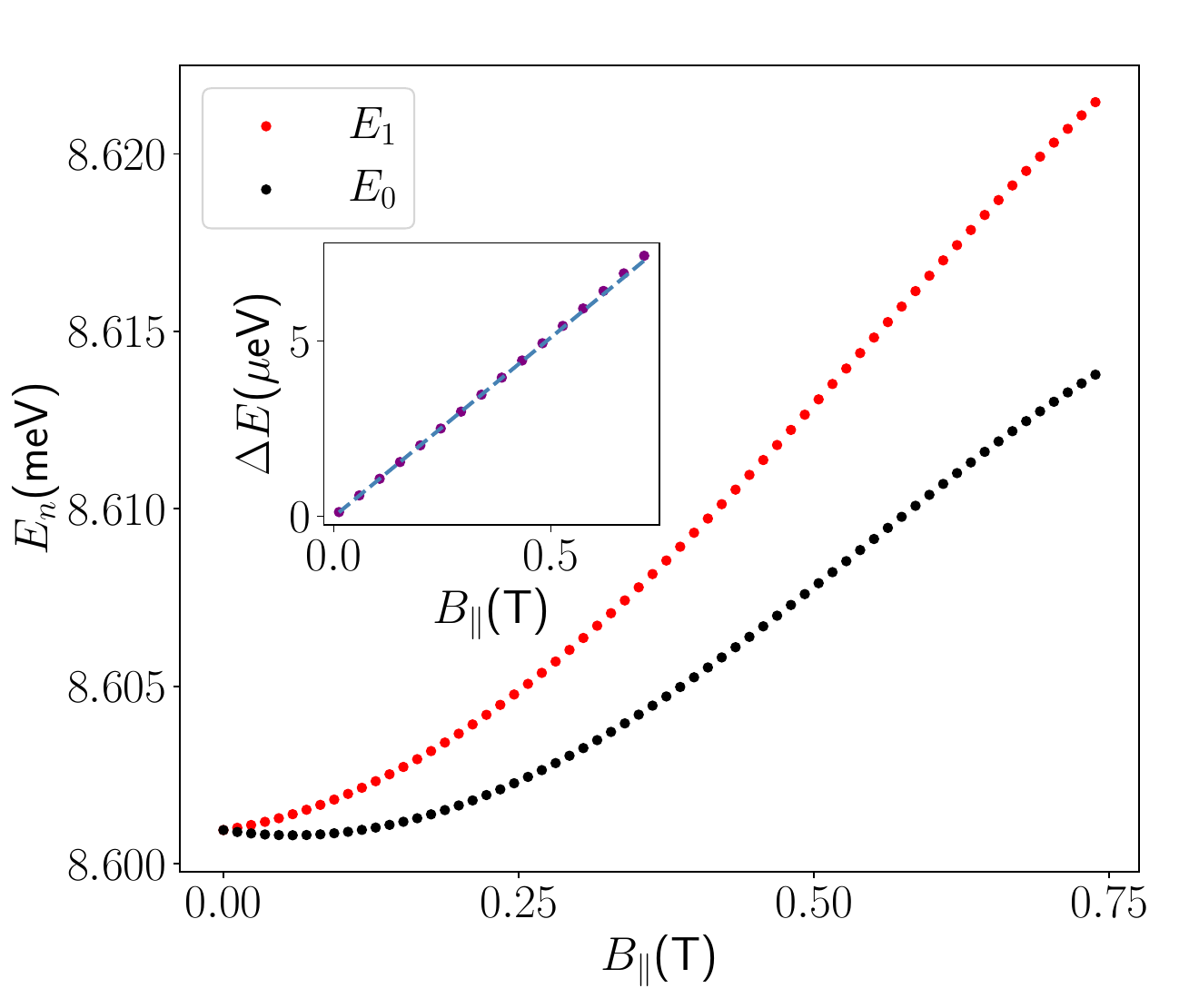}
\caption{Energy spectra of the lowest two spin states with an applied field in the [110] directions. The quantum dot is 40 nm in radius with a germanium layer of 13 nm in thickness and 20\% of silicon in the barriers. The inset shows the corresponding energy difference (purple dotted) as well as the linear fit (blue dashed).}
\label{fig:DEvsB}
\end{figure}

We first calculate the energy spectra of the planar quantum dot in a $\Bvec_\parallel$ field.  The lowest two spin states are shown in Fig.~\ref{fig:DEvsB}.
The inset of Fig.~\ref{fig:DEvsB} is the qubit Zeeman splitting of these lowest two levels as a function of the magnetic field strength. The purple dots are the numerical output, while the blue dashed line shows the corresponding linear trend in the weak-field limit. The spin splitting maintains a good linearity up to at least $0.75$ T.

The total Hamiltonian of the quantum dot \eqref{eq:Htotal} can be written as a series expansion in the $\Bvec$ field \cite{2018VenitucciPRB98}, spin splitting is in general nonlinear in the $\Bvec$ field. For conduction electrons in quantum dots, the nonlinear contributions are too small to make a meaningful contribution in the operating regime of spin qubits. However, the situation appears to be different for holes. The strong HH-LH coupling can give rise to much stronger nonlinear-in-$B$ contributions than in the case of conduction electrons, because a typical $\Delta_{\rm HL}$ is more than an order of magnitude smaller than the band gap in germanium.  Nevertheless, we do not observe any nonlinear response to the $B$ field in our system, at least in the current field range.  Mathematically, the nonlinear contributions originate from orbital mixtures of the HH and the LH sub-bands, which depend inversely on $\Delta_{\rm HL}$. With the strain profile of Fig.~\ref{fig:paradigm1} for the heterostructure we study, $\Delta_{\rm HL}$ is large, thus the nonlinear contributions are almost negligible.  One may expect a dramatic change of results if $\Delta_{\rm HL}$ is much smaller.

\subsection{Effect of quantum dot confinement}
\label{ssec:DEip}

Considering the strength of spin-orbit coupling for holes, it is natural to expect that the confinement potential for the quantum dot will influence the hole spin splitting.  The confinement potential for a planar quantum dot is described by three main parameters: the well width $L_z$ along the growth direction, the silicon concentration $\alpha$ in the barrier, which determines the barrier height, and the in-plane dot size $L_{x,y}$ for the harmonic well.  In this section, we explore these dependencies.

\begin{figure}[htbp] %[H]
    \centering
    \includegraphics[width=0.5\textwidth]{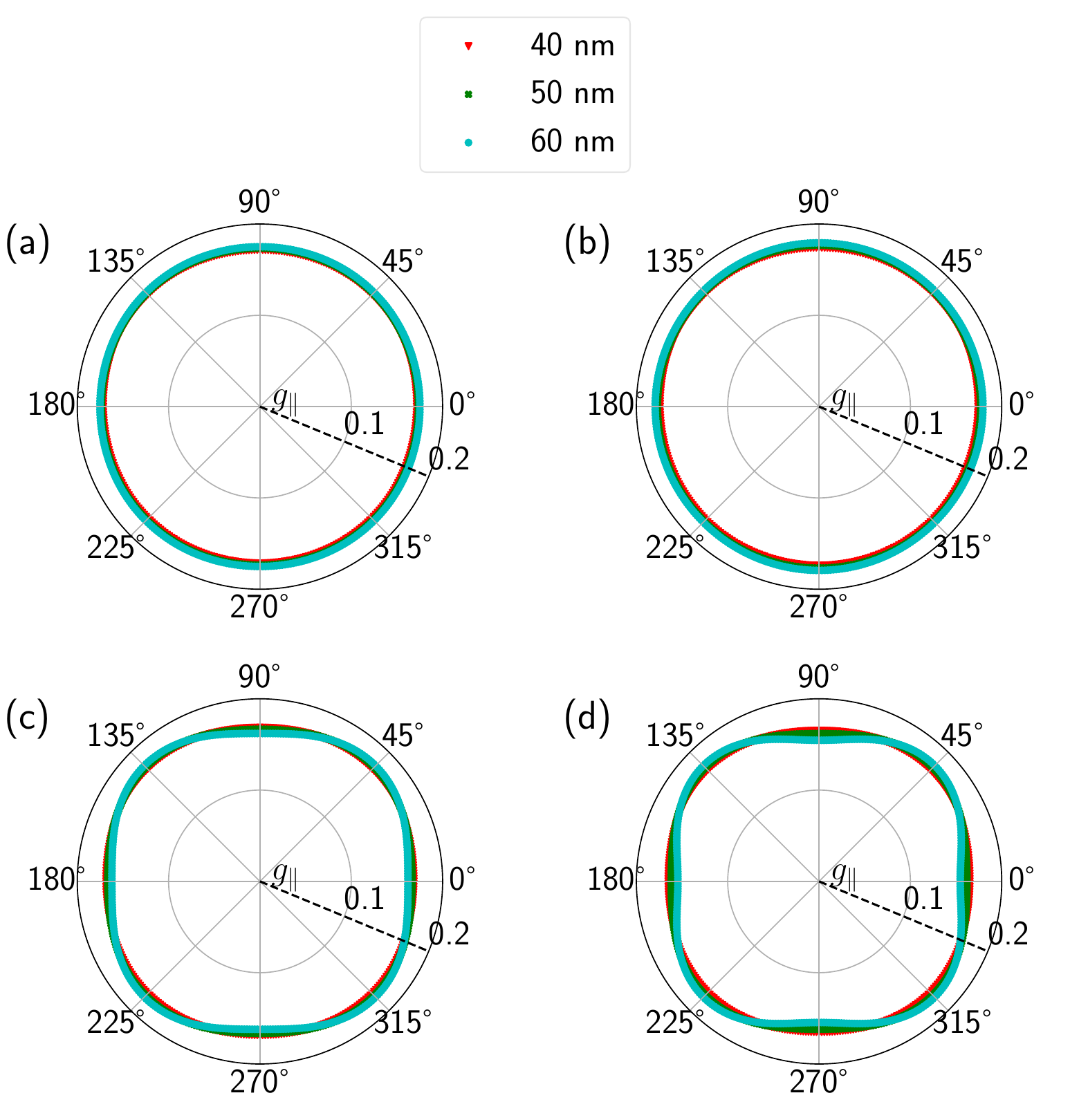}
    \caption{Effective $g_\parallel$ of a circular quantum dot with the Ge/GeSi heterostructure in four different conditions: (a) no electric field or residual tensile strain; (b) residual tensile strain that change 32\% of the GeSi lattice constant; (c) a static electric field of 1 MV/m in the growth direction; (d) combined effect of the residual tensile strain and the static electric field. Here $0^\circ$ refers to the [100] direction.  These results are calculated using the ``shifted-$m^*$'' treatment of boundary conditions.}
    \label{fig:gip360}
\end{figure}

\begin{figure}[t] %[htbp]% [H]
    \centering
    \includegraphics[width=0.48\textwidth]{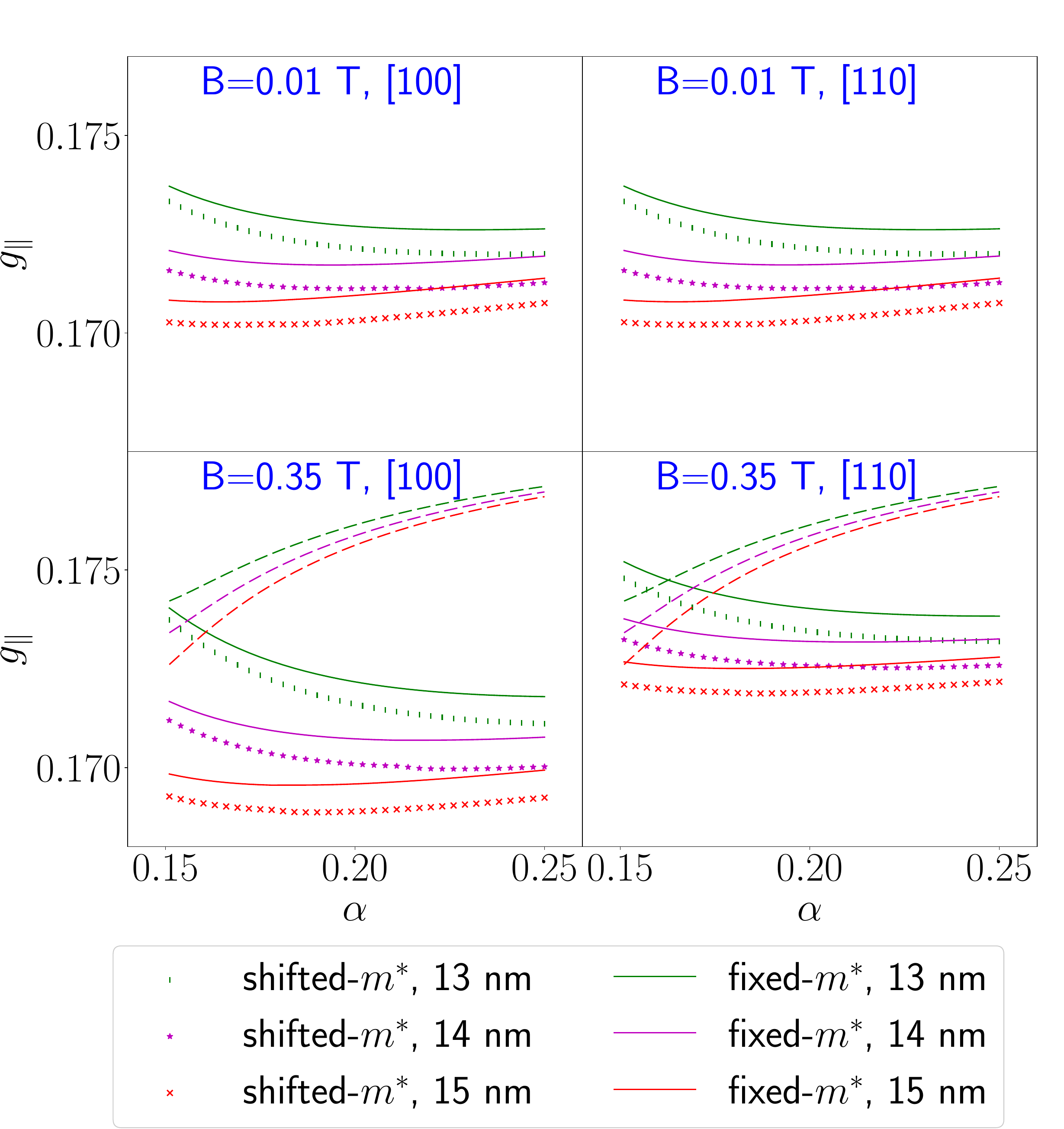}
    \caption{Effective $g_\parallel$ with varying silicon concentration and different well widths for a 40 nm dot in a $\Bvec_\parallel$ field of (upper) 0.01 T; (lower) 0.35 T. The field is in the (left) [100] or (right) [110] directions. The corresponding hard-wall results  (dashed lines) are included in the bottom panels for a reference.}
    \label{fig:gip}
\end{figure}

Figure \ref{fig:gip360}(a) shows the in-plane $g$-factor for dot sizes of 40 nm, 50 nm, and 60 nm with different in-plane directions of the applied field. Clearly, the effective $g_\parallel$ is not sensitive to the dot size, and is nearly isotropic in the $xy$ plane, even though the Burt-Foreman Hamiltonian \eqref{eq:bf4b} is known to have in-plane anisotropy arising from the HH-LH coupling. This coupling is suppressed by the large $\Delta_{\rm HL}$. When examined more closely, a small anisotropy can be seen in Fig.~\ref{fig:gip} on the effective $g_\parallel$ in the [100] and [110] directions when the magnetic field is not small (the lower two panels).

As shown in Fig.~\ref{fig:gip}, when the magnetic field is very small, $g_\parallel$ is almost identical along the two directions; while a notable difference between the two directions appears when the orbital mixing is enhanced by a larger field. The difference is $\approx1\%$ in the current setup of the planar dot, but can be significantly larger if $\Delta_{\rm HL}$ is smaller.
Another feature at the larger applied field is that $g_\parallel$ in the [100] direction is clearly more sensitive to the change of well width compared to the [110] direction.

While most of the studies in Ge quantum dots are based on a Ge/Ge$_{0.8}$Si$_{0.2}$ heterostructure \cite{2023Abadillo-UrielPRL131,2023SarkarPRB108,2024WangNpjQuantumInf10},
one distinct advantage of a finite-barrier model is that the functional dependence of the $g$-factor on silicon concentration in the barrier can be studied accurately. Compared to the hard-wall model where the percentage of silicon only affects the uniaxial strain in the Ge well, the impact to band offset across the boundaries must be considered in the finite-barrier model.
The interplay of band offset and the uniaxial strain affects the sub-band energies and wave vectors in the LCAO basis. Thus, the $g$-factor is expect to have a more non-trivial dependence on silicon concentration in the finite-barrier model, as shown in Fig. \ref{fig:gip}.
With increasing silicon concentration from 0.15 to 0.25, the effective $g_\parallel$ becomes less sensitive to well width.
Figure \ref{fig:gip} also shows numerical results from the ``fixed-$m^*$'' treatment of boundary conditions as a comparison. The two treatments show the same qualitative behavior in the current range of $\alpha$, while a quantitative discrepancy is clear and increases with larger $\alpha$.

\subsection{$g$-factor in the growth direction}
\label{ssec:gz}

\begin{figure}[htbp] %[H]
    \centering
    \includegraphics[width=0.5\textwidth]{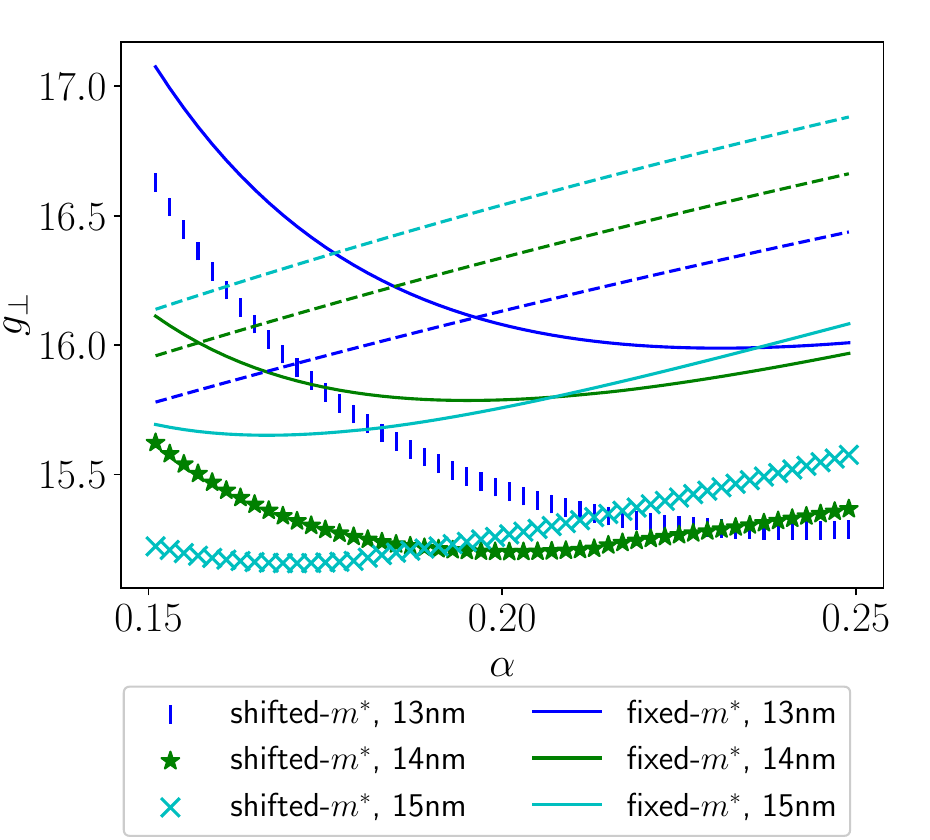}
    \caption{The effective $g_\perp$ as a function of $\alpha$ with different well widths of a 40 nm dot. Sensitivity to well width increases with larger silicon concentration, while the disagreement between two treatments of boundary conditions is almost a constant in the whole range. The corresponding hard-wall results  (dashed lines) are included for a reference.}
    \label{fig:gzwidth}
\end{figure}

Along the growth direction for the Ge/GeSi heterostructure, the hole $g$-factor is known to be much larger compared to the in-plane direction, as the hole spin quantization axis is pinned to the growth direction by the quantum well confinement. Conversely, spin splitting in the quantum well along the growth direction should also be strongly affected by variations in the quantum well parameters.  This observation is indeed clearly seen in our calculations.  For example, Fig.~\ref{fig:gzwidth} shows the effective $g_\perp$ with varying silicon concentration and different well widths. The value of $g_\perp$ is about two orders of magnitude larger than $g_\parallel$, leading to much larger spin splittings in general. 
The sensitivity of $g_\perp$ to the dot size is too small to show. On the other hand, it is more sensitive to the well width, as expected. Similarly, $g_\perp$ shows significant sensitivity to silicon concentration as well. This dependence is clearly not monotonic.  The uptick of $g_\perp$ at larger $\alpha$ is the result of a larger HH-LH splitting. In addition, the two treatments of boundary conditions show nearly identical qualitative behaviors with a constant shift of magnitude.

% Add a short subsection to describe the difference between hard-wall and finite-barrier approaches.

\subsection{Analysis of the finite-barrier model}
\label{ssec:finite}

A comparison between hard-wall (dashed lines) and finite-barrier (solid and dotted lines) in Figs. 5 and 6 clearly shows a qualitative difference between the two approaches in how the obtained $g$-factor depends on the Si concentration $\alpha$. In the hard-wall model, the quantized $k_z$ is determined exclusively by the well width. Increasing the silicon concentration $\alpha$ does not affect $k_z$.  Instead it only increases the uniaxial strain, which makes the energy denominator $\Delta_{\rm HL}$ larger, and results in a monotonic behavior in the $g$-factor. On the other hand, in the realistic situation of finite barriers, the interplay between uniaxial strain and the band offset is included, and the resulting band alignment will affect the value of the quantized $k_z$ (thus the matrix elements that depend on $k_z$) as well as $\Delta_{\rm HL}$. The more complex roles by $\alpha$ in determining the HH-LH mixing then leads to a more nuanced dependence of the $g$-factor on $\alpha$.

The two treatments of boundary conditions, which are only relevant for the finite barrier model we adopt, also lead to non-negligible discrepancies in the calculations of $g_\parallel$ and $g_\perp$.  In essence, these two approaches yield a slight difference in the quantized momentum along the growth direction when solving for the out-of-plane component of the envelope function, therefore affecting the resulting spin-orbit coupling.  Furthermore, the quantized momentum is also modified by varying the silicon concentration $\alpha$ in the barrier. Compared with the hard-wall model, where only $\Delta_{\rm HL}$ is dependent on $\alpha$, our more realistic model contains additional complexity, which in turn leads to different sensitivity of the effective $g_\parallel$ and $g_\perp$ to the choice of boundary conditions and the silicon concentration.

\begin{figure}[htbp] %[H]
    \centering
    \includegraphics[width=0.225\textwidth]{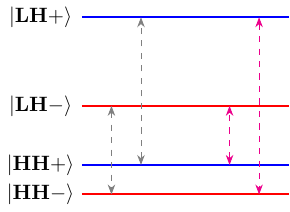}
    \includegraphics[width=0.225\textwidth]{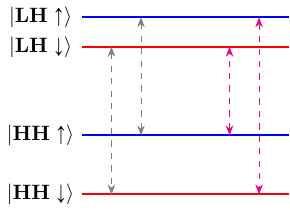}
    \caption{Schematic representation of the HH-LH coupling with an (left) in-plane or (right) out-of-plane magnetic field.}
    \label{fig:HH-LHschematics}
\end{figure}

The different dependencies on the boundary conditions can be understood from the perspective of perturbation theory.  The ``unperturbed'' in-plane $g$-factor provided by the Zeeman term is $g_{\parallel0}=0.18$. This Zeeman effect splits the $\ket{\rm HH+}$ and $\ket{\rm HH-}$ states, where we define
\begin{eqnarray}
    \ket{\rm HH+} &=& \frac{1}{\sqrt2}(\ket{\rm HH\uparrow}+\ket{\rm HH\downarrow})\,,\\
    \ket{\rm HH-} &=& \frac{1}{\sqrt2}(\ket{\rm HH\uparrow}-\ket{\rm HH\downarrow})\,,
\end{eqnarray}
and similarly for the two LH states. The splitting between $\ket{\rm LH+}$ and $\ket{\rm LH-}$ is larger than the $\ket{\rm HH+}$-$\ket{\rm HH+}$ splitting, as shown in Fig.~\ref{fig:HH-LHschematics} (left). 
According to perturbation theory, direct coupling between two states enhances their energy difference. This enhancement is larger when the two states are energetically closer to each other. 
Therefore, the $\ket{\rm HH+}$-$\ket{\rm LH+}$ and $\ket{\rm HH-}$-$\ket{\rm LH-}$ couplings (gray dashed) enhance the spin splitting of the HH states, while the $\ket{\rm HH+}$-$\ket{\rm LH-}$ and $\ket{\rm HH-}$-$\ket{\rm LH+}$ couplings (magenta dashed) decrease it. These two types of direct couplings counteract each other, reducing the sensitivity of the effective $g_\parallel$ to the change of the quantized momentum. Therefore, the effective $g_\parallel$ is not very sensitive to the choice of boundary conditions and to the silicon concentration.

On the other hand, the ``unperturbed'' out-of-plane $g$-factor is $g_{\perp0} \approx 21.27$. The out-of-plane $B$ field splits $\ket{\rm HH\uparrow}$ and $\ket{\rm HH\downarrow}$ a lot more than it does to $\ket{\rm LH\uparrow}$ and $\ket{\rm LH\downarrow}$, as shown in fig. \ref{fig:HH-LHschematics} (right). Both direct coupling mechanisms (gray dashed and magenta dashed) reduce the HH spin splitting. Therefore, this splitting will be more sensitive to the change of $k_z$, thus to the choice of boundary conditions as well as to the silicon concentration.

The above analysis is based on a relatively large HH-LH splitting so that direct couplings play the most important role. A more complicated situation arises if the HH and LH bands are closer to each other.
The non-negligible high-order terms can also contribute to changes in $g_\parallel$ and $g_\perp$, leading to larger discrepancies between the two treatments of boundary conditions, and a more dramatic sensitivity to confinement parameters of the system.

\subsection{Electric field in the growth direction}
\label{ssec:Efield}

For a planar quantum dot in a Ge quantum dot, the in-plane confinement is produced by top gates that are charged.  In addition to providing in-plane confinement, the charged gates also produce a static electric field in the growth direction:
\begin{equation}
    H_Z = eF_z z\,.
\end{equation}
This static electric field induces a Rashba-type spin-orbit coupling by breaking the inversion symmetry of the square well. A systematic study of the effective $g$-factor for a confined hole should thus consider the effect of this electric field. 

\begin{figure}[htbp] %[H]
    \centering
    \includegraphics[width=0.5\textwidth]{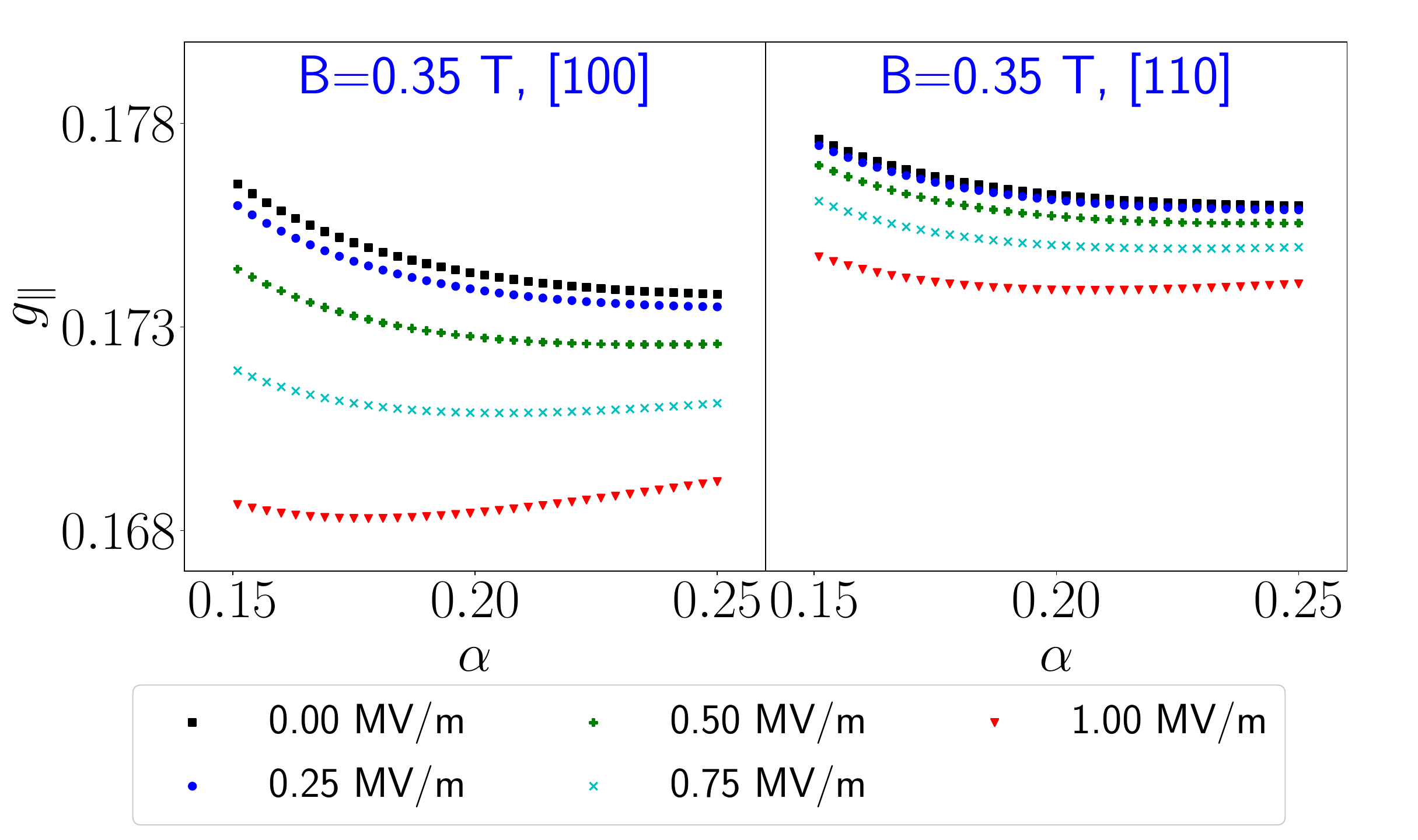}
    \caption{Effective $g_\parallel$ with the ``shifted-$m^*$'' method of a 40 nm dot with a well width of 13 nm. The static electric field is applied in the growth direction with different magnitudes. The difference between $g_\parallel$ in the [100] and the [110] directions are getting larger as the strength of the electric field increases.}
    \label{fig:gipEfield}
\end{figure}

Figure \ref{fig:gipEfield} shows the in-plane $g$-factor of a 40 nm dot with a well width of 13 nm.  With the electric field increased to 1 MV/m, the anisotropic nature of our Hamiltonian becomes clearer. The [100] direction $g$-factor has a larger sensitivity to the electric field, such that a 1 MV/m field can decrease its value by $\approx 4.5\%$, while along the [110] direction, the reduction is $\approx 1.6\%$.

However, the static electric field along $z$ direction has a minimal impact on the $g$-factor along that direction, $g_\perp$. This insensitivity is not really surprising: the out-of-plane magnetic field has already broken the symmetry of the square well by splitting the left-circular and the right-circular orbitals. A further splitting by an electric field thus can only have a limited effect on the spin splitting: about $0.6\%$ decrease in the $g_\perp$ value by increasing the $E$ field from 0 to 1 MV/m.

\subsection{Residual tensile strain}
\label{ssec:ResidualStrain}

The above calculations are based on Fig.~\ref{fig:paradigm1} with the assumption that both Ge$_{1-\alpha}$Si$_\alpha$ barriers are fully relaxed, while the Ge layer is grown lattice-matched to the relaxed Ge$_{1-\alpha}$Si$_\alpha$ and is thus compressively strained so that it acts as a quantum well for the holes. 
Recall that the heterostructure of our concern is grown at finite temperature on a silicon substrate. When cooled down to the experimental temperature in the order of tens of mK, the difference in thermal contraction between Si and Ge leads to a residual tensile strain in the in-plane direction in the Ge$_{1-\alpha}$Si$_\alpha$ barrier \cite{2019SammakAdvFuncMat29,2025BonquetPRAppl23}, which in turn reduces the compressive stress it applies on the Ge well. 
Consequently, the residual tensile strain reduces the HH-LH splitting inside the Ge well \cite{2022MartinezPRB106,2024WangNpjQuantumInf10}. Furthermore, this tensile strain also splits the HH and LH bands in the barriers at the Brillouin zone center.

\begin{figure}[htbp] %[H]
    \centering
    \includegraphics[width=0.5\textwidth]{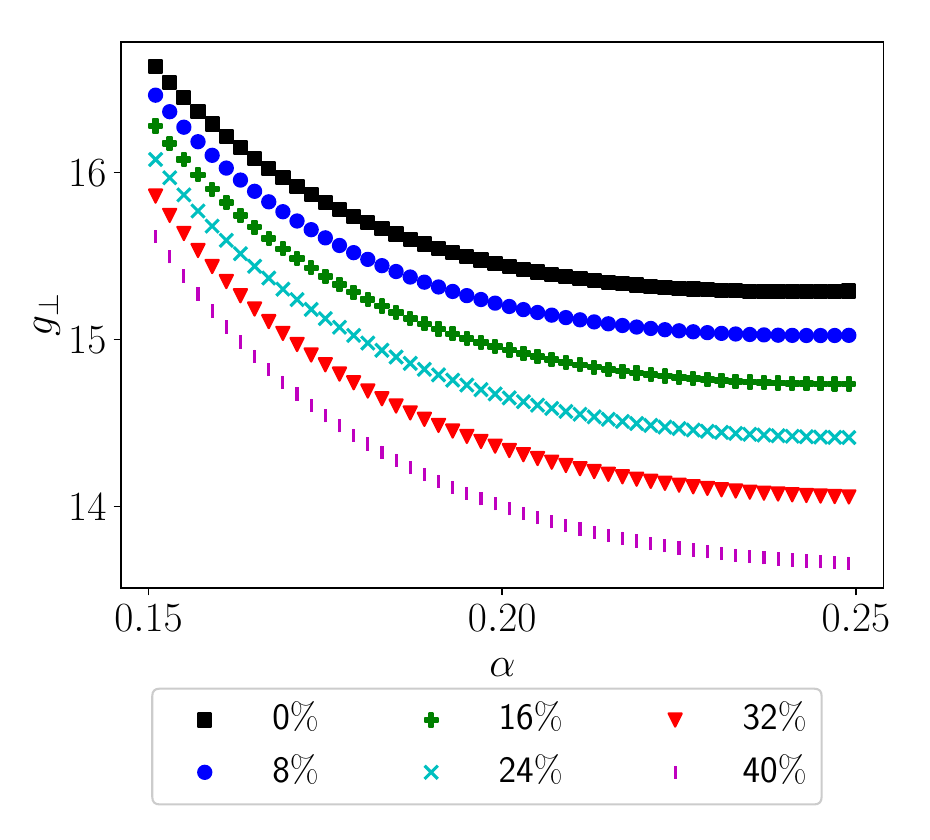}
    \caption{Effective $g_\perp$ with the ``shifted-$m^*$'' method of a 40 nm dot with a well width of 13 nm. The electric field is set to 0 because it has a minimal impact. The residual strain in the barrier shifts the $g_\perp$ vs. $\alpha$ curves downward as it gets larger.}
    \label{fig:gzResidualStrain}
\end{figure}

Here we introduce the ratio $\delta a_0/\Delta a_0[{\rm GeSi}]$ as in Eq.~\eqref{eq:da0} of Appendix \ref{app:parameters} to represent the residual strain in the barrier.  The set of strain elements by most of the studies in the literature is in the range of $\delta a_0/\Delta a_0[{\rm GeSi}]=30\%\sim32\%$ in our calculations. Nevertheless, we vary the residual strain from 0 to 40 percent to evaluate its impact on spin splitting. Based on our calculations, the in-plane $g$-factor changes only minimally, while the out-of-plane $g$-factor is more sensitive to the residual strain, as shown in Fig.~\ref{fig:gzResidualStrain}. In the realistic case of 32\% change of lattice constant, the effective $g_\perp$ has a reduction of $\approx 8\%$ compared with the ideal case of fully relaxed barriers.

\subsection{Combined effects on the in-plane anisotropy}
\label{ssec:combined}

In a realistic Ge/GeSi heterostructure, both residual strain and $z$ direction electric field are present.  The combination helps enhance spin-orbit interaction for the confined hole.  Figure \ref{fig:gip360}(b) shows the effective $g_\parallel$ in the whole $x$-$y$ plane affected by residual tensile strain that changes the GeSi lattice constant by 32\%. The result is slightly more sensitive to the dot size compared with the ideal model counterpart \ref{fig:gip360}(a). However, when a static electric field is applied, the same amount of residual strain will enhance the sensitivity to the dot size more notably, as seen by comparing panels (c) and (d) of Figs.~\ref{fig:gip360}.

\section{Conclusion and Discussion}
\label{sec:summary}

We have studied spin splitting of the ground Kramers doublet in a planar Ge quantum dot with a magnetic field in the in-plane or out-of-plane direction. We adopted a realistic model of finite barriers in the growth direction and performed self-consistent calculations.  Boundary conditions with or without a sudden change in effective mass across the barrier are compared.
We find that in an ideal model of a circular planar dot in a square-well heterostructure with two fully relaxed barriers, the qubit spin splitting has a relatively weak dependence on confinement parameters, especially for the $g$-factor in the in-plane direction. Our results indicate that the effective spin-orbit coupling of such a system may not be large enough for fast electrical control. On the other hand, realistic experimental effects, including the top-gate electric field and the residual strain in the barriers, will both enhance the effective spin-orbit coupling.

Our results show that there is often a non-negligible discrepancy between our two choices of boundary conditions on the effective mass, which should be considered when optimizing the qubit.  While it seems more appropriate to include the change of effective mass in boundary conditions, our calculations provide benchmarks for the two treatments, and emphasize an underlying uncertainty of the formalism.  Experimental measurements will have to be the ultimate judge on which approach is the better reflection of reality.  Qualitatively, in the treatment of such heterostructures, the $\kdotp$ formalism assumes a global lattice periodicity, which contradicts the change of material properties in different layers.  In fact, such a contradiction arises from the effective-mass approximation of a few-band model. If we start with the free-electron mass and use a large enough set of bands to perform $\kdotp$ calculations, the ambiguity of boundary conditions should disappear. But this is computationally challenging and might be better suited for first-principle calculations.

Furthermore, a comparison of hard-wall and finite-barrier results shows a substantial difference in predicting the functional dependence on the silicon concentration. The reason is the hard-wall model falls short in capturing the interplay between band offset and strain. The hard-wall model also tends to overestimate the energy differences between single-band orbitals, leading to an underestimate of the effective spin-orbit coupling. Thus, the in-plane anisotropy intrinsically determined by the symmetry of the $\kdotp$ Hamiltonian is less obvious in the hard-wall results. While some of these differences may be quantitatively small, the more qualitative differences show that the simpler hard-wall potential misses some physical features that could be important for the long-term viability of hole spin qubit.  

From a technical point of view, gauge invariance is another feature within the $\kdotp$ formalism where contradiction could arise.
If we start with the lattice Lagrangian and derive the equation of motion, gauge invariance should be preserved in the system. However, the $\kdotp$ formalism neglects the effect of the vector potential on the Bloch states. In this case, gauge invariance may \textit{in principle} not be preserved for the total effective mass Hamiltonian \eqref{eq:Htotal}. 
The problem of gauge invariance arises because of the HH-LH coupling in the $\kdotp$ Hamiltonian. When the splitting between HH and LH bands is large, the system shows reasonable gauge invariance in the weak-field limit ($B\approx 0$ T) by comparing the Landau gauge and the symmetric gauge in $x$-$y$ directions, as pointed out in \cite{2023Abadillo-UrielPRL131}. However, when the applied field is as large as $B=0.35$ T, a clear discrepancy (5$\sim$10\%) between two gauges is observed. 
This discrepancy can be significantly enhanced if the HH-LH splitting is much smaller.

In other words, the actual gauge-invariant Hamiltonian contains terms beyond the scope of the current effective mass Hamiltonian derived in the $\kdotp$ formalism.
This problem can potentially also be addressed by first-principle calculations. The present calculations are carried out with a symmetric gauge because it contributes in different directions evenly.

Another important aspect of the present treatment is the accuracy of the LCAO technique. The truncation of the number of sub-bands could limit the accuracy of the energy eigenstates for the system.
A given finite well does not have an infinite number of sub-bands. 
The scattering states in the growth direction have a continuous spectrum and cannot be included in the LCAO calculation directly. 
From a perturbative point of view, these states will modify the energy levels of all the confined out-of-plane orbitals through the Burt-Foreman Hamiltonian. However, considering that the highest two heavy-hole states already have only a very small effect on our results, we believe the effects of these higher-energy scattering states are negligible in our current calculations. On the other hand, for a much narrower or shallower quantum well, the scattering states may indeed become important, and a truncated basis set without including the dressing from the continuum may fail to predict accurate spin properties.
One way to account for the scattering states is to add a hard-wall confinement outside the finite barriers to quantize these states and include their effects in the calculations, as was done in Ref.~\onlinecite{2024WangNpjQuantumInf10}. The present work can be viewed as the starting point for the long-term goal of finding a viable way to avoid the hard-wall approximation for the study of more complicated heterostructures.

This paper aims to facilitate the engineering of GeSi alloy for a reliable Ge qubit. 
In Ge/GeSi heterostructure, an adequate treatment of the realistic valence-band offset can help better understand how confining layers affect qubit properties.
The dependence of the qubit spin splitting on silicon concentration shows a possible magnitude of noise coming from the alloy manufacturing. 
On the other hand, varying silicon concentration can also change the sensitivity of the $g$-factor to the confinement. This sensitivity is not monotonic with the percentage of silicon in the barriers, and therefore, can be explored both theoretically and experimentally.
The engineering of confining alloys is thus another degree of freedom for reliable qubits.
Stability of qubit properties is wanted even with imperfections in the manufacturing process. Predictable operating conditions are desired, including dopant concentration.
In addition, this work assumed an alloy with evenly distributed Si and Ge, which is an ideal assumption.
Effect of disorders in the actual alloys may be a fruitful direction of future research.\\

\section*{Acknowledgments} 
We acknowledge financial support from the UB CAS Dean's Office.  HFF thanks support from the National Science Foundation under grants NSF PHY-2014023 and NSF OAC-1931367, and XH thanks support from US AFOSR through grant FA9550-23-1-0710 and US ARO through grant W911NF2310018. We also thank the valuable discussions with Dimitrie Culcer's group at UNSW in Sydney, Australia, and Maximilian Russ' group at TU Delft in Delft, Netherlands.

\begin{widetext}

\appendix

\section{A sign difference in the Burt-Foreman Hamiltonian}
\label{app:signBF}

The different sign in our Burt-Foreman Hamiltonian eq.\eqref{eq:bf4b} compared to the original one by Foreman \cite{1993ForemanPRB48} comes from the use of zone-center basis states. If we use the standard $\ket{j,j_z}$ states for $j=3/2$, the matrix representations of $\Jvec$ components are
\begin{eqnarray}\label{eq:Jmatrices}
    J_x = \begin{pmatrix}
        0 & \frac{\sqrt3}{2} & 0 & 0\\
        \frac{\sqrt3}{2} & 0 & 1 & 0\\
        0 & 1 & 0 & \frac{\sqrt3}{2}\\
        0 & 0 & \frac{\sqrt3}{2} & 0
    \end{pmatrix}\,, \quad
    J_y = \begin{pmatrix}
        0 & -\I\frac{\sqrt3}{2} & 0 & 0\\
        \I\frac{\sqrt3}{2} & 0 & -\I & 0\\
        0 & \I & 0 & -\I\frac{\sqrt3}{2}\\
        0 & 0 & \I\frac{\sqrt3}{2} & 0
    \end{pmatrix}\,, \quad
    J_z = \begin{pmatrix}
        \frac{3}{2} & 0 & 0 & 0\\
        0 & \frac{1}{2} & 0 & 0\\
        0 & 0 & -\frac{1}{2} & 0\\
        0 & 0 & 0 & -\frac{3}{2} 
    \end{pmatrix}\,.
\end{eqnarray}
The $p$-like orbital states $\ket{l,l_z}$, on the other hand, are defined as linear combinations of $\ket{X}$, $\ket{Y}$, $\ket{Z}$, \cite{2010YuCardona}
\begin{eqnarray}
    \ket{1,\phantom{-}1} &=& -\frac{1}{\sqrt2}\ket{X+\I Y}\,,\nonumber\\
    \ket{1,\phantom{-}0} &=& \ket{Z}\,,\\
    \ket{1,-1} &=& \frac{1}{\sqrt2}\ket{X-\I Y}\,.\nonumber
\end{eqnarray}
Then, we can evaluate all $\ket{j,j_z}$ states in the $\ket{l,l_z;s,s_z}$ basis by calculating all the Clebsch-Gordan coefficients. We end up with
\begin{eqnarray}
    \ket{3/2,\phantom{-}3/2} &=& -\frac{1}{\sqrt2}\ket{(X+\I Y),\uparrow}\,,\nonumber\\
    \ket{3/2,\phantom{-}1/2} &=& -\frac{1}{\sqrt6}\ket{(X+\I Y),\downarrow} + \sqrt{\frac{2}{3}}\ket{Z,\uparrow}\,,\nonumber\\
    \ket{3/2,-1/2} &=& \phantom{-}\frac{1}{\sqrt6}\ket{(X-\I Y),\uparrow} + \sqrt{\frac{2}{3}}\ket{Z,\downarrow}\,, \label{eq:j3half}\\
    \ket{3/2,-3/2} &=& \phantom{-}\frac{1}{\sqrt2}\ket{(X-\I Y),\downarrow}\,.\nonumber
\end{eqnarray}
Deriving the Burt-Foreman model with such a basis, we will eventually get \eqref{eq:bf4b} and \eqref{eq:bfelements}. Giving a global phase to the basis states \eqref{eq:j3half} doesn't change the matrix representation. However, Foreman's basis states are not in a consistent phase with \eqref{eq:j3half}. It is still a complete basis which doesn't affect the calculation of bulk or heterostructure properties, but it can cause non-negligible error in qubit Zeeman splitting if $H_{\rm BF}$ and $\Jvec$ are not in the same basis.

\section{Material properties}
\label{app:parameters}

The Foreman parameters are defined as $\bar \gamma=(\gamma_2+\gamma_3)/2$, $\mu=(\gamma_3-\gamma_2)/2$, $\delta=(1+\gamma_1+\gamma_2-3\gamma_3)/9$, $\sigma=\bar \gamma - \delta/2$, $\pi=\mu + 3\delta/2$.

For all the material parameters, we followed \cite{1964DismukesJoPC68} and the supplemental material of \cite{2023Abadillo-UrielPRL131} to calculate them with \textit{linear interpolation}. The strain tensor has only diagonal terms for uniaxial strain. In our calculations, 
\begin{equation}\label{eq:exx}
    \varepsilon_{xx}=\varepsilon_{yy}=\frac{\Delta a_0^2}{2a_0^2},
\end{equation} 
and $\varepsilon_{zz}=-\frac{2c_{12}}{c_{11}}\varepsilon_{xx}$ \cite{2024WangNpjQuantumInf10, 2023SarkarPRB108}. Here $\Delta a_0$ is the difference of lattice constant between GeSi and Ge, and Vegard's law is used to calculate the lattice constant of GeSi
\begin{equation}
    a_0[{\rm Ge}_{1-\alpha}{\rm Si}_\alpha] = a_0[{\rm Ge}](1-\alpha) + a_0[{\rm Si}]\alpha - 0.027\alpha(1-\alpha)\,,
\end{equation}
as well as the variance
\begin{equation}\label{eq:Da02}
    \Delta a_0^2 = a_0[{\rm Ge}_{1-\alpha}{\rm Si}_\alpha]^2 - a_0[{\rm Ge}]^2\,.
\end{equation}

\begin{table}[H]
\begin{center}
\begin{tabular}{|c|*{11}{P{1.25cm}}|}
 \hline
  & $a_0$(\AA) & $\gamma_1 $ & $\gamma_2$ & $\gamma_3$ & $a_v$(eV) & $b_v$(eV) & $c_{11}$(GPa) & $c_{12}$(GPa) & $\kappa$ & $q$ & $V_0$(eV)\\
\hline
 Ge & 5.658 & 13.380 & 4.240 & 5.690 & 2.00 & -2.160 & 131.0 & 49.0 & 3.410 & 0.060 & -\\
 \hline
 Si & 5.430 & 4.285 & 0.339 & 1.446 & 2.10 & -2.330 & 167.5 & 64.9 & -0.420 & 0.010 & -\\
 \hline
 Ge$_{0.85}$Si$_{0.15}$ & 5.620 & 12.016 & 3.655 & 5.053 & 2.015 & -2.186 & 136.5 & 51.4 & 2.836 & 0.053 & 0.104 \\
 \hline
 Ge$_{0.8}$Si$_{0.2}$ & 5.608 & 11.561 & 3.460 & 4.841 & 2.02 & -2.194 & 138.3 & 52.2 & 2.644 & 0.050 & 0.132\\
 \hline
 Ge$_{0.75}$Si$_{0.25}$ & 5.596 & 11.106 & 3.265 & 4.629 & 2.025 & -2.203 & 140.1 & 53.0 & 2.453 & 0.048 & 0.156 \\
 \hline
\end{tabular}
\caption{Material properties of the Ge/GeSi heterostructure.}
\label{tab:parameters}
\end{center}
\end{table}
The barrier height $V_0$ is determined by the valence-band offset $\Delta E^{v}$ of the heterostructure. A theoretical prediction of strained Ge on a relaxed substrate of Ge$_{1-\alpha}$Si$_{\alpha}$ alloy is \cite{2000SylvieSemiSciTech15}
\begin{equation}\label{eq:DEv}
    \Delta E^{v}(\alpha) = \alpha[0.6-0.7(2-\alpha)]\ {\rm eV}\,.
\end{equation}

The calculation of the strain tensor is based on eq. \eqref{eq:exx}. The difference is that both the Ge layer and the two GeSi alloys will have variances.
\begin{eqnarray}
    \Delta a_0^2[\rm Ge] &=& a_0^{*2} - a_0[{\rm Ge}]^2\,,\\
    \Delta a_0^2[\rm GeSi] &=& a_0^{*2} - a_0[{\rm GeSi}]^2\,,
\end{eqnarray}
where $a_0^*$ is the lattice constant of the heterostructure. For convenience, we define the change of lattice constant from the $a_0$[GeSi] as
\begin{equation}\label{eq:da0}
    \delta a_0/\Delta a_0[{\rm GeSi}] = (a_0^*-a_0[{\rm GeSi}])/ (a_0[{\rm Ge}]-a_0[{\rm GeSi}])\,.
\end{equation}
In table \ref{tab:straintensor}, we show all the elements of the strain tensor in the case where there is 20\% silicon in the barriers.

\begin{table}[H]
\begin{center}
\begin{tabular}{|c|*{5}{P{2.5cm}}|}
\hline
$\delta a_0/\Delta a_0$[Ge$_{0.8}$Si$_{0.2}$] & $a_0^*$(\AA) & $\varepsilon_\parallel$[Ge] & $\varepsilon_{zz}$[Ge] & $\varepsilon_\parallel$[Ge$_{0.8}$Si$_{0.2}$] & $\varepsilon_{zz}$[Ge$_{0.8}$Si$_{0.2}$] \\
 \hline
0$\%$ & 5.608    &  -0.878$\%$   &  0.657$\%$  & 0$\%$  &   0$\%$ \\
\hline
8$\%$ & 5.612    &  -0.808$\%$   &  0.605$\%$  & 0.071$\%$  &   -0.054$\%$ \\ 
\hline
16$\%$ & 5.616    &  -0.738$\%$   &  0.552$\%$  & 0.143$\%$  &   -0.108$\%$ \\
\hline
24$\%$ & 5.620    &  -0.668$\%$   &  0.500$\%$  & 0.214$\%$  &   -0.161$\%$ \\
\hline
32$\%$ & 5.624    &  -0.598$\%$   &  0.447$\%$  & 0.285$\%$  &   -0.215$\%$ \\
\hline
40$\%$ & 5.628    &  -0.528$\%$   &  0.395$\%$  & 0.357$\%$  &   -0.269$\%$ \\
\hline
\end{tabular}
\caption{Elements of the strain tensor for the Ge/Ge$_{0.8}$Si$_{0.2}$ heterostructure with different changes of lattice constant in the barrier.}
\label{tab:straintensor}
\end{center}
\end{table}

\section{Coupling of in-plane states with different effective masses}
\label{app:cpinplane}

The electrostatic confinement generates the same harmonic potential for energy bands 1 and 2 of a single particle
\begin{equation}
    V_\parallel(x) = \frac{1}{2}m_{1}\omega_{1}x^2 = \frac{1}{2}m_{2}\omega_{2}x^2\,,
\end{equation}
where $m_{1(2)}$, $\omega_{1(2)}$ are effective mass and frequency of band 1(2). The effective confinement size $L_{1(2)} = \sqrt{\hbar/(m_{1(2)}\omega_{1(2)})}$ should be different if such a particle has different effective masses in these two bands. In this case, the corresponding wave functions $\psi_{1n}$, $\psi_{2m}$ will not follow the orthonormal property $\ovlp{\psi_{1n}}{\psi_{2m}}=\delta_{nm}$. Thus, additional care must be taken when evaluating the matrix representation of terms coupling the two bands. Here, we provide an analytical benchmark for the calculation of matrix elements.

\subsection{Harmonic oscillators}
\label{app:cpharmonic}

In the harmonic oscillator basis, the wave function is in the form
\begin{equation}
    \psi_{1(2)n}=\frac{1}{\sqrt{2^n n!}}\left(\frac{1}{\pi L_{1(2)}^2}\right)^\frac{1}{4} e^{-\frac{x^2}{2L_1^2}} H_n\left(\frac{x}{L_{1(2)}}\right)\,.
\end{equation}
One then needs to evaluate the integral
\begin{equation}
    \ovlp{\psi_{1n}}{\psi_{2m}}=\frac{1}{\sqrt{2^{n+m} n!m!}}\left(\frac{1}{\pi L_1 L_2}\right)^\frac{1}{2}
    \int dx e^{-\frac{x^2}{2L_1^2}-\frac{x^2}{2L_2^2}} H_n\left(\frac{x}{L_1}\right)  H_m\left(\frac{x}{L_2}\right)\,,\label{eq:cpho}
\end{equation}
where $L_1\ne L_2$. This integral can be evaluated using the explicit expression of Hermite polynomials
\begin{equation}
    H_n(x) = n!\sum_{l=0}^{\left[\frac{n}{2}\right]}\frac{(-1)^l}{l!(n-2l)!}(2x)^{n-2l}\,,
\end{equation}
where $[n]$ is a floor function of $n$. With a change of variables, we eventually need to evaluate an integral of the form
\begin{equation}
    \int du e^{-u^2} H_n(au) H_m(bu)\,,
\end{equation}
where the coefficients $a$, $b$ are
\begin{equation}
    a = \frac{\sqrt2 L_{2}}{\sqrt{L_{1}^2 + L_{2}^2}}\,, \quad
    b = \frac{\sqrt2 L_{1}}{\sqrt{L_{1}^2 + L_{2}^2}}\,,
\end{equation}
and the new variable
\begin{equation}
    u = \frac{\sqrt{L_{1}^2 + L_{2}^2}}{\sqrt{2 L_{1} L_{2}}}x\,.
\end{equation}
Applying a useful integral
\begin{equation}
    \mathcal{I}(n,m,r,s)=\int_{-\infty}^\infty e^{-u^2} u^{n+m-2r-2s} du = \frac{\Gamma\left(\frac{n}{2}+\frac{m}{2}-r-s+\frac{1}{2}\right)\left(1+(-1)^{n+m}\right)}{2}\,,
\end{equation}
we can eventually get the expression of the coupling integral \eqref{eq:cpho} in a closed form
\begin{equation}
    \begin{split}
        \ovlp{\psi_{1n}}{\psi_{2m}}=\frac{1}{\sqrt{2^{n+m} n!m!}}\left(\frac{1}{\pi}\right)^\frac{1}{2} \frac{\sqrt{2 L_1 L_2}}{\sqrt{L_1^2 + L_2^2}}
        n!m!2^{n+m}
        \sum_{r=0}^{\left[\frac{n}{2}\right]}\sum_{s=0}^{\left[\frac{m}{2}\right]}\frac{(-1)^{r+s}2^{-2r-2s}}{r!(n-2r)!s!(m-2s)!} a^{n-2r} b^{m-2s}
        \mathcal{I}(n,m,r,s)\,.
    \end{split}
\end{equation}

\subsection{Fock-Darwin states}
\label{app:cpFockDarwin}

In the Fock-Darwin basis, we need to deal with a similar integral. The general expression of circular Fock-Darwin state is 
\begin{equation}
    \psi_{nm}(\rho,\phi) = N R_n(\rho) e^{\I m \phi}\,,
\end{equation}
Here $n$ is the principal quantum number of the Fock-Darwin states $n=n_+ + n_-$, and $m$ is the magnetic quantum number $m=n_+ - n_-$,
with $N$ the normalization factor, and $R_n(\rho)$ the radial part of the wave function. A useful quantum number $n_r$ is defined as $n_r = (n-|m|)/2$ for the radial function
\begin{equation}
    R_n(\rho) = \sqrt{\frac{2n_r!}{(n_r+|m|)!}}\left( \frac{\rho}{L_b} \right)^{|m|} e^{-\frac{\rho^2}{2L_b^2}} \mathcal{L}_{n_r}^{|m|}\left( \frac{\rho^2}{L_b^2} \right)\,,
\end{equation}
where $\mathcal{L}_{n_r}^{|m|}$ is the associated Laguerre polynomial that has a closed form
\begin{equation}
    \mathcal{L}_{n_r}^{|m|}(x) = \sum_{i=0}^{n_r} (-1)^{i} 
    \begin{pmatrix} n_r+|m| \\ n_r-i \end{pmatrix} \frac{x^i}{i!}\,.
\end{equation}
Normalization of the wave function
\begin{equation}
        |N|^2 2\pi \int_0^\infty \frac{n_r!}{(n_1+|m|)!} \left(\frac{\rho^2}{L_b^2}\right)^{|m|} e^{-\frac{\rho^2}{L_b^2}} \left[\mathcal{L}_{n_r}^{|m|}\left(\frac{\rho^2}{L_b^2}\right)\right]^2 d\rho^2=|N|^2 2\pi L_b^2=1\,
\end{equation}
can lead to
\[
    N = \frac{1}{\sqrt{2\pi}L_b}\,.
\]
The effective confinement $L_b$ is determined similarly to the case of the harmonic oscillator and thus varies with the effective masses. We will end up with a similar coupling integral
\begin{equation}
    \begin{split}
        &\int_0^{2\pi} \int_0^{\infty} \psi_{n_{1}m_1}^*(\rho, \phi) \psi_{n_{2} m_2}(\rho,\phi) \rho d\rho d\phi = \frac{1}{2\pi L_{b1}L_{b2}} \int_0^{2\pi} e^{\I(m_2-m_1)\phi} d\phi \int_0^{\infty} R^*_{1,n_{1}}(\rho) R_{2,n_{2}}(\rho) \rho d\rho \\
        =& \frac{1}{2L_{b1}L_{b2}} \int_0^{\infty} \sqrt{\frac{4n_{1r}!n_{2r}!}{(n_{1r}+|m|)!(n_{2r}+|m|)!}} \left( \frac{\rho^2}{L_{b1} L_{b2}}\right)^{|m|} e^{-\frac{\rho^2}{2L_{b1}^2}-\frac{\rho^2}{2L_{b2}^2}} \mathcal{L}_{n_{1r}}^{|m|}\left(\frac{\rho^2}{L_{b1}^2}\right) \mathcal{L}_{n_{2r}}^{|m|}\left(\frac{\rho^2}{L_{b2}^2}\right) d\rho^2\\
        =&  \sqrt{\frac{n_{1r}!n_{2r}!}{(n_{1r}+|m|)!(n_{2r}+|m|)!}} (L_{b1}L_{b2})^{-|m|-1} \int_0^\infty \rho^{2|m|} e^{-\frac{\rho^2}{2L_{b1}^2}-\frac{\rho^2}{2L_{b2}^2}} \mathcal{L}_{n_{1r}}^{|m|}\left(\frac{\rho^2}{L_{b1}^2}\right) \mathcal{L}_{n_{2r}}^{|m|}\left(\frac{\rho^2}{L_{b2}^2}\right) d\rho^2\\
        =& \sqrt{\frac{n_{1r}!n_{2r}!}{(n_{1r}+|m|)!(n_{2r}+|m|)!}} (L_{b1}L_{b2})^{-|m|-1} \sum_{i=0}^{n_{1r}}\sum_{j=0}^{n_{2r}}\frac{(-1)^{i+j}}{i!j!}
        \begin{pmatrix} n_{1r}+|m|\\n_{1r}-i \end{pmatrix}
        \begin{pmatrix} n_{2r}+|m|\\n_{2r}-j \end{pmatrix}\\
        \times&
        \int_0^\infty \rho^{2|m|}e^{-\frac{\rho^2}{2L_{b1}^2}-\frac{\rho^2}{2L_{b2}^2}} \left(\frac{\rho^2}{L_{b1}^2}\right)^i \left(\frac{\rho^2}{L_{b2}^2}\right)^j d\rho^2\\
        =& \sqrt{\frac{n_{1r}!n_{2r}!}{(n_{1r}+|m|)!(n_{2r}+|m|)!}} \sum_{i=0}^{n_{1r}}\sum_{j=0}^{n_{2r}}\frac{(-1)^{i+j}}{i!j!}\begin{pmatrix} n_{1r}+|m|\\n_{1r}-i \end{pmatrix}
        \begin{pmatrix} n_{2r}+|m|\\n_{2r}-j \end{pmatrix}\\
        \times&
        (L_{b1})^{2j+|m|+1} (L_{b2})^{2i+|m|+1} 2^{i+j+|m|+1} (L_{b1}^2+L_{b2}^2)^{-1-i-j-|m|} \Gamma(i+j+|m|+1)\,.
    \end{split}
\end{equation}

\section{Quantum well sub-bands with the coupled boundary conditions}
\label{app:cpsubbands}

Generally, the total envelope function constructed by single-band envelope functions
\begin{equation}
    \ket{\Psi} = \sum_{j_z}\sum_{lmn}\ket{\psi_{j_z}^{mn}(x,y) \phi_{j_z}^l(z)} 
\end{equation}
does not satisfy the coupled boundary conditions. Here we introduce a treatment to modify the quantum-well envelope functions. The idea is: if each basis state $\ket{\phi(z)}$ satisfies the coupled boundary conditions, the total envelope function constructed by these states should satisfy the same boundary conditions automatically. This treatment is to estimate the correction to the quantum well sub-bands by using different boundary conditions. For convenience, we apply this treatment to the Luttinger-Kohn Hamiltonian in order to avoid the complication of LH-LH coupling. In this case, the derivative operator $\mathcal{D}$ can be reduced to the $2\times2$ form in the $\ket{|3/2|}$, $\ket{|1/2|}$ representation
\begin{equation}\label{eq:D_2band}
    \mathcal{D}^{2\times2}_{\rm LK} = \begin{pmatrix}
        (\gamma_1 - 2\gamma_2)\frac{\partial }{\partial z} &  -\I\sqrt3 \gamma_3(k_x - \I k_y) \\
        \I \sqrt3 \gamma_3(k_x + \I k_y) & (\gamma_1 + 2\gamma_2)\frac{\partial }{\partial z}   
    \end{pmatrix}
\end{equation}
We now assume the quantum well sub-bands are in the form of HH-LH mixture. For example, the even HH sub-bands can be in the form
\begin{equation}\label{eq:phi2}
    \ket{\phi_{hw}(z)} = \begin{pmatrix}
        A_{h1} \cos(k_h z) \\  A_{h2} \sin(k_h z)
    \end{pmatrix}\,, \quad 
    \ket{\phi_{hb}(z)} = \begin{pmatrix}
        B_{h1} e^{\pm\chi_h z}\\  B_{h2} e^{\pm\chi_h z}
    \end{pmatrix}\,.
\end{equation}
Component in the $\ket{|3/2|}$ band, $A_{h1} \cos(k_h z)$ in the well and $B_{h1} e^{\pm\chi_h z}$ in the barrier, is the ``main'' state. The corresponding component in the $\ket{|1/2|}$ band is then the ``mixture''.
This means, for each eigenmode $k_h$ ($\chi_h$) in the HH band, there is a coupling to the LH band with an opposite parity. This state should satisfy the continuity of the envelope function in each channel \eqref{eq:1stbc}, and the matrix equation
\begin{equation}\label{eq:cpsubbands}
    \begin{pmatrix}
        (\gamma_1^{(w)} - 2\gamma_2^{(w)})\frac{\partial}{\partial z} & -\I\sqrt{3}\gamma_3^{(w)}(k_x - \I k_y)\\
        \I\sqrt{3}\gamma_3^{(w)}(k_x + \I k_y) & (\gamma_1^{(w)} + 2\gamma_2^{(w)})\frac{\partial}{\partial z}
    \end{pmatrix}\ket{\phi_{hw}} = \begin{pmatrix}
        (\gamma_1^{(b)} - 2\gamma_2^{(b)})\frac{\partial}{\partial z} & -\I\sqrt{3}\gamma_3^{(b)}(k_x - \I k_y)\\
        \I\sqrt{3}\gamma_3^{(b)}(k_x + \I k_y) & (\gamma_1^{(b)} + 2\gamma_2^{(b)})\frac{\partial}{\partial z}
    \end{pmatrix}\ket{\phi_{hb}}\,.
\end{equation}
On the other hand, applying the uncoupled Hamiltonian of the quantum well to sub-band states \eqref{eq:cpsubbands}
\begin{equation}
    \left(-\frac{\hbar^2}{2m_{\perp}^{h(l)}}\frac{\partial^2}{\partial z^2} + V_{\rm eff}^{h(l)}(z) \right)\ket{\phi_{h(l)}(z)} = E^{h(l)}\ket{\phi_{h(l)}(z)}\,
\end{equation}
leads to the same energy equation \eqref{eq:Energyeq}, where $m_{\perp}^{h(l)}$, $V_{\rm eff}^{h(l)}(z)$ and $E^{h(l)}$ refer to the heavy-hole (light-hole) effective mass, effective barrier and sub-band energy in the quantum well.

The solutions of these equations are modified finite-well sub-bands. They do not refer to any physical states because the uncoupled Hamiltonian and the matrix equation \eqref{eq:cpsubbands} are not consistent with each other. Instead, they are artificial states in order to construct the total envelope function required by the coupled boundary conditions.

Table \ref{tab:Mixing} shows how large the mixture amplitude is compared to its main state, for example, $|A_{h2}/A_{h_1}|$ for \eqref{eq:phi2}. The quantum well is set to be 10 nm wide and the in-plane wave vector is 0.002 \AA$^{-1}$ (inverse of a 50 nm confinement) with $20\%$ silicon in the barrier. We can see the correction to the envelope function is very small; thus, the correction to energy is almost negligible. However, the correction can be significant if the in-plane confinement is comparable to the out-of-plane confinement.
\begin{table}[h]
\begin{center}
\begin{tabular}{ |c|c|c| } 
 \hline
    & $|A_{\rm mixture}/A_{\rm main}|$\\
  \hline
  HH1 & $0.07\%$ \\
  \hline
  HH2 & $0.14\%$ \\
  \hline
  HH3 & $0.19\%$ \\
  \hline
  LH1 & $1.47\%$ \\
  \hline
  LH2 & $2.52\%$ \\ 
 \hline
\end{tabular}
\caption{Mixture amplitude as a percentage of its main state in a Ge/Ge$_{0.8}$Si$_{0.2}$ heterostructure.}
\label{tab:Mixing}
\end{center}
\end{table}

The above calculation is to provide a way to examine the effect of the multi-band boundary conditions. This artificial treatment to the coupled boundary conditions was not applied to the $g$-factor calculations because 
there is not a self-consistent way to combine LCAO and the coupled boundary conditions. One can only estimate the impact to the single-band basis states, in a relatively crude way.
In an accurate quantum dot calculation, the in-plane wave vectors are operators instead of numbers. Such a calculation is computationally challenging which is beyond the scope of this work.

\section{Spin splitting with a different HH-LH splitting}
\label{app:paradigm2}

\begin{figure}[ht]%[htbp]
    \centering
    \includegraphics[width=0.4\textwidth]{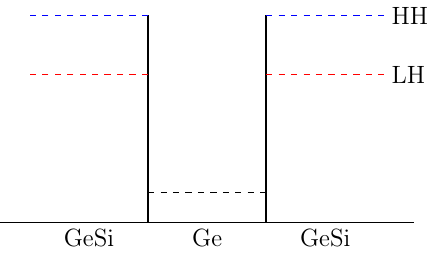}
    \caption{A fictitious band alignment of the Ge/GeSi heterostructure if the GeSi barriers are strained and the Ge layer is relaxed.}
    \label{fig:paradigm2}
\end{figure}

The purpose of this section is to explore the theoretical possibility of the HH-LH splitting, $\Delta_{\rm HL}$, being smaller. Here we consider the scenario of the germanium layer being relaxed while the two GeSi barriers are strained by the lattice matching mechanism. The resulting band alignment will be as pictured in Fig. \ref{fig:paradigm2}. Comparing the results of this scenario to the realistic case, we can see how the size of $\Delta_{\rm HL}$ affects the qubit Zeeman splitting. Therefore, we reproduced the same type of calculations with these assumptions.

\begin{figure}[ht] %[htbp]
    \centering
    \includegraphics*[width=8.0cm]{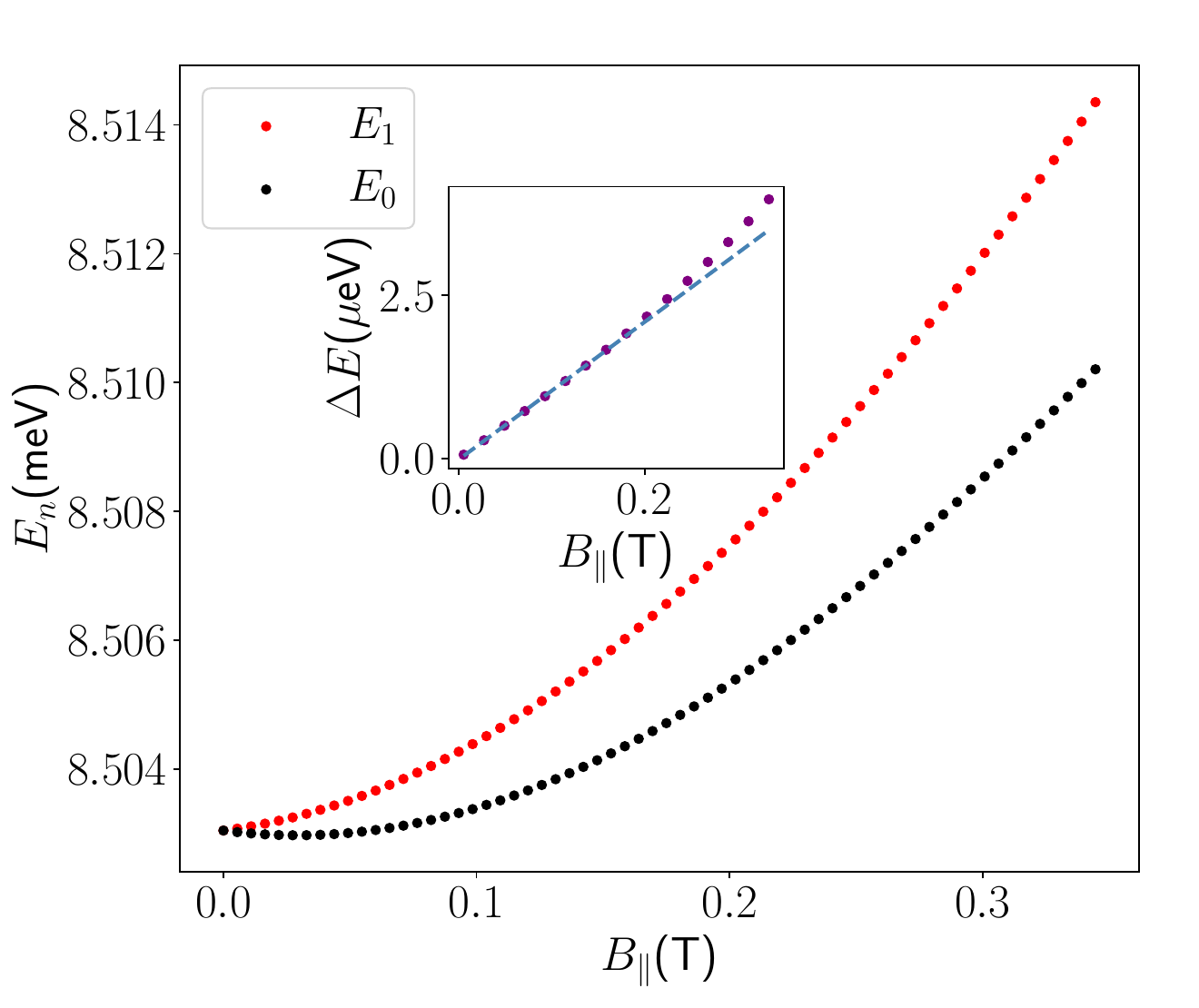}
    \includegraphics*[width=8.0cm]{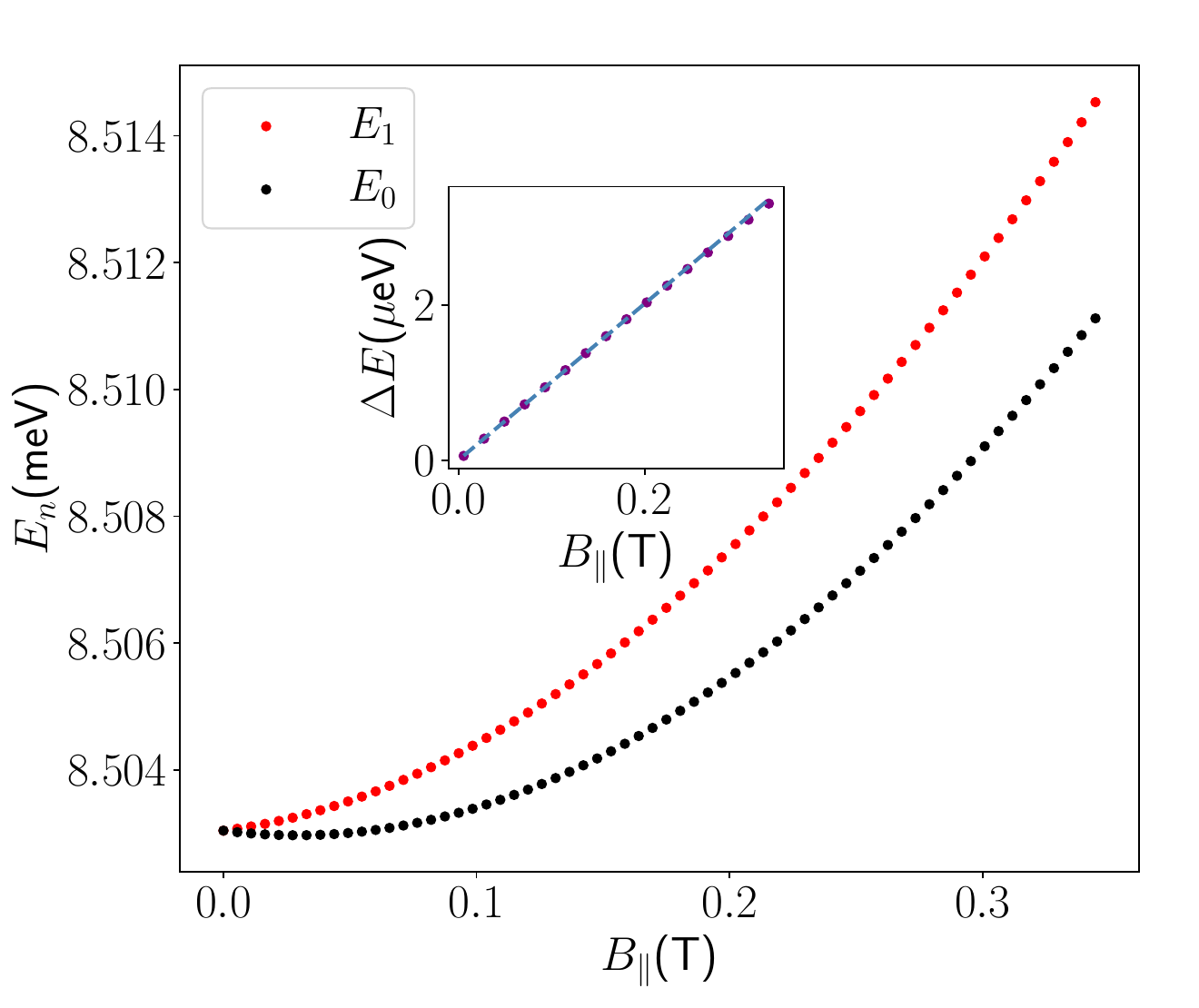}
    \caption{Energy spectra of the same setup of quantum dot as Fig. \ref{fig:DEvsB} in a different strain paradigm with an applied field in the (left) [100] and (right) [110] directions. Nonlinear behavior can already be seen for the spin splitting when the field is in the [100] direction with a magnitude larger than 0.2 T.}
    \label{fig:DEvsB_paradigm2}
\end{figure}

The energy spectra in Fig. \ref{fig:DEip_paradigm2} show an obvious nonlinear splitting if the applied magnetic field is in the [100] direction when the field is larger than 0.2 T. The nonlinear effect is less dramatic in the [110] direction, and thus, a strong anisotropy in the $x$-$y$ plane can be expected, as shown in the left plot of Fig. \ref{fig:DEip_paradigm2}, where we applied a 0.35 T in-plane field. Here we abandoned the concept of ``effective $g$-factor'' but showed energy splitting instead because we no longer have a field-independent $g$-factor. 

\begin{figure}[htbp] %[H]
    \centering
    \includegraphics[width=0.5\textwidth]{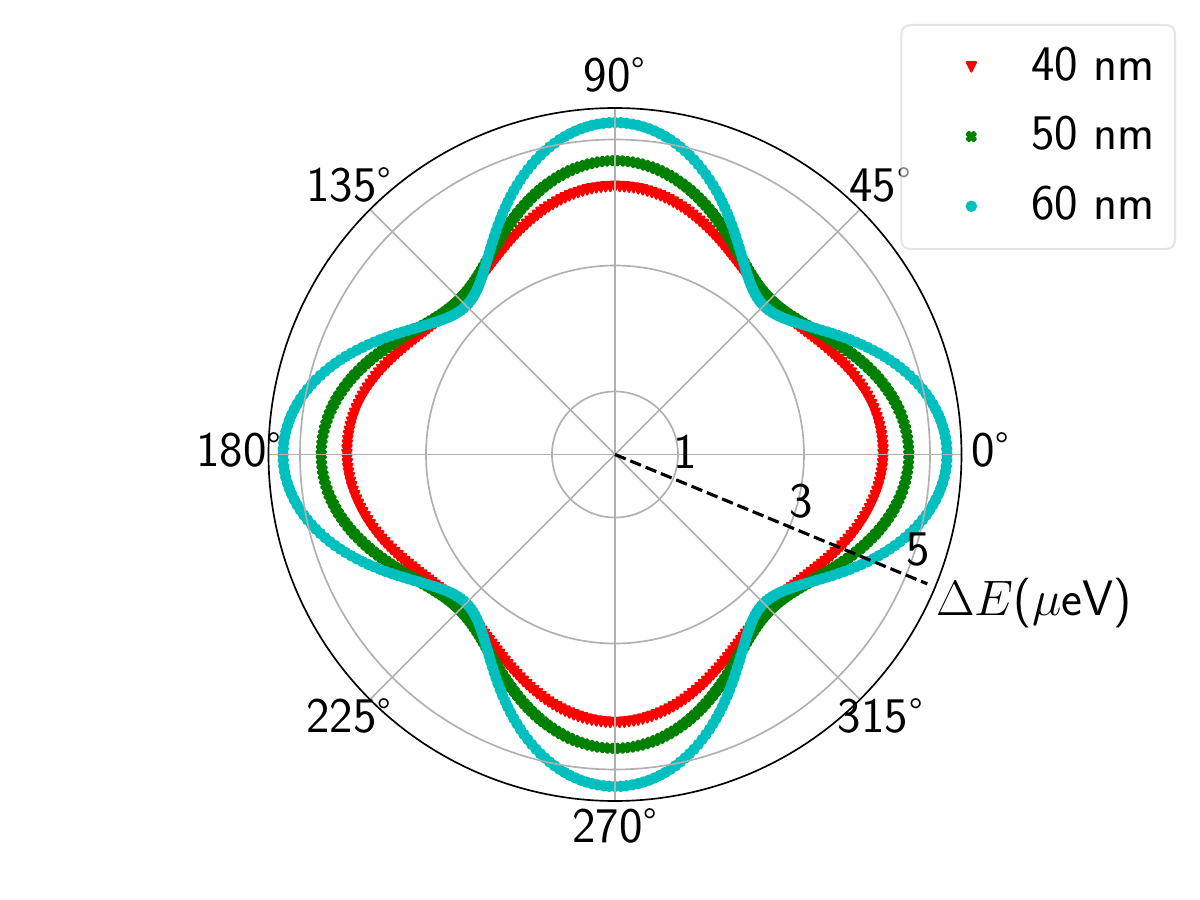}
    \includegraphics[width=0.48\textwidth]{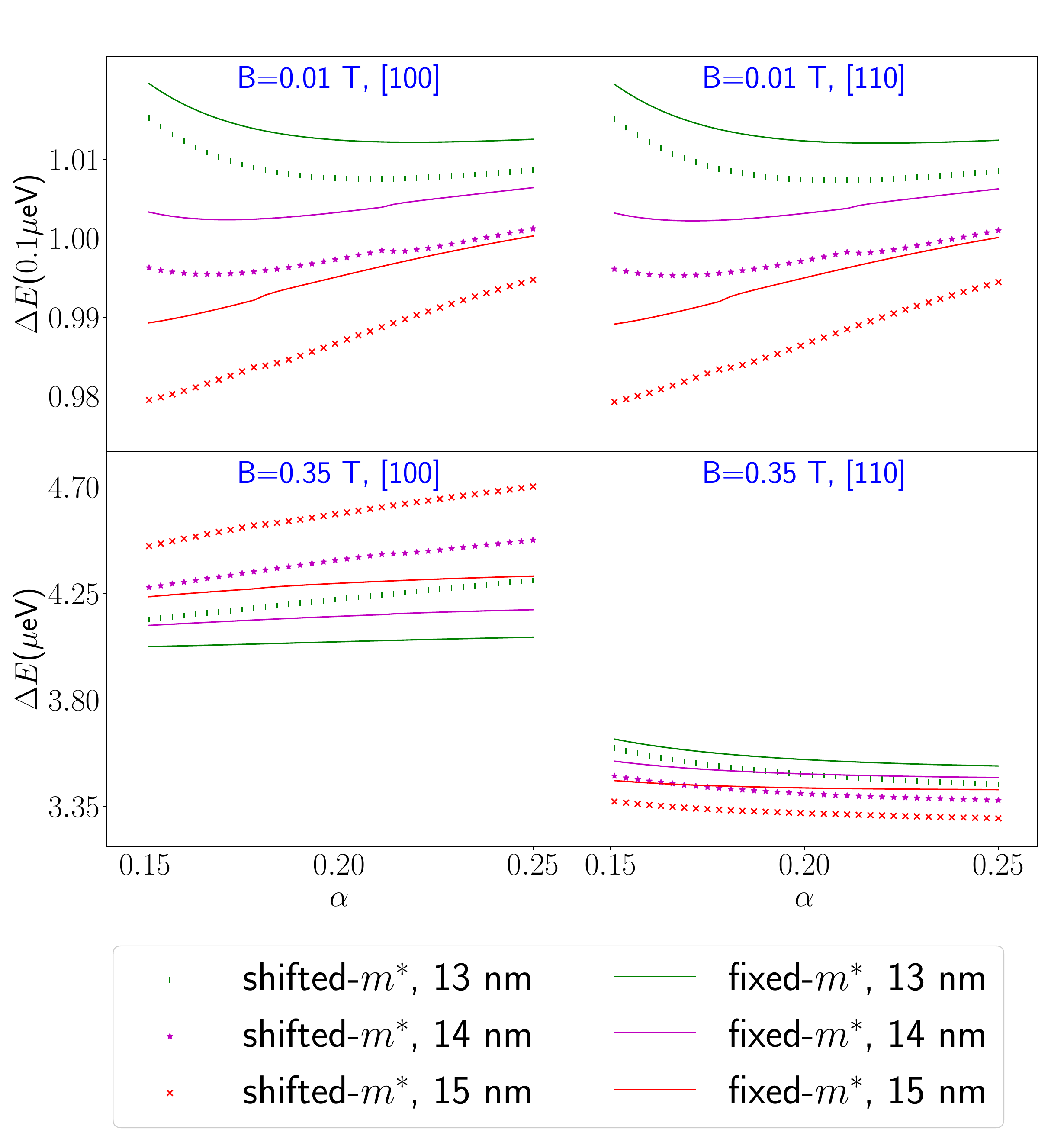}
    \caption{(Left) Spin splitting of a dot with a 13-nm-wide quantum well and an in-plane magnetic field applied in all different directions. The [100] direction is when $\theta=0^\circ$. (Right) Spin splitting of a 40 nm dot as a function of silicon concentration $\alpha$ in the [100] and [110] directions. Results include the weak-field limit (0.01 T) and a regular $B$ field (0.35 T), with three different well widths, in two treatments of boundary conditions.}
    \label{fig:DEip_paradigm2}
\end{figure}

The in-plane anisotropy is clearly sensitive to the size of the dot, where a larger dot has a stronger anisotropy. From the right plot of Fig. \ref{fig:DEip_paradigm2}, we can see the nonlinear effect can change dramatically how the well width affects the spin splitting. It also increases the sensitivity to the well width as well as the discrepancy between the two treatments of boundary conditions.

\begin{figure}[htbp] %[H]
    \centering
    \includegraphics[width=0.5\textwidth]{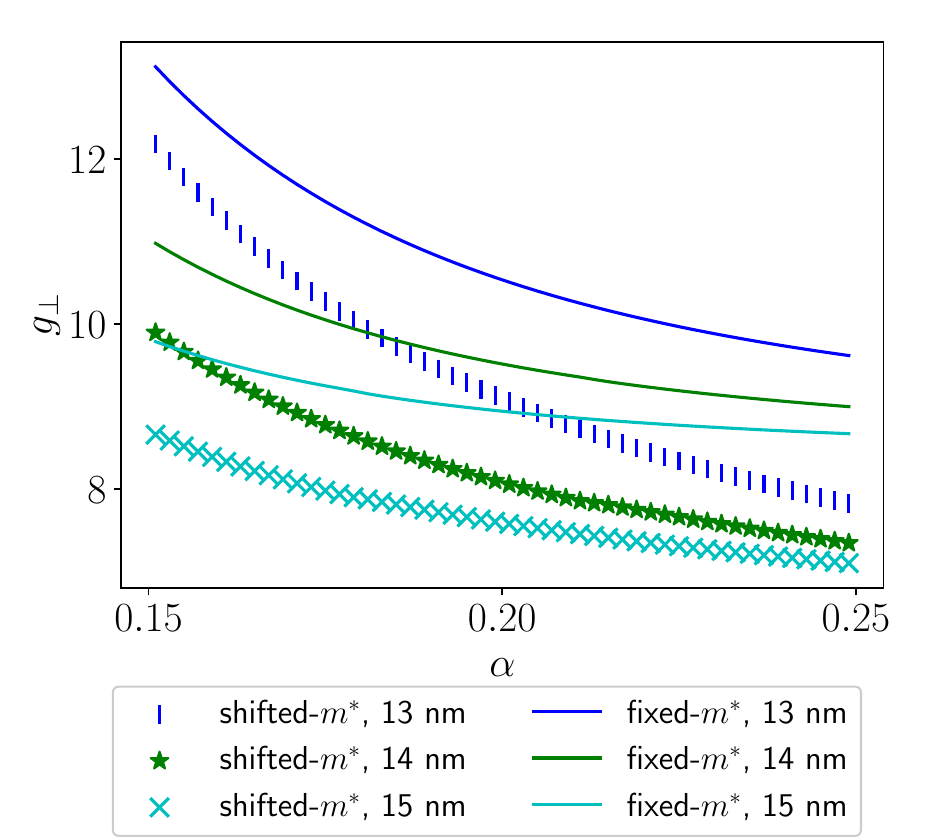}
    \caption{Similar to Fig. \ref{fig:gzwidth} but with a different band alignment. The effective $g_\perp$ has a stronger reduction in this paradigm.}
    \label{fig:gzwidth_paradigm2}
\end{figure}

The stronger HH-LH coupling reduces the value of $g_\perp$. Compared with the original paradigm, the effective $g_\perp$ becomes less sensitive to well width with the increase of silicon concentration. Other than that, the value of $g_\perp$ seems to converge to a certain value with growing $\alpha$. The reason is that the HH-LH splitting comes exclusively from the different energies of the lowest sub-bands of HH and LH. As $\alpha$ increases, the quantum wells of HH and LH will become closer and closer to the hard-wall scenario.

\end{widetext}
\bibliography{cite}

%apsrev4-2.bst 2019-01-14 (MD) hand-edited version of apsrev4-1.bst
%Control: key (0)
%Control: author (8) initials jnrlst
%Control: editor formatted (1) identically to author
%Control: production of article title (0) allowed
%Control: page (0) single
%Control: year (1) truncated
%Control: production of eprint (0) enabled
\begin{thebibliography}{77}%
\makeatletter
\providecommand \@ifxundefined [1]{%
 \@ifx{#1\undefined}
}%
\providecommand \@ifnum [1]{%
 \ifnum #1\expandafter \@firstoftwo
 \else \expandafter \@secondoftwo
 \fi
}%
\providecommand \@ifx [1]{%
 \ifx #1\expandafter \@firstoftwo
 \else \expandafter \@secondoftwo
 \fi
}%
\providecommand \natexlab [1]{#1}%
\providecommand \enquote  [1]{``#1''}%
\providecommand \bibnamefont  [1]{#1}%
\providecommand \bibfnamefont [1]{#1}%
\providecommand \citenamefont [1]{#1}%
\providecommand \href@noop [0]{\@secondoftwo}%
\providecommand \href [0]{\begingroup \@sanitize@url \@href}%
\providecommand \@href[1]{\@@startlink{#1}\@@href}%
\providecommand \@@href[1]{\endgroup#1\@@endlink}%
\providecommand \@sanitize@url [0]{\catcode `\\12\catcode `\$12\catcode
  `\&12\catcode `\#12\catcode `\^12\catcode `\_12\catcode `\%12\relax}%
\providecommand \@@startlink[1]{}%
\providecommand \@@endlink[0]{}%
\providecommand \url  [0]{\begingroup\@sanitize@url \@url }%
\providecommand \@url [1]{\endgroup\@href {#1}{\urlprefix }}%
\providecommand \urlprefix  [0]{URL }%
\providecommand \Eprint [0]{\href }%
\providecommand \doibase [0]{https://doi.org/}%
\providecommand \selectlanguage [0]{\@gobble}%
\providecommand \bibinfo  [0]{\@secondoftwo}%
\providecommand \bibfield  [0]{\@secondoftwo}%
\providecommand \translation [1]{[#1]}%
\providecommand \BibitemOpen [0]{}%
\providecommand \bibitemStop [0]{}%
\providecommand \bibitemNoStop [0]{.\EOS\space}%
\providecommand \EOS [0]{\spacefactor3000\relax}%
\providecommand \BibitemShut  [1]{\csname bibitem#1\endcsname}%
\let\auto@bib@innerbib\@empty
%</preamble>
\bibitem [{\citenamefont {Loss}\ and\ \citenamefont
  {DiVincenzo}(1998)}]{1998LossDiVincenzo}%
  \BibitemOpen
  \bibfield  {author} {\bibinfo {author} {\bibfnamefont {D.}~\bibnamefont
  {Loss}}\ and\ \bibinfo {author} {\bibfnamefont {D.~P.}\ \bibnamefont
  {DiVincenzo}},\ }\bibfield  {title} {\bibinfo {title} {Quantum computation
  with quantum dots},\ }\href {https://doi.org/10.1103/PhysRevA.57.120}
  {\bibfield  {journal} {\bibinfo  {journal} {Phys. Rev. A}\ }\textbf {\bibinfo
  {volume} {57}},\ \bibinfo {pages} {120} (\bibinfo {year} {1998})}\BibitemShut
  {NoStop}%
\bibitem [{\citenamefont {Kane}(1998)}]{1998KaneNat393}%
  \BibitemOpen
  \bibfield  {author} {\bibinfo {author} {\bibfnamefont {B.~E.}\ \bibnamefont
  {Kane}},\ }\bibfield  {title} {\bibinfo {title} {A silicon-based nuclear spin
  quantum computer},\ }\href {https://doi.org/10.1038/30156} {\bibfield
  {journal} {\bibinfo  {journal} {Nature}\ }\textbf {\bibinfo {volume} {393}},\
  \bibinfo {pages} {133} (\bibinfo {year} {1998})}\BibitemShut {NoStop}%
\bibitem [{\citenamefont {Petta}\ \emph {et~al.}(2005)\citenamefont {Petta},
  \citenamefont {Johnson}, \citenamefont {Taylor}, \citenamefont {Laird},
  \citenamefont {Yacoby}, \citenamefont {Lukin}, \citenamefont {Marcus},
  \citenamefont {Hanson},\ and\ \citenamefont {Gossard}}]{2005PettaSci309}%
  \BibitemOpen
  \bibfield  {author} {\bibinfo {author} {\bibfnamefont {J.~R.}\ \bibnamefont
  {Petta}}, \bibinfo {author} {\bibfnamefont {A.~C.}\ \bibnamefont {Johnson}},
  \bibinfo {author} {\bibfnamefont {J.~M.}\ \bibnamefont {Taylor}}, \bibinfo
  {author} {\bibfnamefont {E.~A.}\ \bibnamefont {Laird}}, \bibinfo {author}
  {\bibfnamefont {A.}~\bibnamefont {Yacoby}}, \bibinfo {author} {\bibfnamefont
  {M.~D.}\ \bibnamefont {Lukin}}, \bibinfo {author} {\bibfnamefont {C.~M.}\
  \bibnamefont {Marcus}}, \bibinfo {author} {\bibfnamefont {M.~P.}\
  \bibnamefont {Hanson}},\ and\ \bibinfo {author} {\bibfnamefont {A.~C.}\
  \bibnamefont {Gossard}},\ }\bibfield  {title} {\bibinfo {title} {Coherent
  manipulation of coupled electron spins in semiconductor quantum dots},\
  }\href {https://doi.org/10.1126/science.1116955} {\bibfield  {journal}
  {\bibinfo  {journal} {Science}\ }\textbf {\bibinfo {volume} {309}},\ \bibinfo
  {pages} {2180} (\bibinfo {year} {2005})}\BibitemShut {NoStop}%
\bibitem [{\citenamefont {Hanson}\ \emph {et~al.}(2007)\citenamefont {Hanson},
  \citenamefont {Kouwenhoven}, \citenamefont {Petta}, \citenamefont {Tarucha},\
  and\ \citenamefont {Vandersypen}}]{2007HansonRMP79}%
  \BibitemOpen
  \bibfield  {author} {\bibinfo {author} {\bibfnamefont {R.}~\bibnamefont
  {Hanson}}, \bibinfo {author} {\bibfnamefont {L.~P.}\ \bibnamefont
  {Kouwenhoven}}, \bibinfo {author} {\bibfnamefont {J.~R.}\ \bibnamefont
  {Petta}}, \bibinfo {author} {\bibfnamefont {S.}~\bibnamefont {Tarucha}},\
  and\ \bibinfo {author} {\bibfnamefont {L.~M.~K.}\ \bibnamefont
  {Vandersypen}},\ }\bibfield  {title} {\bibinfo {title} {Spins in few-electron
  quantum dots},\ }\href {https://doi.org/10.1103/RevModPhys.79.1217}
  {\bibfield  {journal} {\bibinfo  {journal} {Rev. Mod. Phys.}\ }\textbf
  {\bibinfo {volume} {79}},\ \bibinfo {pages} {1217} (\bibinfo {year}
  {2007})}\BibitemShut {NoStop}%
\bibitem [{\citenamefont {Hanson}\ and\ \citenamefont
  {Awschalom}(2008)}]{2008HansonNat453}%
  \BibitemOpen
  \bibfield  {author} {\bibinfo {author} {\bibfnamefont {R.}~\bibnamefont
  {Hanson}}\ and\ \bibinfo {author} {\bibfnamefont {D.~D.}\ \bibnamefont
  {Awschalom}},\ }\bibfield  {title} {\bibinfo {title} {Coherent manipulation
  of single spins in semiconductors},\ }\href
  {https://doi.org/10.1038/nature07129} {\bibfield  {journal} {\bibinfo
  {journal} {Nature}\ }\textbf {\bibinfo {volume} {453}},\ \bibinfo {pages}
  {1043} (\bibinfo {year} {2008})}\BibitemShut {NoStop}%
\bibitem [{\citenamefont {Zwanenburg}\ \emph {et~al.}(2013)\citenamefont
  {Zwanenburg}, \citenamefont {Dzurak}, \citenamefont {Morello}, \citenamefont
  {Simmons}, \citenamefont {Hollenberg}, \citenamefont {Klimeck}, \citenamefont
  {Rogge}, \citenamefont {Coppersmith},\ and\ \citenamefont
  {Eriksson}}]{2013ZwanenburgRMP85}%
  \BibitemOpen
  \bibfield  {author} {\bibinfo {author} {\bibfnamefont {F.~A.}\ \bibnamefont
  {Zwanenburg}}, \bibinfo {author} {\bibfnamefont {A.~S.}\ \bibnamefont
  {Dzurak}}, \bibinfo {author} {\bibfnamefont {A.}~\bibnamefont {Morello}},
  \bibinfo {author} {\bibfnamefont {M.~Y.}\ \bibnamefont {Simmons}}, \bibinfo
  {author} {\bibfnamefont {L.~C.~L.}\ \bibnamefont {Hollenberg}}, \bibinfo
  {author} {\bibfnamefont {G.}~\bibnamefont {Klimeck}}, \bibinfo {author}
  {\bibfnamefont {S.}~\bibnamefont {Rogge}}, \bibinfo {author} {\bibfnamefont
  {S.~N.}\ \bibnamefont {Coppersmith}},\ and\ \bibinfo {author} {\bibfnamefont
  {M.~A.}\ \bibnamefont {Eriksson}},\ }\bibfield  {title} {\bibinfo {title}
  {Silicon quantum electronics},\ }\href
  {https://doi.org/10.1103/RevModPhys.85.961} {\bibfield  {journal} {\bibinfo
  {journal} {Rev. Mod. Phys.}\ }\textbf {\bibinfo {volume} {85}},\ \bibinfo
  {pages} {961} (\bibinfo {year} {2013})}\BibitemShut {NoStop}%
\bibitem [{\citenamefont {Chatterjee}\ \emph {et~al.}(2021)\citenamefont
  {Chatterjee}, \citenamefont {Stevenson}, \citenamefont {De~Franceschi},
  \citenamefont {Morello}, \citenamefont {de~Leon},\ and\ \citenamefont
  {Kuemmeth}}]{2021ChatterjeeNatRevPhys3}%
  \BibitemOpen
  \bibfield  {author} {\bibinfo {author} {\bibfnamefont {A.}~\bibnamefont
  {Chatterjee}}, \bibinfo {author} {\bibfnamefont {P.}~\bibnamefont
  {Stevenson}}, \bibinfo {author} {\bibfnamefont {S.}~\bibnamefont
  {De~Franceschi}}, \bibinfo {author} {\bibfnamefont {A.}~\bibnamefont
  {Morello}}, \bibinfo {author} {\bibfnamefont {N.~P.}\ \bibnamefont
  {de~Leon}},\ and\ \bibinfo {author} {\bibfnamefont {F.}~\bibnamefont
  {Kuemmeth}},\ }\bibfield  {title} {\bibinfo {title} {Semiconductor qubits in
  practice},\ }\href {https://doi.org/10.1038/s42254-021-00283-9} {\bibfield
  {journal} {\bibinfo  {journal} {Nature Reviews Physics}\ }\textbf {\bibinfo
  {volume} {3}},\ \bibinfo {pages} {157} (\bibinfo {year} {2021})}\BibitemShut
  {NoStop}%
\bibitem [{\citenamefont {Scappucci}\ \emph {et~al.}(2021)\citenamefont
  {Scappucci}, \citenamefont {Kloeffel}, \citenamefont {Zwanenburg},
  \citenamefont {Loss}, \citenamefont {Myronov}, \citenamefont {Zhang},
  \citenamefont {De~Franceschi}, \citenamefont {Katsaros},\ and\ \citenamefont
  {Veldhorst}}]{2021ScappucciNatRevMat6}%
  \BibitemOpen
  \bibfield  {author} {\bibinfo {author} {\bibfnamefont {G.}~\bibnamefont
  {Scappucci}}, \bibinfo {author} {\bibfnamefont {C.}~\bibnamefont {Kloeffel}},
  \bibinfo {author} {\bibfnamefont {F.~A.}\ \bibnamefont {Zwanenburg}},
  \bibinfo {author} {\bibfnamefont {D.}~\bibnamefont {Loss}}, \bibinfo {author}
  {\bibfnamefont {M.}~\bibnamefont {Myronov}}, \bibinfo {author} {\bibfnamefont
  {J.-J.}\ \bibnamefont {Zhang}}, \bibinfo {author} {\bibfnamefont
  {S.}~\bibnamefont {De~Franceschi}}, \bibinfo {author} {\bibfnamefont
  {G.}~\bibnamefont {Katsaros}},\ and\ \bibinfo {author} {\bibfnamefont
  {M.}~\bibnamefont {Veldhorst}},\ }\bibfield  {title} {\bibinfo {title} {The
  germanium quantum information route},\ }\href
  {https://doi.org/10.1038/s41578-020-00262-z} {\bibfield  {journal} {\bibinfo
  {journal} {Nature Reviews Materials}\ }\textbf {\bibinfo {volume} {6}},\
  \bibinfo {pages} {926} (\bibinfo {year} {2021})}\BibitemShut {NoStop}%
\bibitem [{\citenamefont {Fang}\ \emph {et~al.}(2023)\citenamefont {Fang},
  \citenamefont {Philippopoulos}, \citenamefont {Culcer}, \citenamefont
  {Coish},\ and\ \citenamefont {Chesi}}]{2023FangMatQuantTech3}%
  \BibitemOpen
  \bibfield  {author} {\bibinfo {author} {\bibfnamefont {Y.}~\bibnamefont
  {Fang}}, \bibinfo {author} {\bibfnamefont {P.}~\bibnamefont
  {Philippopoulos}}, \bibinfo {author} {\bibfnamefont {D.}~\bibnamefont
  {Culcer}}, \bibinfo {author} {\bibfnamefont {W.~A.}\ \bibnamefont {Coish}},\
  and\ \bibinfo {author} {\bibfnamefont {S.}~\bibnamefont {Chesi}},\ }\bibfield
   {title} {\bibinfo {title} {Recent advances in hole-spin qubits},\ }\href
  {https://doi.org/10.1088/2633-4356/acb87e} {\bibfield  {journal} {\bibinfo
  {journal} {Materials for Quantum Technology}\ }\textbf {\bibinfo {volume}
  {3}},\ \bibinfo {pages} {012003} (\bibinfo {year} {2023})}\BibitemShut
  {NoStop}%
\bibitem [{\citenamefont {Rashba}(1965)}]{1965Rashba}%
  \BibitemOpen
  \bibfield  {author} {\bibinfo {author} {\bibfnamefont {{\'E}.~I.}\
  \bibnamefont {Rashba}},\ }\bibfield  {title} {\bibinfo {title} {Combined
  resonance in semiconductors},\ }\href
  {https://doi.org/10.1070/PU1965v007n06ABEH003687} {\bibfield  {journal}
  {\bibinfo  {journal} {Soviet Physics Uspekhi}\ }\textbf {\bibinfo {volume}
  {7}},\ \bibinfo {pages} {823} (\bibinfo {year} {1965})}\BibitemShut {NoStop}%
\bibitem [{\citenamefont {Yoneda}\ \emph {et~al.}(2018)\citenamefont {Yoneda},
  \citenamefont {Takeda}, \citenamefont {Otsuka}, \citenamefont {Nakajima},
  \citenamefont {Delbecq}, \citenamefont {Allison}, \citenamefont {Honda},
  \citenamefont {Kodera}, \citenamefont {Oda}, \citenamefont {Hoshi},
  \citenamefont {Usami}, \citenamefont {Itoh},\ and\ \citenamefont
  {Tarucha}}]{2018YonedaNatNanoTech13}%
  \BibitemOpen
  \bibfield  {author} {\bibinfo {author} {\bibfnamefont {J.}~\bibnamefont
  {Yoneda}}, \bibinfo {author} {\bibfnamefont {K.}~\bibnamefont {Takeda}},
  \bibinfo {author} {\bibfnamefont {T.}~\bibnamefont {Otsuka}}, \bibinfo
  {author} {\bibfnamefont {T.}~\bibnamefont {Nakajima}}, \bibinfo {author}
  {\bibfnamefont {M.~R.}\ \bibnamefont {Delbecq}}, \bibinfo {author}
  {\bibfnamefont {G.}~\bibnamefont {Allison}}, \bibinfo {author} {\bibfnamefont
  {T.}~\bibnamefont {Honda}}, \bibinfo {author} {\bibfnamefont
  {T.}~\bibnamefont {Kodera}}, \bibinfo {author} {\bibfnamefont
  {S.}~\bibnamefont {Oda}}, \bibinfo {author} {\bibfnamefont {Y.}~\bibnamefont
  {Hoshi}}, \bibinfo {author} {\bibfnamefont {N.}~\bibnamefont {Usami}},
  \bibinfo {author} {\bibfnamefont {K.~M.}\ \bibnamefont {Itoh}},\ and\
  \bibinfo {author} {\bibfnamefont {S.}~\bibnamefont {Tarucha}},\ }\bibfield
  {title} {\bibinfo {title} {A quantum-dot spin qubit with coherence limited by
  charge noise and fidelity higher than 99.9\%},\ }\href
  {https://doi.org/10.1038/s41565-017-0014-x} {\bibfield  {journal} {\bibinfo
  {journal} {Nature Nanotechnology}\ }\textbf {\bibinfo {volume} {13}},\
  \bibinfo {pages} {102} (\bibinfo {year} {2018})}\BibitemShut {NoStop}%
\bibitem [{\citenamefont {Zajac}\ \emph {et~al.}(2018)\citenamefont {Zajac},
  \citenamefont {Sigillito}, \citenamefont {Russ}, \citenamefont {Borjans},
  \citenamefont {Taylor}, \citenamefont {Burkard},\ and\ \citenamefont
  {Petta}}]{2018ZajacSci359}%
  \BibitemOpen
  \bibfield  {author} {\bibinfo {author} {\bibfnamefont {D.~M.}\ \bibnamefont
  {Zajac}}, \bibinfo {author} {\bibfnamefont {A.~J.}\ \bibnamefont
  {Sigillito}}, \bibinfo {author} {\bibfnamefont {M.}~\bibnamefont {Russ}},
  \bibinfo {author} {\bibfnamefont {F.}~\bibnamefont {Borjans}}, \bibinfo
  {author} {\bibfnamefont {J.~M.}\ \bibnamefont {Taylor}}, \bibinfo {author}
  {\bibfnamefont {G.}~\bibnamefont {Burkard}},\ and\ \bibinfo {author}
  {\bibfnamefont {J.~R.}\ \bibnamefont {Petta}},\ }\bibfield  {title} {\bibinfo
  {title} {Resonantly driven cnot gate for electron spins},\ }\href
  {https://doi.org/10.1126/science.aao5965} {\bibfield  {journal} {\bibinfo
  {journal} {Science}\ }\textbf {\bibinfo {volume} {359}},\ \bibinfo {pages}
  {439} (\bibinfo {year} {2018})}\BibitemShut {NoStop}%
\bibitem [{\citenamefont {Noiri}\ \emph {et~al.}(2022)\citenamefont {Noiri},
  \citenamefont {Takeda}, \citenamefont {Nakajima}, \citenamefont {Kobayashi},
  \citenamefont {Sammak}, \citenamefont {Scappucci},\ and\ \citenamefont
  {Tarucha}}]{2022NoiriNat601}%
  \BibitemOpen
  \bibfield  {author} {\bibinfo {author} {\bibfnamefont {A.}~\bibnamefont
  {Noiri}}, \bibinfo {author} {\bibfnamefont {K.}~\bibnamefont {Takeda}},
  \bibinfo {author} {\bibfnamefont {T.}~\bibnamefont {Nakajima}}, \bibinfo
  {author} {\bibfnamefont {T.}~\bibnamefont {Kobayashi}}, \bibinfo {author}
  {\bibfnamefont {A.}~\bibnamefont {Sammak}}, \bibinfo {author} {\bibfnamefont
  {G.}~\bibnamefont {Scappucci}},\ and\ \bibinfo {author} {\bibfnamefont
  {S.}~\bibnamefont {Tarucha}},\ }\bibfield  {title} {\bibinfo {title} {Fast
  universal quantum gate above the fault-tolerance threshold in silicon},\
  }\href {https://doi.org/10.1038/s41586-021-04182-y} {\bibfield  {journal}
  {\bibinfo  {journal} {Nature}\ }\textbf {\bibinfo {volume} {601}},\ \bibinfo
  {pages} {338} (\bibinfo {year} {2022})}\BibitemShut {NoStop}%
\bibitem [{\citenamefont {Mills}\ \emph {et~al.}(2022)\citenamefont {Mills},
  \citenamefont {Guinn}, \citenamefont {Gullans}, \citenamefont {Sigillito},
  \citenamefont {Feldman}, \citenamefont {Nielsen},\ and\ \citenamefont
  {Petta}}]{2022MillsSciAdv8}%
  \BibitemOpen
  \bibfield  {author} {\bibinfo {author} {\bibfnamefont {A.~R.}\ \bibnamefont
  {Mills}}, \bibinfo {author} {\bibfnamefont {C.~R.}\ \bibnamefont {Guinn}},
  \bibinfo {author} {\bibfnamefont {M.~J.}\ \bibnamefont {Gullans}}, \bibinfo
  {author} {\bibfnamefont {A.~J.}\ \bibnamefont {Sigillito}}, \bibinfo {author}
  {\bibfnamefont {M.~M.}\ \bibnamefont {Feldman}}, \bibinfo {author}
  {\bibfnamefont {E.}~\bibnamefont {Nielsen}},\ and\ \bibinfo {author}
  {\bibfnamefont {J.~R.}\ \bibnamefont {Petta}},\ }\bibfield  {title} {\bibinfo
  {title} {Two-qubit silicon quantum processor with operation fidelity
  exceeding 99\%},\ }\href {https://doi.org/10.1126/sciadv.abn5130} {\bibfield
  {journal} {\bibinfo  {journal} {Science Advances}\ }\textbf {\bibinfo
  {volume} {8}},\ \bibinfo {pages} {eabn5130} (\bibinfo {year}
  {2022})}\BibitemShut {NoStop}%
\bibitem [{\citenamefont {Xue}\ \emph {et~al.}(2022)\citenamefont {Xue},
  \citenamefont {Russ}, \citenamefont {Samkharadze}, \citenamefont {Undseth},
  \citenamefont {Sammak}, \citenamefont {Scappucci},\ and\ \citenamefont
  {Vandersypen}}]{2022XueNat601}%
  \BibitemOpen
  \bibfield  {author} {\bibinfo {author} {\bibfnamefont {X.}~\bibnamefont
  {Xue}}, \bibinfo {author} {\bibfnamefont {M.}~\bibnamefont {Russ}}, \bibinfo
  {author} {\bibfnamefont {N.}~\bibnamefont {Samkharadze}}, \bibinfo {author}
  {\bibfnamefont {B.}~\bibnamefont {Undseth}}, \bibinfo {author} {\bibfnamefont
  {A.}~\bibnamefont {Sammak}}, \bibinfo {author} {\bibfnamefont
  {G.}~\bibnamefont {Scappucci}},\ and\ \bibinfo {author} {\bibfnamefont
  {L.~M.~K.}\ \bibnamefont {Vandersypen}},\ }\bibfield  {title} {\bibinfo
  {title} {Quantum logic with spin qubits crossing the surface code
  threshold},\ }\href {https://doi.org/10.1038/s41586-021-04273-w} {\bibfield
  {journal} {\bibinfo  {journal} {Nature}\ }\textbf {\bibinfo {volume} {601}},\
  \bibinfo {pages} {343} (\bibinfo {year} {2022})}\BibitemShut {NoStop}%
\bibitem [{\citenamefont {Burkard}\ \emph {et~al.}(2023)\citenamefont
  {Burkard}, \citenamefont {Ladd}, \citenamefont {Pan}, \citenamefont
  {Nichol},\ and\ \citenamefont {Petta}}]{2023BurkardRMP95}%
  \BibitemOpen
  \bibfield  {author} {\bibinfo {author} {\bibfnamefont {G.}~\bibnamefont
  {Burkard}}, \bibinfo {author} {\bibfnamefont {T.~D.}\ \bibnamefont {Ladd}},
  \bibinfo {author} {\bibfnamefont {A.}~\bibnamefont {Pan}}, \bibinfo {author}
  {\bibfnamefont {J.~M.}\ \bibnamefont {Nichol}},\ and\ \bibinfo {author}
  {\bibfnamefont {J.~R.}\ \bibnamefont {Petta}},\ }\bibfield  {title} {\bibinfo
  {title} {Semiconductor spin qubits},\ }\href
  {https://doi.org/10.1103/RevModPhys.95.025003} {\bibfield  {journal}
  {\bibinfo  {journal} {Reviews of Modern Physics}\ }\textbf {\bibinfo {volume}
  {95}},\ \bibinfo {pages} {025003} (\bibinfo {year} {2023})},\ \bibinfo {note}
  {rMP}\BibitemShut {NoStop}%
\bibitem [{\citenamefont {Woods}\ \emph {et~al.}(2004)\citenamefont {Woods},
  \citenamefont {Reinecke},\ and\ \citenamefont {Kotlyar}}]{2004WoodsPRB69}%
  \BibitemOpen
  \bibfield  {author} {\bibinfo {author} {\bibfnamefont {L.~M.}\ \bibnamefont
  {Woods}}, \bibinfo {author} {\bibfnamefont {T.~L.}\ \bibnamefont
  {Reinecke}},\ and\ \bibinfo {author} {\bibfnamefont {R.}~\bibnamefont
  {Kotlyar}},\ }\bibfield  {title} {\bibinfo {title} {Hole spin relaxation in
  quantum dots},\ }\href {https://doi.org/10.1103/PhysRevB.69.125330}
  {\bibfield  {journal} {\bibinfo  {journal} {Phys. Rev. B}\ }\textbf {\bibinfo
  {volume} {69}},\ \bibinfo {pages} {125330} (\bibinfo {year}
  {2004})}\BibitemShut {NoStop}%
\bibitem [{\citenamefont {Bulaev}\ and\ \citenamefont
  {Loss}(2005)}]{2005BulaevPRL95}%
  \BibitemOpen
  \bibfield  {author} {\bibinfo {author} {\bibfnamefont {D.~V.}\ \bibnamefont
  {Bulaev}}\ and\ \bibinfo {author} {\bibfnamefont {D.}~\bibnamefont {Loss}},\
  }\bibfield  {title} {\bibinfo {title} {Spin relaxation and decoherence of
  holes in quantum dots},\ }\href
  {https://doi.org/10.1103/PhysRevLett.95.076805} {\bibfield  {journal}
  {\bibinfo  {journal} {Phys. Rev. Lett.}\ }\textbf {\bibinfo {volume} {95}},\
  \bibinfo {pages} {076805} (\bibinfo {year} {2005})}\BibitemShut {NoStop}%
\bibitem [{\citenamefont {Danneau}\ \emph {et~al.}(2006)\citenamefont
  {Danneau}, \citenamefont {Klochan}, \citenamefont {Clarke}, \citenamefont
  {Ho}, \citenamefont {Micolich}, \citenamefont {Simmons}, \citenamefont
  {Hamilton}, \citenamefont {Pepper}, \citenamefont {Ritchie},\ and\
  \citenamefont {Zülicke}}]{2006DanneauPRL97}%
  \BibitemOpen
  \bibfield  {author} {\bibinfo {author} {\bibfnamefont {R.}~\bibnamefont
  {Danneau}}, \bibinfo {author} {\bibfnamefont {O.}~\bibnamefont {Klochan}},
  \bibinfo {author} {\bibfnamefont {W.~R.}\ \bibnamefont {Clarke}}, \bibinfo
  {author} {\bibfnamefont {L.~H.}\ \bibnamefont {Ho}}, \bibinfo {author}
  {\bibfnamefont {A.~P.}\ \bibnamefont {Micolich}}, \bibinfo {author}
  {\bibfnamefont {M.~Y.}\ \bibnamefont {Simmons}}, \bibinfo {author}
  {\bibfnamefont {A.~R.}\ \bibnamefont {Hamilton}}, \bibinfo {author}
  {\bibfnamefont {M.}~\bibnamefont {Pepper}}, \bibinfo {author} {\bibfnamefont
  {D.~A.}\ \bibnamefont {Ritchie}},\ and\ \bibinfo {author} {\bibfnamefont
  {U.}~\bibnamefont {Zülicke}},\ }\bibfield  {title} {\bibinfo {title} {Zeeman
  splitting in ballistic hole quantum wires},\ }\href
  {https://doi.org/10.1103/PhysRevLett.97.026403} {\bibfield  {journal}
  {\bibinfo  {journal} {Physical Review Letters}\ }\textbf {\bibinfo {volume}
  {97}},\ \bibinfo {pages} {026403} (\bibinfo {year} {2006})},\ \bibinfo {note}
  {pRL}\BibitemShut {NoStop}%
\bibitem [{\citenamefont {Gvozdić}\ and\ \citenamefont
  {Ekenberg}(2006)}]{2006GvozdicPhysScr2006}%
  \BibitemOpen
  \bibfield  {author} {\bibinfo {author} {\bibfnamefont {D.~M.}\ \bibnamefont
  {Gvozdić}}\ and\ \bibinfo {author} {\bibfnamefont {U.}~\bibnamefont
  {Ekenberg}},\ }\bibfield  {title} {\bibinfo {title} {Superiority of p-type
  spin transistors},\ }\href {https://doi.org/10.1088/0031-8949/2006/T126/005}
  {\bibfield  {journal} {\bibinfo  {journal} {Physica Scripta}\ }\textbf
  {\bibinfo {volume} {2006}},\ \bibinfo {pages} {21} (\bibinfo {year}
  {2006})}\BibitemShut {NoStop}%
\bibitem [{\citenamefont {Bulaev}\ and\ \citenamefont
  {Loss}(2007)}]{2007BulaevPRL98}%
  \BibitemOpen
  \bibfield  {author} {\bibinfo {author} {\bibfnamefont {D.~V.}\ \bibnamefont
  {Bulaev}}\ and\ \bibinfo {author} {\bibfnamefont {D.}~\bibnamefont {Loss}},\
  }\bibfield  {title} {\bibinfo {title} {Electric dipole spin resonance for
  heavy holes in quantum dots},\ }\href
  {https://doi.org/10.1103/PhysRevLett.98.097202} {\bibfield  {journal}
  {\bibinfo  {journal} {Phys. Rev. Lett.}\ }\textbf {\bibinfo {volume} {98}},\
  \bibinfo {pages} {097202} (\bibinfo {year} {2007})}\BibitemShut {NoStop}%
\bibitem [{\citenamefont {Heiss}\ \emph {et~al.}(2007)\citenamefont {Heiss},
  \citenamefont {Schaeck}, \citenamefont {Huebl}, \citenamefont {Bichler},
  \citenamefont {Abstreiter}, \citenamefont {Finley}, \citenamefont {Bulaev},\
  and\ \citenamefont {Loss}}]{2007HeissPRB76}%
  \BibitemOpen
  \bibfield  {author} {\bibinfo {author} {\bibfnamefont {D.}~\bibnamefont
  {Heiss}}, \bibinfo {author} {\bibfnamefont {S.}~\bibnamefont {Schaeck}},
  \bibinfo {author} {\bibfnamefont {H.}~\bibnamefont {Huebl}}, \bibinfo
  {author} {\bibfnamefont {M.}~\bibnamefont {Bichler}}, \bibinfo {author}
  {\bibfnamefont {G.}~\bibnamefont {Abstreiter}}, \bibinfo {author}
  {\bibfnamefont {J.~J.}\ \bibnamefont {Finley}}, \bibinfo {author}
  {\bibfnamefont {D.~V.}\ \bibnamefont {Bulaev}},\ and\ \bibinfo {author}
  {\bibfnamefont {D.}~\bibnamefont {Loss}},\ }\bibfield  {title} {\bibinfo
  {title} {Observation of extremely slow hole spin relaxation in self-assembled
  quantum dots},\ }\href {https://doi.org/10.1103/PhysRevB.76.241306}
  {\bibfield  {journal} {\bibinfo  {journal} {Phys. Rev. B}\ }\textbf {\bibinfo
  {volume} {76}},\ \bibinfo {pages} {241306} (\bibinfo {year}
  {2007})}\BibitemShut {NoStop}%
\bibitem [{\citenamefont {Klauser}\ \emph {et~al.}(2007)\citenamefont
  {Klauser}, \citenamefont {Bulaev}, \citenamefont {Coish},\ and\ \citenamefont
  {Loss}}]{2007KlauserSemiQuantBits}%
  \BibitemOpen
  \bibfield  {author} {\bibinfo {author} {\bibfnamefont {D.}~\bibnamefont
  {Klauser}}, \bibinfo {author} {\bibfnamefont {D.~V.}\ \bibnamefont {Bulaev}},
  \bibinfo {author} {\bibfnamefont {W.~A.}\ \bibnamefont {Coish}},\ and\
  \bibinfo {author} {\bibfnamefont {D.}~\bibnamefont {Loss}},\ }\bibinfo
  {title} {Electron and hole spin dynamics and decoherence in quantum dots},\
  in\ \href {https://arxiv.org/abs/0706.1514} {\emph {\bibinfo {booktitle}
  {Semiconductor Quantum Bits}}}\ (\bibinfo  {publisher} {World Scientific},\
  \bibinfo {year} {2007})\ Chap.~\bibinfo {chapter} {10}\BibitemShut {NoStop}%
\bibitem [{\citenamefont {Fischer}\ \emph {et~al.}(2008)\citenamefont
  {Fischer}, \citenamefont {Coish}, \citenamefont {Bulaev},\ and\ \citenamefont
  {Loss}}]{2008FischerPRB78}%
  \BibitemOpen
  \bibfield  {author} {\bibinfo {author} {\bibfnamefont {J.}~\bibnamefont
  {Fischer}}, \bibinfo {author} {\bibfnamefont {W.~A.}\ \bibnamefont {Coish}},
  \bibinfo {author} {\bibfnamefont {D.~V.}\ \bibnamefont {Bulaev}},\ and\
  \bibinfo {author} {\bibfnamefont {D.}~\bibnamefont {Loss}},\ }\bibfield
  {title} {\bibinfo {title} {Spin decoherence of a heavy hole coupled to
  nuclear spins in a quantum dot},\ }\href
  {https://doi.org/10.1103/PhysRevB.78.155329} {\bibfield  {journal} {\bibinfo
  {journal} {Phys. Rev. B}\ }\textbf {\bibinfo {volume} {78}},\ \bibinfo
  {pages} {155329} (\bibinfo {year} {2008})}\BibitemShut {NoStop}%
\bibitem [{\citenamefont {Roddaro}\ \emph {et~al.}(2008)\citenamefont
  {Roddaro}, \citenamefont {Fuhrer}, \citenamefont {Brusheim}, \citenamefont
  {Fasth}, \citenamefont {Xu}, \citenamefont {Samuelson}, \citenamefont
  {Xiang},\ and\ \citenamefont {Lieber}}]{2008RoddaroPRL101}%
  \BibitemOpen
  \bibfield  {author} {\bibinfo {author} {\bibfnamefont {S.}~\bibnamefont
  {Roddaro}}, \bibinfo {author} {\bibfnamefont {A.}~\bibnamefont {Fuhrer}},
  \bibinfo {author} {\bibfnamefont {P.}~\bibnamefont {Brusheim}}, \bibinfo
  {author} {\bibfnamefont {C.}~\bibnamefont {Fasth}}, \bibinfo {author}
  {\bibfnamefont {H.~Q.}\ \bibnamefont {Xu}}, \bibinfo {author} {\bibfnamefont
  {L.}~\bibnamefont {Samuelson}}, \bibinfo {author} {\bibfnamefont
  {J.}~\bibnamefont {Xiang}},\ and\ \bibinfo {author} {\bibfnamefont {C.~M.}\
  \bibnamefont {Lieber}},\ }\bibfield  {title} {\bibinfo {title} {Spin states
  of holes in $\mathrm{Ge}/\mathrm{Si}$ nanowire quantum dots},\ }\href
  {https://doi.org/10.1103/PhysRevLett.101.186802} {\bibfield  {journal}
  {\bibinfo  {journal} {Physical Review Letters}\ }\textbf {\bibinfo {volume}
  {101}},\ \bibinfo {pages} {186802} (\bibinfo {year} {2008})},\ \bibinfo
  {note} {pRL}\BibitemShut {NoStop}%
\bibitem [{\citenamefont {Hsieh}\ \emph {et~al.}(2009)\citenamefont {Hsieh},
  \citenamefont {Cheriton}, \citenamefont {Korkusinski},\ and\ \citenamefont
  {Hawrylak}}]{2009HsiehPRB80}%
  \BibitemOpen
  \bibfield  {author} {\bibinfo {author} {\bibfnamefont {C.-Y.}\ \bibnamefont
  {Hsieh}}, \bibinfo {author} {\bibfnamefont {R.}~\bibnamefont {Cheriton}},
  \bibinfo {author} {\bibfnamefont {M.}~\bibnamefont {Korkusinski}},\ and\
  \bibinfo {author} {\bibfnamefont {P.}~\bibnamefont {Hawrylak}},\ }\bibfield
  {title} {\bibinfo {title} {Valence holes as luttinger spinor based qubits in
  quantum dots},\ }\href {https://doi.org/10.1103/PhysRevB.80.235320}
  {\bibfield  {journal} {\bibinfo  {journal} {Phys. Rev. B}\ }\textbf {\bibinfo
  {volume} {80}},\ \bibinfo {pages} {235320} (\bibinfo {year}
  {2009})}\BibitemShut {NoStop}%
\bibitem [{\citenamefont {Zwanenburg}\ \emph {et~al.}(2009)\citenamefont
  {Zwanenburg}, \citenamefont {van Rijmenam}, \citenamefont {Fang},
  \citenamefont {Lieber},\ and\ \citenamefont
  {Kouwenhoven}}]{2009ZwanenburgNanoLett9}%
  \BibitemOpen
  \bibfield  {author} {\bibinfo {author} {\bibfnamefont {F.~A.}\ \bibnamefont
  {Zwanenburg}}, \bibinfo {author} {\bibfnamefont {C.~E. W.~M.}\ \bibnamefont
  {van Rijmenam}}, \bibinfo {author} {\bibfnamefont {Y.}~\bibnamefont {Fang}},
  \bibinfo {author} {\bibfnamefont {C.~M.}\ \bibnamefont {Lieber}},\ and\
  \bibinfo {author} {\bibfnamefont {L.~P.}\ \bibnamefont {Kouwenhoven}},\
  }\bibfield  {title} {\bibinfo {title} {Spin states of the first four holes in
  a silicon nanowire quantum dot},\ }\href {https://doi.org/10.1021/nl803440s}
  {\bibfield  {journal} {\bibinfo  {journal} {Nano Letters}\ }\textbf {\bibinfo
  {volume} {9}},\ \bibinfo {pages} {1071} (\bibinfo {year} {2009})},\ \bibinfo
  {note} {doi: 10.1021/nl803440s}\BibitemShut {NoStop}%
\bibitem [{\citenamefont {Kloeffel}\ \emph {et~al.}(2013)\citenamefont
  {Kloeffel}, \citenamefont {Trif}, \citenamefont {Stano},\ and\ \citenamefont
  {Loss}}]{2013KloeffelPRB88}%
  \BibitemOpen
  \bibfield  {author} {\bibinfo {author} {\bibfnamefont {C.}~\bibnamefont
  {Kloeffel}}, \bibinfo {author} {\bibfnamefont {M.}~\bibnamefont {Trif}},
  \bibinfo {author} {\bibfnamefont {P.}~\bibnamefont {Stano}},\ and\ \bibinfo
  {author} {\bibfnamefont {D.}~\bibnamefont {Loss}},\ }\bibfield  {title}
  {\bibinfo {title} {Circuit qed with hole-spin qubits in ge/si nanowire
  quantum dots},\ }\href {https://doi.org/10.1103/PhysRevB.88.241405}
  {\bibfield  {journal} {\bibinfo  {journal} {Phys. Rev. B}\ }\textbf {\bibinfo
  {volume} {88}},\ \bibinfo {pages} {241405} (\bibinfo {year}
  {2013})}\BibitemShut {NoStop}%
\bibitem [{\citenamefont {Watzinger}\ \emph {et~al.}(2016)\citenamefont
  {Watzinger}, \citenamefont {Kloeffel}, \citenamefont {Vukušić},
  \citenamefont {Rossell}, \citenamefont {Sessi}, \citenamefont {Kukučka},
  \citenamefont {Kirchschlager}, \citenamefont {Lausecker}, \citenamefont
  {Truhlar}, \citenamefont {Glaser}, \citenamefont {Rastelli}, \citenamefont
  {Fuhrer}, \citenamefont {Loss},\ and\ \citenamefont
  {Katsaros}}]{2016WatzingerNanoLett16}%
  \BibitemOpen
  \bibfield  {author} {\bibinfo {author} {\bibfnamefont {H.}~\bibnamefont
  {Watzinger}}, \bibinfo {author} {\bibfnamefont {C.}~\bibnamefont {Kloeffel}},
  \bibinfo {author} {\bibfnamefont {L.}~\bibnamefont {Vukušić}}, \bibinfo
  {author} {\bibfnamefont {M.~D.}\ \bibnamefont {Rossell}}, \bibinfo {author}
  {\bibfnamefont {V.}~\bibnamefont {Sessi}}, \bibinfo {author} {\bibfnamefont
  {J.}~\bibnamefont {Kukučka}}, \bibinfo {author} {\bibfnamefont
  {R.}~\bibnamefont {Kirchschlager}}, \bibinfo {author} {\bibfnamefont
  {E.}~\bibnamefont {Lausecker}}, \bibinfo {author} {\bibfnamefont
  {A.}~\bibnamefont {Truhlar}}, \bibinfo {author} {\bibfnamefont
  {M.}~\bibnamefont {Glaser}}, \bibinfo {author} {\bibfnamefont
  {A.}~\bibnamefont {Rastelli}}, \bibinfo {author} {\bibfnamefont
  {A.}~\bibnamefont {Fuhrer}}, \bibinfo {author} {\bibfnamefont
  {D.}~\bibnamefont {Loss}},\ and\ \bibinfo {author} {\bibfnamefont
  {G.}~\bibnamefont {Katsaros}},\ }\bibfield  {title} {\bibinfo {title}
  {Heavy-hole states in germanium hut wires},\ }\href
  {https://doi.org/10.1021/acs.nanolett.6b02715} {\bibfield  {journal}
  {\bibinfo  {journal} {Nano Letters}\ }\textbf {\bibinfo {volume} {16}},\
  \bibinfo {pages} {6879} (\bibinfo {year} {2016})},\ \bibinfo {note} {pMID:
  27656760}\BibitemShut {NoStop}%
\bibitem [{\citenamefont {Kloeffel}\ \emph {et~al.}(2018)\citenamefont
  {Kloeffel}, \citenamefont {Ran\ifmmode \check{c}\else
  \v{c}\fi{}i\ifmmode~\acute{c}\else \'{c}\fi{}},\ and\ \citenamefont
  {Loss}}]{2018KloeffelPRB97}%
  \BibitemOpen
  \bibfield  {author} {\bibinfo {author} {\bibfnamefont {C.}~\bibnamefont
  {Kloeffel}}, \bibinfo {author} {\bibfnamefont {M.~J.}\ \bibnamefont
  {Ran\ifmmode \check{c}\else \v{c}\fi{}i\ifmmode~\acute{c}\else \'{c}\fi{}}},\
  and\ \bibinfo {author} {\bibfnamefont {D.}~\bibnamefont {Loss}},\ }\bibfield
  {title} {\bibinfo {title} {Direct rashba spin-orbit interaction in si and ge
  nanowires with different growth directions},\ }\href
  {https://doi.org/10.1103/PhysRevB.97.235422} {\bibfield  {journal} {\bibinfo
  {journal} {Phys. Rev. B}\ }\textbf {\bibinfo {volume} {97}},\ \bibinfo
  {pages} {235422} (\bibinfo {year} {2018})}\BibitemShut {NoStop}%
\bibitem [{\citenamefont {Abadillo-Uriel}\ \emph {et~al.}(2018)\citenamefont
  {Abadillo-Uriel}, \citenamefont {Salfi}, \citenamefont {Hu}, \citenamefont
  {Rogge}, \citenamefont {Calderón},\ and\ \citenamefont
  {Culcer}}]{2018Abadillo-UrielAPL113}%
  \BibitemOpen
  \bibfield  {author} {\bibinfo {author} {\bibfnamefont {J.~C.}\ \bibnamefont
  {Abadillo-Uriel}}, \bibinfo {author} {\bibfnamefont {J.}~\bibnamefont
  {Salfi}}, \bibinfo {author} {\bibfnamefont {X.}~\bibnamefont {Hu}}, \bibinfo
  {author} {\bibfnamefont {S.}~\bibnamefont {Rogge}}, \bibinfo {author}
  {\bibfnamefont {M.~J.}\ \bibnamefont {Calderón}},\ and\ \bibinfo {author}
  {\bibfnamefont {D.}~\bibnamefont {Culcer}},\ }\bibfield  {title} {\bibinfo
  {title} {{Entanglement control and magic angles for acceptor qubits in Si}},\
  }\href {https://doi.org/10.1063/1.5036521} {\bibfield  {journal} {\bibinfo
  {journal} {Applied Physics Letters}\ }\textbf {\bibinfo {volume} {113}},\
  \bibinfo {pages} {012102} (\bibinfo {year} {2018})}\BibitemShut {NoStop}%
\bibitem [{\citenamefont {Machnikowski}\ \emph {et~al.}(2019)\citenamefont
  {Machnikowski}, \citenamefont {Gawarecki},\ and\ \citenamefont
  {Cywi\ifmmode~\acute{n}\else \'{n}\fi{}ski}}]{2019MachnikowskiPRB100}%
  \BibitemOpen
  \bibfield  {author} {\bibinfo {author} {\bibfnamefont {P.}~\bibnamefont
  {Machnikowski}}, \bibinfo {author} {\bibfnamefont {K.}~\bibnamefont
  {Gawarecki}},\ and\ \bibinfo {author} {\bibfnamefont {L.}~\bibnamefont
  {Cywi\ifmmode~\acute{n}\else \'{n}\fi{}ski}},\ }\bibfield  {title} {\bibinfo
  {title} {Hyperfine interaction for holes in quantum dots:
  $k\ifmmode\cdot\else\textperiodcentered\fi{}p$ model},\ }\href
  {https://doi.org/10.1103/PhysRevB.100.085305} {\bibfield  {journal} {\bibinfo
   {journal} {Phys. Rev. B}\ }\textbf {\bibinfo {volume} {100}},\ \bibinfo
  {pages} {085305} (\bibinfo {year} {2019})}\BibitemShut {NoStop}%
\bibitem [{\citenamefont {Wang}\ \emph {et~al.}(2022)\citenamefont {Wang},
  \citenamefont {Xu}, \citenamefont {Gao}, \citenamefont {Liu}, \citenamefont
  {Ma}, \citenamefont {Zhang}, \citenamefont {Wang}, \citenamefont {Cao},
  \citenamefont {Wang}, \citenamefont {Zhang}, \citenamefont {Culcer},
  \citenamefont {Hu}, \citenamefont {Jiang}, \citenamefont {Li}, \citenamefont
  {Guo},\ and\ \citenamefont {Guo}}]{2022WangNatComm13}%
  \BibitemOpen
  \bibfield  {author} {\bibinfo {author} {\bibfnamefont {K.}~\bibnamefont
  {Wang}}, \bibinfo {author} {\bibfnamefont {G.}~\bibnamefont {Xu}}, \bibinfo
  {author} {\bibfnamefont {F.}~\bibnamefont {Gao}}, \bibinfo {author}
  {\bibfnamefont {H.}~\bibnamefont {Liu}}, \bibinfo {author} {\bibfnamefont
  {R.-L.}\ \bibnamefont {Ma}}, \bibinfo {author} {\bibfnamefont
  {X.}~\bibnamefont {Zhang}}, \bibinfo {author} {\bibfnamefont
  {Z.}~\bibnamefont {Wang}}, \bibinfo {author} {\bibfnamefont {G.}~\bibnamefont
  {Cao}}, \bibinfo {author} {\bibfnamefont {T.}~\bibnamefont {Wang}}, \bibinfo
  {author} {\bibfnamefont {J.-J.}\ \bibnamefont {Zhang}}, \bibinfo {author}
  {\bibfnamefont {D.}~\bibnamefont {Culcer}}, \bibinfo {author} {\bibfnamefont
  {X.}~\bibnamefont {Hu}}, \bibinfo {author} {\bibfnamefont {H.-W.}\
  \bibnamefont {Jiang}}, \bibinfo {author} {\bibfnamefont {H.-O.}\ \bibnamefont
  {Li}}, \bibinfo {author} {\bibfnamefont {G.-C.}\ \bibnamefont {Guo}},\ and\
  \bibinfo {author} {\bibfnamefont {G.-P.}\ \bibnamefont {Guo}},\ }\bibfield
  {title} {\bibinfo {title} {Ultrafast coherent control of a hole spin qubit in
  a germanium quantum dot},\ }\href
  {https://doi.org/10.1038/s41467-021-27880-7} {\bibfield  {journal} {\bibinfo
  {journal} {Nature Communications}\ }\textbf {\bibinfo {volume} {13}},\
  \bibinfo {pages} {206} (\bibinfo {year} {2022})}\BibitemShut {NoStop}%
\bibitem [{\citenamefont {Rimbach-Russ}\ \emph {et~al.}(2024)\citenamefont
  {Rimbach-Russ}, \citenamefont {John}, \citenamefont {van Straaten},\ and\
  \citenamefont {Bosco}}]{2024RussSpinless}%
  \BibitemOpen
  \bibfield  {author} {\bibinfo {author} {\bibfnamefont {M.}~\bibnamefont
  {Rimbach-Russ}}, \bibinfo {author} {\bibfnamefont {V.}~\bibnamefont {John}},
  \bibinfo {author} {\bibfnamefont {B.}~\bibnamefont {van Straaten}},\ and\
  \bibinfo {author} {\bibfnamefont {S.}~\bibnamefont {Bosco}},\ }\href
  {https://arxiv.org/abs/2412.13658} {\bibinfo {title} {A spinless spin qubit}}
  (\bibinfo {year} {2024}),\ \Eprint {https://arxiv.org/abs/2412.13658}
  {arXiv:2412.13658 [cond-mat.mes-hall]} \BibitemShut {NoStop}%
\bibitem [{\citenamefont {Fischer}\ and\ \citenamefont
  {Loss}(2010)}]{2010FischerPRL105}%
  \BibitemOpen
  \bibfield  {author} {\bibinfo {author} {\bibfnamefont {J.}~\bibnamefont
  {Fischer}}\ and\ \bibinfo {author} {\bibfnamefont {D.}~\bibnamefont {Loss}},\
  }\bibfield  {title} {\bibinfo {title} {Hybridization and spin decoherence in
  heavy-hole quantum dots},\ }\href
  {https://doi.org/10.1103/PhysRevLett.105.266603} {\bibfield  {journal}
  {\bibinfo  {journal} {Phys. Rev. Lett.}\ }\textbf {\bibinfo {volume} {105}},\
  \bibinfo {pages} {266603} (\bibinfo {year} {2010})}\BibitemShut {NoStop}%
\bibitem [{\citenamefont {Szumniak}\ \emph {et~al.}(2012)\citenamefont
  {Szumniak}, \citenamefont {Bednarek}, \citenamefont {Partoens},\ and\
  \citenamefont {Peeters}}]{2012SzumniakPRL109}%
  \BibitemOpen
  \bibfield  {author} {\bibinfo {author} {\bibfnamefont {P.}~\bibnamefont
  {Szumniak}}, \bibinfo {author} {\bibfnamefont {S.}~\bibnamefont {Bednarek}},
  \bibinfo {author} {\bibfnamefont {B.}~\bibnamefont {Partoens}},\ and\
  \bibinfo {author} {\bibfnamefont {F.~M.}\ \bibnamefont {Peeters}},\
  }\bibfield  {title} {\bibinfo {title} {Spin-orbit-mediated manipulation of
  heavy-hole spin qubits in gated semiconductor nanodevices},\ }\href
  {https://doi.org/10.1103/PhysRevLett.109.107201} {\bibfield  {journal}
  {\bibinfo  {journal} {Phys. Rev. Lett.}\ }\textbf {\bibinfo {volume} {109}},\
  \bibinfo {pages} {107201} (\bibinfo {year} {2012})}\BibitemShut {NoStop}%
\bibitem [{\citenamefont {Kobayashi}\ \emph {et~al.}(2021)\citenamefont
  {Kobayashi}, \citenamefont {Salfi}, \citenamefont {Chua}, \citenamefont
  {van~der Heijden}, \citenamefont {House}, \citenamefont {Culcer},
  \citenamefont {Hutchison}, \citenamefont {Johnson}, \citenamefont {McCallum},
  \citenamefont {Riemann}, \citenamefont {Abrosimov}, \citenamefont {Becker},
  \citenamefont {Pohl}, \citenamefont {Simmons},\ and\ \citenamefont
  {Rogge}}]{2021KobayashiNatMat20}%
  \BibitemOpen
  \bibfield  {author} {\bibinfo {author} {\bibfnamefont {T.}~\bibnamefont
  {Kobayashi}}, \bibinfo {author} {\bibfnamefont {J.}~\bibnamefont {Salfi}},
  \bibinfo {author} {\bibfnamefont {C.}~\bibnamefont {Chua}}, \bibinfo {author}
  {\bibfnamefont {J.}~\bibnamefont {van~der Heijden}}, \bibinfo {author}
  {\bibfnamefont {M.~G.}\ \bibnamefont {House}}, \bibinfo {author}
  {\bibfnamefont {D.}~\bibnamefont {Culcer}}, \bibinfo {author} {\bibfnamefont
  {W.~D.}\ \bibnamefont {Hutchison}}, \bibinfo {author} {\bibfnamefont {B.~C.}\
  \bibnamefont {Johnson}}, \bibinfo {author} {\bibfnamefont {J.~C.}\
  \bibnamefont {McCallum}}, \bibinfo {author} {\bibfnamefont {H.}~\bibnamefont
  {Riemann}}, \bibinfo {author} {\bibfnamefont {N.~V.}\ \bibnamefont
  {Abrosimov}}, \bibinfo {author} {\bibfnamefont {P.}~\bibnamefont {Becker}},
  \bibinfo {author} {\bibfnamefont {H.-J.}\ \bibnamefont {Pohl}}, \bibinfo
  {author} {\bibfnamefont {M.~Y.}\ \bibnamefont {Simmons}},\ and\ \bibinfo
  {author} {\bibfnamefont {S.}~\bibnamefont {Rogge}},\ }\bibfield  {title}
  {\bibinfo {title} {Engineering long spin coherence times of spin–orbit
  qubits in silicon},\ }\href {https://doi.org/10.1038/s41563-020-0743-3}
  {\bibfield  {journal} {\bibinfo  {journal} {Nature Materials}\ }\textbf
  {\bibinfo {volume} {20}},\ \bibinfo {pages} {38} (\bibinfo {year}
  {2021})}\BibitemShut {NoStop}%
\bibitem [{\citenamefont {Hendrickx}\ \emph {et~al.}(2021)\citenamefont
  {Hendrickx}, \citenamefont {Lawrie}, \citenamefont {Russ}, \citenamefont {van
  Riggelen}, \citenamefont {de~Snoo}, \citenamefont {Schouten}, \citenamefont
  {Sammak}, \citenamefont {Scappucci},\ and\ \citenamefont
  {Veldhorst}}]{2021HendrickxNat591}%
  \BibitemOpen
  \bibfield  {author} {\bibinfo {author} {\bibfnamefont {N.~W.}\ \bibnamefont
  {Hendrickx}}, \bibinfo {author} {\bibfnamefont {W.~I.~L.}\ \bibnamefont
  {Lawrie}}, \bibinfo {author} {\bibfnamefont {M.}~\bibnamefont {Russ}},
  \bibinfo {author} {\bibfnamefont {F.}~\bibnamefont {van Riggelen}}, \bibinfo
  {author} {\bibfnamefont {S.~L.}\ \bibnamefont {de~Snoo}}, \bibinfo {author}
  {\bibfnamefont {R.~N.}\ \bibnamefont {Schouten}}, \bibinfo {author}
  {\bibfnamefont {A.}~\bibnamefont {Sammak}}, \bibinfo {author} {\bibfnamefont
  {G.}~\bibnamefont {Scappucci}},\ and\ \bibinfo {author} {\bibfnamefont
  {M.}~\bibnamefont {Veldhorst}},\ }\bibfield  {title} {\bibinfo {title} {A
  four-qubit germanium quantum processor},\ }\href
  {https://doi.org/10.1038/s41586-021-03332-6} {\bibfield  {journal} {\bibinfo
  {journal} {Nature}\ }\textbf {\bibinfo {volume} {591}},\ \bibinfo {pages}
  {580} (\bibinfo {year} {2021})}\BibitemShut {NoStop}%
\bibitem [{\citenamefont {Borsoi}\ \emph {et~al.}(2024)\citenamefont {Borsoi},
  \citenamefont {Hendrickx}, \citenamefont {John}, \citenamefont {Meyer},
  \citenamefont {Motz}, \citenamefont {van Riggelen}, \citenamefont {Sammak},
  \citenamefont {de~Snoo}, \citenamefont {Scappucci},\ and\ \citenamefont
  {Veldhorst}}]{2024BorsoiNatNanoTech19}%
  \BibitemOpen
  \bibfield  {author} {\bibinfo {author} {\bibfnamefont {F.}~\bibnamefont
  {Borsoi}}, \bibinfo {author} {\bibfnamefont {N.~W.}\ \bibnamefont
  {Hendrickx}}, \bibinfo {author} {\bibfnamefont {V.}~\bibnamefont {John}},
  \bibinfo {author} {\bibfnamefont {M.}~\bibnamefont {Meyer}}, \bibinfo
  {author} {\bibfnamefont {S.}~\bibnamefont {Motz}}, \bibinfo {author}
  {\bibfnamefont {F.}~\bibnamefont {van Riggelen}}, \bibinfo {author}
  {\bibfnamefont {A.}~\bibnamefont {Sammak}}, \bibinfo {author} {\bibfnamefont
  {S.~L.}\ \bibnamefont {de~Snoo}}, \bibinfo {author} {\bibfnamefont
  {G.}~\bibnamefont {Scappucci}},\ and\ \bibinfo {author} {\bibfnamefont
  {M.}~\bibnamefont {Veldhorst}},\ }\bibfield  {title} {\bibinfo {title}
  {Shared control of a 16 semiconductor quantum dot crossbar array},\ }\href
  {https://doi.org/10.1038/s41565-023-01491-3} {\bibfield  {journal} {\bibinfo
  {journal} {Nature Nanotechnology}\ }\textbf {\bibinfo {volume} {19}},\
  \bibinfo {pages} {21} (\bibinfo {year} {2024})}\BibitemShut {NoStop}%
\bibitem [{\citenamefont {Yu}\ \emph {et~al.}(2023)\citenamefont {Yu},
  \citenamefont {Zihlmann}, \citenamefont {Abadillo-Uriel}, \citenamefont
  {Michal}, \citenamefont {Rambal}, \citenamefont {Niebojewski}, \citenamefont
  {Bedecarrats}, \citenamefont {Vinet}, \citenamefont {Dumur}, \citenamefont
  {Filippone}, \citenamefont {Bertrand}, \citenamefont {De~Franceschi},
  \citenamefont {Niquet},\ and\ \citenamefont {Maurand}}]{2023YuNatNanoTech18}%
  \BibitemOpen
  \bibfield  {author} {\bibinfo {author} {\bibfnamefont {C.~X.}\ \bibnamefont
  {Yu}}, \bibinfo {author} {\bibfnamefont {S.}~\bibnamefont {Zihlmann}},
  \bibinfo {author} {\bibfnamefont {J.~C.}\ \bibnamefont {Abadillo-Uriel}},
  \bibinfo {author} {\bibfnamefont {V.~P.}\ \bibnamefont {Michal}}, \bibinfo
  {author} {\bibfnamefont {N.}~\bibnamefont {Rambal}}, \bibinfo {author}
  {\bibfnamefont {H.}~\bibnamefont {Niebojewski}}, \bibinfo {author}
  {\bibfnamefont {T.}~\bibnamefont {Bedecarrats}}, \bibinfo {author}
  {\bibfnamefont {M.}~\bibnamefont {Vinet}}, \bibinfo {author} {\bibfnamefont
  {{\'E}.}~\bibnamefont {Dumur}}, \bibinfo {author} {\bibfnamefont
  {M.}~\bibnamefont {Filippone}}, \bibinfo {author} {\bibfnamefont
  {B.}~\bibnamefont {Bertrand}}, \bibinfo {author} {\bibfnamefont
  {S.}~\bibnamefont {De~Franceschi}}, \bibinfo {author} {\bibfnamefont {Y.-M.}\
  \bibnamefont {Niquet}},\ and\ \bibinfo {author} {\bibfnamefont
  {R.}~\bibnamefont {Maurand}},\ }\bibfield  {title} {\bibinfo {title} {Strong
  coupling between a photon and a hole spin in silicon},\ }\href
  {https://doi.org/10.1038/s41565-023-01332-3} {\bibfield  {journal} {\bibinfo
  {journal} {Nature Nanotechnology}\ }\textbf {\bibinfo {volume} {18}},\
  \bibinfo {pages} {741} (\bibinfo {year} {2023})}\BibitemShut {NoStop}%
\bibitem [{\citenamefont {Zhang}\ \emph {et~al.}(2024)\citenamefont {Zhang},
  \citenamefont {Morozova}, \citenamefont {Rimbach-Russ}, \citenamefont
  {Jirovec}, \citenamefont {Hsiao}, \citenamefont {Fariña}, \citenamefont
  {Wang}, \citenamefont {Oosterhout}, \citenamefont {Sammak}, \citenamefont
  {Scappucci}, \citenamefont {Veldhorst},\ and\ \citenamefont
  {Vandersypen}}]{2024ZhangNatNanoTechnol}%
  \BibitemOpen
  \bibfield  {author} {\bibinfo {author} {\bibfnamefont {X.}~\bibnamefont
  {Zhang}}, \bibinfo {author} {\bibfnamefont {E.}~\bibnamefont {Morozova}},
  \bibinfo {author} {\bibfnamefont {M.}~\bibnamefont {Rimbach-Russ}}, \bibinfo
  {author} {\bibfnamefont {D.}~\bibnamefont {Jirovec}}, \bibinfo {author}
  {\bibfnamefont {T.-K.}\ \bibnamefont {Hsiao}}, \bibinfo {author}
  {\bibfnamefont {P.~C.}\ \bibnamefont {Fariña}}, \bibinfo {author}
  {\bibfnamefont {C.-A.}\ \bibnamefont {Wang}}, \bibinfo {author}
  {\bibfnamefont {S.~D.}\ \bibnamefont {Oosterhout}}, \bibinfo {author}
  {\bibfnamefont {A.}~\bibnamefont {Sammak}}, \bibinfo {author} {\bibfnamefont
  {G.}~\bibnamefont {Scappucci}}, \bibinfo {author} {\bibfnamefont
  {M.}~\bibnamefont {Veldhorst}},\ and\ \bibinfo {author} {\bibfnamefont
  {L.~M.~K.}\ \bibnamefont {Vandersypen}},\ }\bibfield  {title} {\bibinfo
  {title} {Universal control of four singlet–triplet qubits},\ }\bibfield
  {journal} {\bibinfo  {journal} {Nature Nanotechnology}\ }\href
  {https://doi.org/10.1038/s41565-024-01817-9} {10.1038/s41565-024-01817-9}
  (\bibinfo {year} {2024})\BibitemShut {NoStop}%
\bibitem [{\citenamefont {Kloeffel}\ \emph {et~al.}(2011)\citenamefont
  {Kloeffel}, \citenamefont {Trif},\ and\ \citenamefont
  {Loss}}]{2011KloeffelPRB84}%
  \BibitemOpen
  \bibfield  {author} {\bibinfo {author} {\bibfnamefont {C.}~\bibnamefont
  {Kloeffel}}, \bibinfo {author} {\bibfnamefont {M.}~\bibnamefont {Trif}},\
  and\ \bibinfo {author} {\bibfnamefont {D.}~\bibnamefont {Loss}},\ }\bibfield
  {title} {\bibinfo {title} {Strong spin-orbit interaction and helical hole
  states in ge/si nanowires},\ }\href
  {https://doi.org/10.1103/PhysRevB.84.195314} {\bibfield  {journal} {\bibinfo
  {journal} {Phys. Rev. B}\ }\textbf {\bibinfo {volume} {84}},\ \bibinfo
  {pages} {195314} (\bibinfo {year} {2011})}\BibitemShut {NoStop}%
\bibitem [{\citenamefont {Terrazos}\ \emph {et~al.}(2021)\citenamefont
  {Terrazos}, \citenamefont {Marcellina}, \citenamefont {Wang}, \citenamefont
  {Coppersmith}, \citenamefont {Friesen}, \citenamefont {Hamilton},
  \citenamefont {Hu}, \citenamefont {Koiller}, \citenamefont {Saraiva},
  \citenamefont {Culcer},\ and\ \citenamefont {Capaz}}]{2021TerrazosPRB103}%
  \BibitemOpen
  \bibfield  {author} {\bibinfo {author} {\bibfnamefont {L.~A.}\ \bibnamefont
  {Terrazos}}, \bibinfo {author} {\bibfnamefont {E.}~\bibnamefont
  {Marcellina}}, \bibinfo {author} {\bibfnamefont {Z.~N.}\ \bibnamefont
  {Wang}}, \bibinfo {author} {\bibfnamefont {S.~N.}\ \bibnamefont
  {Coppersmith}}, \bibinfo {author} {\bibfnamefont {M.}~\bibnamefont
  {Friesen}}, \bibinfo {author} {\bibfnamefont {A.~R.}\ \bibnamefont
  {Hamilton}}, \bibinfo {author} {\bibfnamefont {X.~D.}\ \bibnamefont {Hu}},
  \bibinfo {author} {\bibfnamefont {B.}~\bibnamefont {Koiller}}, \bibinfo
  {author} {\bibfnamefont {A.~L.}\ \bibnamefont {Saraiva}}, \bibinfo {author}
  {\bibfnamefont {D.}~\bibnamefont {Culcer}},\ and\ \bibinfo {author}
  {\bibfnamefont {R.~B.}\ \bibnamefont {Capaz}},\ }\bibfield  {title} {\bibinfo
  {title} {Theory of hole-spin qubits in strained germanium quantum dots},\
  }\bibfield  {journal} {\bibinfo  {journal} {Physical Review B}\ }\textbf
  {\bibinfo {volume} {103}},\ \href {https://doi.org/ARTN 125201
  10.1103/PhysRevB.103.125201} {ARTN 125201 10.1103/PhysRevB.103.125201}
  (\bibinfo {year} {2021}),\ \bibinfo {note} {rv9xo Times Cited:57 Cited
  References Count:59}\BibitemShut {NoStop}%
\bibitem [{\citenamefont {Bosco}\ and\ \citenamefont
  {Loss}(2022)}]{2022BoscoPRAppl18}%
  \BibitemOpen
  \bibfield  {author} {\bibinfo {author} {\bibfnamefont {S.}~\bibnamefont
  {Bosco}}\ and\ \bibinfo {author} {\bibfnamefont {D.}~\bibnamefont {Loss}},\
  }\bibfield  {title} {\bibinfo {title} {Hole spin qubits in thin curved
  quantum wells},\ }\href {https://doi.org/10.1103/PhysRevApplied.18.044038}
  {\bibfield  {journal} {\bibinfo  {journal} {Phys. Rev. Appl.}\ }\textbf
  {\bibinfo {volume} {18}},\ \bibinfo {pages} {044038} (\bibinfo {year}
  {2022})}\BibitemShut {NoStop}%
\bibitem [{\citenamefont {Wang}\ \emph {et~al.}(2021)\citenamefont {Wang},
  \citenamefont {Marcellina}, \citenamefont {Hamilton}, \citenamefont {Cullen},
  \citenamefont {Rogge}, \citenamefont {Salfi},\ and\ \citenamefont
  {Culcer}}]{2021WangNpjQuantInf7}%
  \BibitemOpen
  \bibfield  {author} {\bibinfo {author} {\bibfnamefont {Z.}~\bibnamefont
  {Wang}}, \bibinfo {author} {\bibfnamefont {E.}~\bibnamefont {Marcellina}},
  \bibinfo {author} {\bibfnamefont {A.~R.}\ \bibnamefont {Hamilton}}, \bibinfo
  {author} {\bibfnamefont {J.~H.}\ \bibnamefont {Cullen}}, \bibinfo {author}
  {\bibfnamefont {S.}~\bibnamefont {Rogge}}, \bibinfo {author} {\bibfnamefont
  {J.}~\bibnamefont {Salfi}},\ and\ \bibinfo {author} {\bibfnamefont
  {D.}~\bibnamefont {Culcer}},\ }\bibfield  {title} {\bibinfo {title} {Optimal
  operation points for ultrafast, highly coherent ge hole spin-orbit qubits},\
  }\href {https://doi.org/10.1038/s41534-021-00386-2} {\bibfield  {journal}
  {\bibinfo  {journal} {npj Quantum Information}\ }\textbf {\bibinfo {volume}
  {7}},\ \bibinfo {pages} {54} (\bibinfo {year} {2021})}\BibitemShut {NoStop}%
\bibitem [{\citenamefont {Sarkar}\ \emph {et~al.}(2023)\citenamefont {Sarkar},
  \citenamefont {Wang}, \citenamefont {Rendell}, \citenamefont {Hendrickx},
  \citenamefont {Veldhorst}, \citenamefont {Scappucci}, \citenamefont
  {Khalifa}, \citenamefont {Salfi}, \citenamefont {Saraiva}, \citenamefont
  {Dzurak}, \citenamefont {Hamilton},\ and\ \citenamefont
  {Culcer}}]{2023SarkarPRB108}%
  \BibitemOpen
  \bibfield  {author} {\bibinfo {author} {\bibfnamefont {A.}~\bibnamefont
  {Sarkar}}, \bibinfo {author} {\bibfnamefont {Z.}~\bibnamefont {Wang}},
  \bibinfo {author} {\bibfnamefont {M.}~\bibnamefont {Rendell}}, \bibinfo
  {author} {\bibfnamefont {N.~W.}\ \bibnamefont {Hendrickx}}, \bibinfo {author}
  {\bibfnamefont {M.}~\bibnamefont {Veldhorst}}, \bibinfo {author}
  {\bibfnamefont {G.}~\bibnamefont {Scappucci}}, \bibinfo {author}
  {\bibfnamefont {M.}~\bibnamefont {Khalifa}}, \bibinfo {author} {\bibfnamefont
  {J.}~\bibnamefont {Salfi}}, \bibinfo {author} {\bibfnamefont
  {A.}~\bibnamefont {Saraiva}}, \bibinfo {author} {\bibfnamefont {A.~S.}\
  \bibnamefont {Dzurak}}, \bibinfo {author} {\bibfnamefont {A.~R.}\
  \bibnamefont {Hamilton}},\ and\ \bibinfo {author} {\bibfnamefont
  {D.}~\bibnamefont {Culcer}},\ }\bibfield  {title} {\bibinfo {title}
  {Electrical operation of planar ge hole spin qubits in an in-plane magnetic
  field},\ }\href {https://doi.org/10.1103/PhysRevB.108.245301} {\bibfield
  {journal} {\bibinfo  {journal} {Phys. Rev. B}\ }\textbf {\bibinfo {volume}
  {108}},\ \bibinfo {pages} {245301} (\bibinfo {year} {2023})}\BibitemShut
  {NoStop}%
\bibitem [{\citenamefont {Wang}\ \emph
  {et~al.}(2024{\natexlab{a}})\citenamefont {Wang}, \citenamefont {Sarkar},
  \citenamefont {Liles}, \citenamefont {Saraiva}, \citenamefont {Dzurak},
  \citenamefont {Hamilton},\ and\ \citenamefont {Culcer}}]{2024WangPRB109}%
  \BibitemOpen
  \bibfield  {author} {\bibinfo {author} {\bibfnamefont {Z.~N.}\ \bibnamefont
  {Wang}}, \bibinfo {author} {\bibfnamefont {A.}~\bibnamefont {Sarkar}},
  \bibinfo {author} {\bibfnamefont {S.~D.}\ \bibnamefont {Liles}}, \bibinfo
  {author} {\bibfnamefont {A.}~\bibnamefont {Saraiva}}, \bibinfo {author}
  {\bibfnamefont {A.~S.}\ \bibnamefont {Dzurak}}, \bibinfo {author}
  {\bibfnamefont {A.~R.}\ \bibnamefont {Hamilton}},\ and\ \bibinfo {author}
  {\bibfnamefont {D.}~\bibnamefont {Culcer}},\ }\bibfield  {title} {\bibinfo
  {title} {Electrical operation of hole spin qubits in planar mos silicon
  quantum dots},\ }\bibfield  {journal} {\bibinfo  {journal} {Physical Review
  B}\ }\textbf {\bibinfo {volume} {109}},\ \href {https://doi.org/ARTN
  07542710.1103/PhysRevB.109.075427} {ARTN 07542710.1103/PhysRevB.109.075427}
  (\bibinfo {year} {2024}{\natexlab{a}}),\ \bibinfo {note} {ne0i8 Times Cited:1
  Cited References Count:242}\BibitemShut {NoStop}%
\bibitem [{\citenamefont {Wang}\ \emph
  {et~al.}(2024{\natexlab{b}})\citenamefont {Wang}, \citenamefont {Ercan},
  \citenamefont {Gyure}, \citenamefont {Scappucci}, \citenamefont {Veldhorst},\
  and\ \citenamefont {Rimbach-Russ}}]{2024WangNpjQuantumInf10}%
  \BibitemOpen
  \bibfield  {author} {\bibinfo {author} {\bibfnamefont {C.-A.}\ \bibnamefont
  {Wang}}, \bibinfo {author} {\bibfnamefont {H.~E.}\ \bibnamefont {Ercan}},
  \bibinfo {author} {\bibfnamefont {M.~F.}\ \bibnamefont {Gyure}}, \bibinfo
  {author} {\bibfnamefont {G.}~\bibnamefont {Scappucci}}, \bibinfo {author}
  {\bibfnamefont {M.}~\bibnamefont {Veldhorst}},\ and\ \bibinfo {author}
  {\bibfnamefont {M.}~\bibnamefont {Rimbach-Russ}},\ }\bibfield  {title}
  {\bibinfo {title} {Modeling of planar germanium hole qubits in electric and
  magnetic fields},\ }\href {https://doi.org/10.1038/s41534-024-00897-8}
  {\bibfield  {journal} {\bibinfo  {journal} {npj Quantum Information}\
  }\textbf {\bibinfo {volume} {10}},\ \bibinfo {pages} {102} (\bibinfo {year}
  {2024}{\natexlab{b}})}\BibitemShut {NoStop}%
\bibitem [{\citenamefont {Luttinger}\ and\ \citenamefont
  {Kohn}(1955)}]{1955LuttingerKohn}%
  \BibitemOpen
  \bibfield  {author} {\bibinfo {author} {\bibfnamefont {J.~M.}\ \bibnamefont
  {Luttinger}}\ and\ \bibinfo {author} {\bibfnamefont {W.}~\bibnamefont
  {Kohn}},\ }\bibfield  {title} {\bibinfo {title} {Motion of electrons and
  holes in perturbed periodic fields},\ }\href
  {https://doi.org/10.1103/PhysRev.97.869} {\bibfield  {journal} {\bibinfo
  {journal} {Phys. Rev.}\ }\textbf {\bibinfo {volume} {97}},\ \bibinfo {pages}
  {869} (\bibinfo {year} {1955})}\BibitemShut {NoStop}%
\bibitem [{\citenamefont {Burt}(1987)}]{1987BurtSemiSciTech2}%
  \BibitemOpen
  \bibfield  {author} {\bibinfo {author} {\bibfnamefont {M.~G.}\ \bibnamefont
  {Burt}},\ }\bibfield  {title} {\bibinfo {title} {An exact formulation of the
  envelope function method for the determination of electronic states in
  semiconductor microstructures},\ }\href
  {https://doi.org/10.1088/0268-1242/2/7/012} {\bibfield  {journal} {\bibinfo
  {journal} {Semiconductor Science and Technology}\ }\textbf {\bibinfo {volume}
  {2}},\ \bibinfo {pages} {460} (\bibinfo {year} {1987})}\BibitemShut {NoStop}%
\bibitem [{\citenamefont {Burt}(1988{\natexlab{a}})}]{1988BurtSemiSciTech3a}%
  \BibitemOpen
  \bibfield  {author} {\bibinfo {author} {\bibfnamefont {M.~G.}\ \bibnamefont
  {Burt}},\ }\bibfield  {title} {\bibinfo {title} {An exact formulation of the
  envelope function method for the determination of electronic states in
  semiconductor microstructures},\ }\href
  {https://doi.org/10.1088/0268-1242/3/8/003} {\bibfield  {journal} {\bibinfo
  {journal} {Semiconductor Science and Technology}\ }\textbf {\bibinfo {volume}
  {3}},\ \bibinfo {pages} {739} (\bibinfo {year}
  {1988}{\natexlab{a}})}\BibitemShut {NoStop}%
\bibitem [{\citenamefont {Burt}(1988{\natexlab{b}})}]{1988BurtSemiSciTech3b}%
  \BibitemOpen
  \bibfield  {author} {\bibinfo {author} {\bibfnamefont {M.~G.}\ \bibnamefont
  {Burt}},\ }\bibfield  {title} {\bibinfo {title} {A new effective-mass
  equation for microstructures},\ }\href
  {https://doi.org/10.1088/0268-1242/3/12/013} {\bibfield  {journal} {\bibinfo
  {journal} {Semiconductor Science and Technology}\ }\textbf {\bibinfo {volume}
  {3}},\ \bibinfo {pages} {1224} (\bibinfo {year}
  {1988}{\natexlab{b}})}\BibitemShut {NoStop}%
\bibitem [{\citenamefont {Burt}(1992)}]{1992BurtJoPCondMat4}%
  \BibitemOpen
  \bibfield  {author} {\bibinfo {author} {\bibfnamefont {M.~G.}\ \bibnamefont
  {Burt}},\ }\bibfield  {title} {\bibinfo {title} {The justification for
  applying the effective-mass approximation to microstructures},\ }\href
  {https://doi.org/10.1088/0953-8984/4/32/003} {\bibfield  {journal} {\bibinfo
  {journal} {Journal of Physics: Condensed Matter}\ }\textbf {\bibinfo {volume}
  {4}},\ \bibinfo {pages} {6651} (\bibinfo {year} {1992})}\BibitemShut
  {NoStop}%
\bibitem [{\citenamefont {Foreman}(1993)}]{1993ForemanPRB48}%
  \BibitemOpen
  \bibfield  {author} {\bibinfo {author} {\bibfnamefont {B.~A.}\ \bibnamefont
  {Foreman}},\ }\bibfield  {title} {\bibinfo {title} {Effective-mass
  hamiltonian and boundary conditions for the valence bands of semiconductor
  microstructures},\ }\href {https://doi.org/10.1103/PhysRevB.48.4964}
  {\bibfield  {journal} {\bibinfo  {journal} {Phys. Rev. B}\ }\textbf {\bibinfo
  {volume} {48}},\ \bibinfo {pages} {4964} (\bibinfo {year}
  {1993})}\BibitemShut {NoStop}%
\bibitem [{\citenamefont {Galeriu}(2005)}]{2005GaleriuThesis}%
  \BibitemOpen
  \bibfield  {author} {\bibinfo {author} {\bibfnamefont {C.}~\bibnamefont
  {Galeriu}},\ }\emph {\bibinfo {title} {kp Theory of Semiconductor
  Nanostructures}},\ \href@noop {} {Ph.D. thesis},\ \bibinfo  {school}
  {Worcester Polytechnic Institute} (\bibinfo {year} {2005})\BibitemShut
  {NoStop}%
\bibitem [{\citenamefont {Bastard}(1990)}]{1990BastardHabilitation}%
  \BibitemOpen
  \bibfield  {author} {\bibinfo {author} {\bibfnamefont {G.}~\bibnamefont
  {Bastard}},\ }\href@noop {} {\emph {\bibinfo {title} {Wave mechanics applied
  to semiconductor heterostructures}}}\ (\bibinfo {year} {1990})\BibitemShut
  {NoStop}%
\bibitem [{\citenamefont {Altarelli}(1983)}]{1983AltarelliPRB28}%
  \BibitemOpen
  \bibfield  {author} {\bibinfo {author} {\bibfnamefont {M.}~\bibnamefont
  {Altarelli}},\ }\bibfield  {title} {\bibinfo {title} {Electronic structure
  and semiconductor-semimetal transition in inas-gasb superlattices},\ }\href
  {https://doi.org/10.1103/PhysRevB.28.842} {\bibfield  {journal} {\bibinfo
  {journal} {Phys. Rev. B}\ }\textbf {\bibinfo {volume} {28}},\ \bibinfo
  {pages} {842} (\bibinfo {year} {1983})}\BibitemShut {NoStop}%
\bibitem [{\citenamefont {Eppenga}\ \emph {et~al.}(1987)\citenamefont
  {Eppenga}, \citenamefont {Schuurmans},\ and\ \citenamefont
  {Colak}}]{1987EppengaPRB36}%
  \BibitemOpen
  \bibfield  {author} {\bibinfo {author} {\bibfnamefont {R.}~\bibnamefont
  {Eppenga}}, \bibinfo {author} {\bibfnamefont {M.~F.~H.}\ \bibnamefont
  {Schuurmans}},\ and\ \bibinfo {author} {\bibfnamefont {S.}~\bibnamefont
  {Colak}},\ }\bibfield  {title} {\bibinfo {title} {New k\ensuremath{\cdot}p
  theory for
  gaas/${\mathrm{ga}}_{1\mathrm{\ensuremath{-}}\mathrm{x}}$${\mathrm{a}1}_{\mathrm{x}}$as-type
  quantum wells},\ }\href {https://doi.org/10.1103/PhysRevB.36.1554} {\bibfield
   {journal} {\bibinfo  {journal} {Phys. Rev. B}\ }\textbf {\bibinfo {volume}
  {36}},\ \bibinfo {pages} {1554} (\bibinfo {year} {1987})}\BibitemShut
  {NoStop}%
\bibitem [{\citenamefont {Andreani}\ \emph {et~al.}(1987)\citenamefont
  {Andreani}, \citenamefont {Pasquarello},\ and\ \citenamefont
  {Bassani}}]{1987AndreaniPRB36}%
  \BibitemOpen
  \bibfield  {author} {\bibinfo {author} {\bibfnamefont {L.~C.}\ \bibnamefont
  {Andreani}}, \bibinfo {author} {\bibfnamefont {A.}~\bibnamefont
  {Pasquarello}},\ and\ \bibinfo {author} {\bibfnamefont {F.}~\bibnamefont
  {Bassani}},\ }\bibfield  {title} {\bibinfo {title} {Hole subbands in strained
  gaas-${\mathrm{ga}}_{1\mathrm{\ensuremath{-}}\mathrm{x}}$${\mathrm{al}}_{\mathrm{x}}$as
  quantum wells: Exact solution of the effective-mass equation},\ }\href
  {https://doi.org/10.1103/PhysRevB.36.5887} {\bibfield  {journal} {\bibinfo
  {journal} {Phys. Rev. B}\ }\textbf {\bibinfo {volume} {36}},\ \bibinfo
  {pages} {5887} (\bibinfo {year} {1987})}\BibitemShut {NoStop}%
\bibitem [{\citenamefont {Cartoixà}\ \emph {et~al.}(2003)\citenamefont
  {Cartoixà}, \citenamefont {Ting},\ and\ \citenamefont
  {McGill}}]{2003CartoixaJoAP93}%
  \BibitemOpen
  \bibfield  {author} {\bibinfo {author} {\bibfnamefont {X.}~\bibnamefont
  {Cartoixà}}, \bibinfo {author} {\bibfnamefont {D.~Z.-Y.}\ \bibnamefont
  {Ting}},\ and\ \bibinfo {author} {\bibfnamefont {T.~C.}\ \bibnamefont
  {McGill}},\ }\bibfield  {title} {\bibinfo {title} {{Numerical spurious
  solutions in the effective mass approximation}},\ }\href
  {https://doi.org/10.1063/1.1555833} {\bibfield  {journal} {\bibinfo
  {journal} {Journal of Applied Physics}\ }\textbf {\bibinfo {volume} {93}},\
  \bibinfo {pages} {3974} (\bibinfo {year} {2003})},\ \Eprint
  {https://arxiv.org/abs/https://pubs.aip.org/aip/jap/article-pdf/93/7/3974/19245819/3974\_1\_online.pdf}
  {https://pubs.aip.org/aip/jap/article-pdf/93/7/3974/19245819/3974\_1\_online.pdf}
  \BibitemShut {NoStop}%
\bibitem [{\citenamefont {Bir}\ and\ \citenamefont
  {Pikus}(1974)}]{1974BirPikus}%
  \BibitemOpen
  \bibfield  {author} {\bibinfo {author} {\bibfnamefont {G.}~\bibnamefont
  {Bir}}\ and\ \bibinfo {author} {\bibfnamefont {G.}~\bibnamefont {Pikus}},\
  }\href {https://books.google.com/books?id=38m2QgAACAAJ} {\emph {\bibinfo
  {title} {Symmetry and Strain-induced Effects in Semiconductors}}},\ A Halsted
  Press book\ (\bibinfo  {publisher} {Wiley},\ \bibinfo {year}
  {1974})\BibitemShut {NoStop}%
\bibitem [{\citenamefont {Willatzen}\ and\ \citenamefont
  {Voon}(2009)}]{2009kpMethod}%
  \BibitemOpen
  \bibfield  {author} {\bibinfo {author} {\bibfnamefont {M.}~\bibnamefont
  {Willatzen}}\ and\ \bibinfo {author} {\bibfnamefont {L.~C. L.~Y.}\
  \bibnamefont {Voon}},\ }\href
  {https://link.springer.com/book/10.1007/978-3-540-92872-0} {\emph {\bibinfo
  {title} {The k p Method}}}\ (\bibinfo  {publisher} {Springer Berlin
  Heidelberg},\ \bibinfo {year} {2009})\BibitemShut {NoStop}%
\bibitem [{\citenamefont {Tarucha}\ \emph {et~al.}(1996)\citenamefont
  {Tarucha}, \citenamefont {Austing}, \citenamefont {Honda}, \citenamefont
  {van~der Hage},\ and\ \citenamefont {Kouwenhoven}}]{1996TaruchaPRL77}%
  \BibitemOpen
  \bibfield  {author} {\bibinfo {author} {\bibfnamefont {S.}~\bibnamefont
  {Tarucha}}, \bibinfo {author} {\bibfnamefont {D.~G.}\ \bibnamefont
  {Austing}}, \bibinfo {author} {\bibfnamefont {T.}~\bibnamefont {Honda}},
  \bibinfo {author} {\bibfnamefont {R.~J.}\ \bibnamefont {van~der Hage}},\ and\
  \bibinfo {author} {\bibfnamefont {L.~P.}\ \bibnamefont {Kouwenhoven}},\
  }\bibfield  {title} {\bibinfo {title} {Shell filling and spin effects in a
  few electron quantum dot},\ }\href
  {https://doi.org/10.1103/PhysRevLett.77.3613} {\bibfield  {journal} {\bibinfo
   {journal} {Phys. Rev. Lett.}\ }\textbf {\bibinfo {volume} {77}},\ \bibinfo
  {pages} {3613} (\bibinfo {year} {1996})}\BibitemShut {NoStop}%
\bibitem [{\citenamefont {BenDaniel}\ and\ \citenamefont
  {Duke}(1966)}]{1966BenDanielDukePR152}%
  \BibitemOpen
  \bibfield  {author} {\bibinfo {author} {\bibfnamefont {D.~J.}\ \bibnamefont
  {BenDaniel}}\ and\ \bibinfo {author} {\bibfnamefont {C.~B.}\ \bibnamefont
  {Duke}},\ }\bibfield  {title} {\bibinfo {title} {Space-charge effects on
  electron tunneling},\ }\href {https://doi.org/10.1103/PhysRev.152.683}
  {\bibfield  {journal} {\bibinfo  {journal} {Phys. Rev.}\ }\textbf {\bibinfo
  {volume} {152}},\ \bibinfo {pages} {683} (\bibinfo {year}
  {1966})}\BibitemShut {NoStop}%
\bibitem [{\citenamefont {Bastard}(1981)}]{1981BastardPRB24}%
  \BibitemOpen
  \bibfield  {author} {\bibinfo {author} {\bibfnamefont {G.}~\bibnamefont
  {Bastard}},\ }\bibfield  {title} {\bibinfo {title} {Superlattice band
  structure in the envelope-function approximation},\ }\href
  {https://doi.org/10.1103/PhysRevB.24.5693} {\bibfield  {journal} {\bibinfo
  {journal} {Phys. Rev. B}\ }\textbf {\bibinfo {volume} {24}},\ \bibinfo
  {pages} {5693} (\bibinfo {year} {1981})}\BibitemShut {NoStop}%
\bibitem [{\citenamefont {Ekenberg}\ and\ \citenamefont
  {Altarelli}(1984)}]{1984EkenbergPRB30}%
  \BibitemOpen
  \bibfield  {author} {\bibinfo {author} {\bibfnamefont {U.}~\bibnamefont
  {Ekenberg}}\ and\ \bibinfo {author} {\bibfnamefont {M.}~\bibnamefont
  {Altarelli}},\ }\bibfield  {title} {\bibinfo {title} {Calculation of hole
  subbands at the
  gaas-${\mathrm{al}}_{x}{\mathrm{ga}}_{1\ensuremath{-}x}\mathrm{As}$
  interface},\ }\href {https://doi.org/10.1103/PhysRevB.30.3569} {\bibfield
  {journal} {\bibinfo  {journal} {Phys. Rev. B}\ }\textbf {\bibinfo {volume}
  {30}},\ \bibinfo {pages} {3569} (\bibinfo {year} {1984})}\BibitemShut
  {NoStop}%
\bibitem [{\citenamefont {Schuurmans}\ and\ \citenamefont
  {'t~Hooft}(1985)}]{1985SchuurmansPRB31}%
  \BibitemOpen
  \bibfield  {author} {\bibinfo {author} {\bibfnamefont {M.~F.~H.}\
  \bibnamefont {Schuurmans}}\ and\ \bibinfo {author} {\bibfnamefont {G.~W.}\
  \bibnamefont {'t~Hooft}},\ }\bibfield  {title} {\bibinfo {title} {Simple
  calculations of confinement states in a quantum well},\ }\href
  {https://doi.org/10.1103/PhysRevB.31.8041} {\bibfield  {journal} {\bibinfo
  {journal} {Phys. Rev. B}\ }\textbf {\bibinfo {volume} {31}},\ \bibinfo
  {pages} {8041} (\bibinfo {year} {1985})}\BibitemShut {NoStop}%
\bibitem [{\citenamefont {Broido}\ and\ \citenamefont
  {Sham}(1985)}]{1985BroidoPRB31}%
  \BibitemOpen
  \bibfield  {author} {\bibinfo {author} {\bibfnamefont {D.~A.}\ \bibnamefont
  {Broido}}\ and\ \bibinfo {author} {\bibfnamefont {L.~J.}\ \bibnamefont
  {Sham}},\ }\bibfield  {title} {\bibinfo {title} {Effective masses of holes at
  gaas-algaas heterojunctions},\ }\href
  {https://doi.org/10.1103/PhysRevB.31.888} {\bibfield  {journal} {\bibinfo
  {journal} {Phys. Rev. B}\ }\textbf {\bibinfo {volume} {31}},\ \bibinfo
  {pages} {888} (\bibinfo {year} {1985})}\BibitemShut {NoStop}%
\bibitem [{\citenamefont {Winkler}\ \emph {et~al.}(1996)\citenamefont
  {Winkler}, \citenamefont {Merkler}, \citenamefont {Darnhofer},\ and\
  \citenamefont {R\"ossler}}]{1996WinklerPRB53}%
  \BibitemOpen
  \bibfield  {author} {\bibinfo {author} {\bibfnamefont {R.}~\bibnamefont
  {Winkler}}, \bibinfo {author} {\bibfnamefont {M.}~\bibnamefont {Merkler}},
  \bibinfo {author} {\bibfnamefont {T.}~\bibnamefont {Darnhofer}},\ and\
  \bibinfo {author} {\bibfnamefont {U.}~\bibnamefont {R\"ossler}},\ }\bibfield
  {title} {\bibinfo {title} {Theory for the cyclotron resonance of holes in
  strained asymmetric ge-sige quantum wells},\ }\href
  {https://doi.org/10.1103/PhysRevB.53.10858} {\bibfield  {journal} {\bibinfo
  {journal} {Phys. Rev. B}\ }\textbf {\bibinfo {volume} {53}},\ \bibinfo
  {pages} {10858} (\bibinfo {year} {1996})}\BibitemShut {NoStop}%
\bibitem [{\citenamefont {Venitucci}\ \emph {et~al.}(2018)\citenamefont
  {Venitucci}, \citenamefont {Bourdet}, \citenamefont {Pouzada},\ and\
  \citenamefont {Niquet}}]{2018VenitucciPRB98}%
  \BibitemOpen
  \bibfield  {author} {\bibinfo {author} {\bibfnamefont {B.}~\bibnamefont
  {Venitucci}}, \bibinfo {author} {\bibfnamefont {L.}~\bibnamefont {Bourdet}},
  \bibinfo {author} {\bibfnamefont {D.}~\bibnamefont {Pouzada}},\ and\ \bibinfo
  {author} {\bibfnamefont {Y.-M.}\ \bibnamefont {Niquet}},\ }\bibfield  {title}
  {\bibinfo {title} {Electrical manipulation of semiconductor spin qubits
  within the $g$-matrix formalism},\ }\href
  {https://doi.org/10.1103/PhysRevB.98.155319} {\bibfield  {journal} {\bibinfo
  {journal} {Phys. Rev. B}\ }\textbf {\bibinfo {volume} {98}},\ \bibinfo
  {pages} {155319} (\bibinfo {year} {2018})}\BibitemShut {NoStop}%
\bibitem [{\citenamefont {Abadillo-Uriel}\ \emph {et~al.}(2023)\citenamefont
  {Abadillo-Uriel}, \citenamefont {Rodr\'{\i}guez-Mena}, \citenamefont
  {Martinez},\ and\ \citenamefont {Niquet}}]{2023Abadillo-UrielPRL131}%
  \BibitemOpen
  \bibfield  {author} {\bibinfo {author} {\bibfnamefont {J.~C.}\ \bibnamefont
  {Abadillo-Uriel}}, \bibinfo {author} {\bibfnamefont {E.~A.}\ \bibnamefont
  {Rodr\'{\i}guez-Mena}}, \bibinfo {author} {\bibfnamefont {B.}~\bibnamefont
  {Martinez}},\ and\ \bibinfo {author} {\bibfnamefont {Y.-M.}\ \bibnamefont
  {Niquet}},\ }\bibfield  {title} {\bibinfo {title} {Hole-spin driving by
  strain-induced spin-orbit interactions},\ }\href
  {https://doi.org/10.1103/PhysRevLett.131.097002} {\bibfield  {journal}
  {\bibinfo  {journal} {Phys. Rev. Lett.}\ }\textbf {\bibinfo {volume} {131}},\
  \bibinfo {pages} {097002} (\bibinfo {year} {2023})}\BibitemShut {NoStop}%
\bibitem [{\citenamefont {Sammak}\ \emph {et~al.}(2019)\citenamefont {Sammak},
  \citenamefont {Sabbagh}, \citenamefont {Hendrickx}, \citenamefont {Lodari},
  \citenamefont {Paquelet~Wuetz}, \citenamefont {Tosato}, \citenamefont {Yeoh},
  \citenamefont {Bollani}, \citenamefont {Virgilio}, \citenamefont {Schubert},
  \citenamefont {Zaumseil}, \citenamefont {Capellini}, \citenamefont
  {Veldhorst},\ and\ \citenamefont {Scappucci}}]{2019SammakAdvFuncMat29}%
  \BibitemOpen
  \bibfield  {author} {\bibinfo {author} {\bibfnamefont {A.}~\bibnamefont
  {Sammak}}, \bibinfo {author} {\bibfnamefont {D.}~\bibnamefont {Sabbagh}},
  \bibinfo {author} {\bibfnamefont {N.~W.}\ \bibnamefont {Hendrickx}}, \bibinfo
  {author} {\bibfnamefont {M.}~\bibnamefont {Lodari}}, \bibinfo {author}
  {\bibfnamefont {B.}~\bibnamefont {Paquelet~Wuetz}}, \bibinfo {author}
  {\bibfnamefont {A.}~\bibnamefont {Tosato}}, \bibinfo {author} {\bibfnamefont
  {L.}~\bibnamefont {Yeoh}}, \bibinfo {author} {\bibfnamefont {M.}~\bibnamefont
  {Bollani}}, \bibinfo {author} {\bibfnamefont {M.}~\bibnamefont {Virgilio}},
  \bibinfo {author} {\bibfnamefont {M.~A.}\ \bibnamefont {Schubert}}, \bibinfo
  {author} {\bibfnamefont {P.}~\bibnamefont {Zaumseil}}, \bibinfo {author}
  {\bibfnamefont {G.}~\bibnamefont {Capellini}}, \bibinfo {author}
  {\bibfnamefont {M.}~\bibnamefont {Veldhorst}},\ and\ \bibinfo {author}
  {\bibfnamefont {G.}~\bibnamefont {Scappucci}},\ }\bibfield  {title} {\bibinfo
  {title} {Shallow and undoped germanium quantum wells: A playground for spin
  and hybrid quantum technology},\ }\href
  {https://doi.org/https://doi.org/10.1002/adfm.201807613} {\bibfield
  {journal} {\bibinfo  {journal} {Advanced Functional Materials}\ }\textbf
  {\bibinfo {volume} {29}},\ \bibinfo {pages} {1807613} (\bibinfo {year}
  {2019})}\BibitemShut {NoStop}%
\bibitem [{\citenamefont {Bouquet}\ \emph {et~al.}(2025)\citenamefont
  {Bouquet}, \citenamefont {Cao},\ and\ \citenamefont
  {Luisier}}]{2025BonquetPRAppl23}%
  \BibitemOpen
  \bibfield  {author} {\bibinfo {author} {\bibfnamefont {I.}~\bibnamefont
  {Bouquet}}, \bibinfo {author} {\bibfnamefont {J.}~\bibnamefont {Cao}},\ and\
  \bibinfo {author} {\bibfnamefont {M.}~\bibnamefont {Luisier}},\ }\bibfield
  {title} {\bibinfo {title} {Simulation of a single hole-spin qubit in a
  strained triangular finfet quantum device},\ }\href
  {https://doi.org/10.1103/PhysRevApplied.23.054030} {\bibfield  {journal}
  {\bibinfo  {journal} {Phys. Rev. Appl.}\ }\textbf {\bibinfo {volume} {23}},\
  \bibinfo {pages} {054030} (\bibinfo {year} {2025})}\BibitemShut {NoStop}%
\bibitem [{\citenamefont {Martinez}\ \emph {et~al.}(2022)\citenamefont
  {Martinez}, \citenamefont {Abadillo-Uriel}, \citenamefont
  {Rodr\'{\i}guez-Mena},\ and\ \citenamefont {Niquet}}]{2022MartinezPRB106}%
  \BibitemOpen
  \bibfield  {author} {\bibinfo {author} {\bibfnamefont {B.}~\bibnamefont
  {Martinez}}, \bibinfo {author} {\bibfnamefont {J.~C.}\ \bibnamefont
  {Abadillo-Uriel}}, \bibinfo {author} {\bibfnamefont {E.~A.}\ \bibnamefont
  {Rodr\'{\i}guez-Mena}},\ and\ \bibinfo {author} {\bibfnamefont {Y.-M.}\
  \bibnamefont {Niquet}},\ }\bibfield  {title} {\bibinfo {title} {Hole spin
  manipulation in inhomogeneous and nonseparable electric fields},\ }\href
  {https://doi.org/10.1103/PhysRevB.106.235426} {\bibfield  {journal} {\bibinfo
   {journal} {Phys. Rev. B}\ }\textbf {\bibinfo {volume} {106}},\ \bibinfo
  {pages} {235426} (\bibinfo {year} {2022})}\BibitemShut {NoStop}%
\bibitem [{\citenamefont {YU}\ and\ \citenamefont
  {Cardona}(2010)}]{2010YuCardona}%
  \BibitemOpen
  \bibfield  {author} {\bibinfo {author} {\bibfnamefont {P.}~\bibnamefont
  {YU}}\ and\ \bibinfo {author} {\bibfnamefont {M.}~\bibnamefont {Cardona}},\
  }\href@noop {} {\emph {\bibinfo {title} {Fundamentals of Semiconductors:
  Physics and Materials Properties}}},\ Graduate Texts in Physics\ (\bibinfo
  {publisher} {Springer Berlin Heidelberg},\ \bibinfo {year}
  {2010})\BibitemShut {NoStop}%
\bibitem [{\citenamefont {Dismukes}\ \emph {et~al.}(1964)\citenamefont
  {Dismukes}, \citenamefont {Ekstrom},\ and\ \citenamefont
  {Paff}}]{1964DismukesJoPC68}%
  \BibitemOpen
  \bibfield  {author} {\bibinfo {author} {\bibfnamefont {J.~P.}\ \bibnamefont
  {Dismukes}}, \bibinfo {author} {\bibfnamefont {L.}~\bibnamefont {Ekstrom}},\
  and\ \bibinfo {author} {\bibfnamefont {R.~J.}\ \bibnamefont {Paff}},\
  }\bibfield  {title} {\bibinfo {title} {Lattice parameter and density in
  germanium-silicon alloys1},\ }\href {https://doi.org/10.1021/j100792a049}
  {\bibfield  {journal} {\bibinfo  {journal} {The Journal of Physical
  Chemistry}\ }\textbf {\bibinfo {volume} {68}},\ \bibinfo {pages} {3021}
  (\bibinfo {year} {1964})}\BibitemShut {NoStop}%
\bibitem [{\citenamefont {Galdin}\ \emph {et~al.}(2000)\citenamefont {Galdin},
  \citenamefont {Dollfus}, \citenamefont {Aubry-Fortuna}, \citenamefont
  {Hesto},\ and\ \citenamefont {Osten}}]{2000SylvieSemiSciTech15}%
  \BibitemOpen
  \bibfield  {author} {\bibinfo {author} {\bibfnamefont {S.}~\bibnamefont
  {Galdin}}, \bibinfo {author} {\bibfnamefont {P.}~\bibnamefont {Dollfus}},
  \bibinfo {author} {\bibfnamefont {V.}~\bibnamefont {Aubry-Fortuna}}, \bibinfo
  {author} {\bibfnamefont {P.}~\bibnamefont {Hesto}},\ and\ \bibinfo {author}
  {\bibfnamefont {H.~J.}\ \bibnamefont {Osten}},\ }\bibfield  {title} {\bibinfo
  {title} {Band offset predictions for strained group iv alloys: Si1-x-ygexcy
  on si(001) and si1-xgex on si1-zgez(001)},\ }\href
  {https://doi.org/10.1088/0268-1242/15/6/314} {\bibfield  {journal} {\bibinfo
  {journal} {Semiconductor Science and Technology}\ }\textbf {\bibinfo {volume}
  {15}},\ \bibinfo {pages} {565} (\bibinfo {year} {2000})}\BibitemShut
  {NoStop}%
\end{thebibliography}%

\end{document}